# Robust Inference for Causal Mediation Analysis of Recurrent Event Data

Yan-Lin Chen[1], Yan-Hong Chen[2], Pei-Fang Su[3], Huang-Tz Ou[4], and An-Shun Tai[3]*

[1] Institute of Statistics, National Yang Ming Chiao Tung University, Hsin-Chu, Taiwan
[2] Institute of Statistical Science, Academia Sinica, Taipei, Taiwan
[3] Department of Statistics, National Cheng Kung University, Tainan, Taiwan.
[4] Department of Pharmacy, College of Medicine, National Cheng Kung University, Tainan, Taiwan

**\*Corresponding author**
An-Shun Tai
Department of Statistics, National Cheng Kung University, Tainan, Taiwan.
E-mail: ashtai@gs.ncku.edu.tw

SUMMARY: Recurrent events, including cardiovascular events, are commonly observed in biomedical studies. Researchers must understand the effects of various treatments on recurrent events and investigate the underlying mediation mechanisms by which treatments may reduce the frequency of recurrent events are crucial. Although causal inference methods for recurrent event data have been proposed, they cannot be used to assess mediation. This study proposed a novel methodology of causal mediation analysis that accommodates recurrent outcomes of interest in a given individual. A formal definition of causal estimands (direct and indirect effects) within a counterfactual framework is given, empirical expressions for these effects are identified. To estimate these effects, a semiparametric estimator with triple robustness against model misspecification was developed. The proposed methodology was demonstrated in a real-world application. The method was applied to measure the effects of two diabetes drugs on the recurrence of cardiovascular disease and to examine the mediating role of kidney function in this process.

# 1. Introduction

## 1.1 *Mediation problem with recurrent event data*

Advances in medical science in recent decades have led to a shift in the leading causes of human mortality from infectious diseases to noninfectious and chronic diseases (Grove and Hetzel, 1968; Heron, 2021). Chronic diseases, including cardiovascular disease, chronic obstructive pulmonary disease, and diabetes, are typically recurrent in nature and eventually lead to progressive debilitation or death. Consequently, these diseases have emerged as a central focus in contemporary health care. When recurrent event data are being examined, a common approach involves modeling the timing of the first event without regarding subsequent event data; however, this approach may not make efficient use of all available information. To address this issue, statisticians have devoted effort toward the development of methodologies for analyzing recurrent event data, yielding fruitful outcomes (Cook and Lawless, 2007).

Statistical models are typically developed to provide evidence of correlation and have found extensive use in medical research. However, as data science progresses, many medical studies have shifted their focus towards assessing the causal effect of a treatment on the outcome of interest and investigating the causal mediation mechanism that involves an intermediate variable. For example, an ongoing retrospective cohort study of diabetic patients with cardiovascular disease aims to investigate the causal mediation mechanism involving renal function. Specifically, the retrospective study examined the effects of two oral diabetes medications (dipeptidyl peptidase-4 [DPP-4] inhibitors and sodium-glucose cotransporter 2 [SGLT2] inhibitors) on type 2 diabetes mellitus in adults. SGLT2 inhibitors have been shown to provide therapeutic benefits in the treatment of cardiovascular disease (Davies et al., 2018; Savarese et al., 2016). This is particularly significant given that cardiovascular disease represents the primary cause of mortality in patients with diabetes (Sowers, Epstein and Frohlich, 2001; Strain and Paldánius, 2018). Additionally, dipeptidyl peptidase-4 (DPP-4) inhibitors have been demonstrated to exhibit cardiovascular safety (Davies et al., 2018;

Savarese et al., 2016). Thus, these classes of glucose-lowering medications contribute to addressing the cardiovascular concerns associated with diabetes management. Even though the causal effects of diabetes medications on the recurrence of cardiovascular disease have been confirmed, the underlying mediation mechanism is still understudied. Out of numerous clinical variables, renal function is considered a potential mediator of the effects of diabetes medication on cardiovascular risk. This is because several previous studies have identified a relationship between renal function, the use of diabetes medication, and cardiovascular risk. (Go et al., 2004; Kanasaki et al., 2014; Kojima et al., 2013; Ojima et al., 2015; Sharkovska et al., 2014; Vavrinec et al., 2014). Consequently, our primary research question is as follows: "To what extent does renal function mediate the rate of cardiovascular disease recurrence among patients with diabetes and a history of cardiovascular disease who are taking two diabetes medications?" To the best of our knowledge, no causal study has assessed the role of renal function in cardiovascular disease recurrence in patients taking two diabetes medications. This paper presents a novel methodology that applies mediation analysis to recurrent event outcomes.

### 1.2 *Related works*

Two related research frameworks exist for causal inferences involving time-to-event outcomes: causal inference models with recurrent event data and causal mediation methods with survival outcomes. In the first framework, three methods have been developed to estimate causal effects using recurrent event data. Gao and Zheng (2016) proposed a causal proportional intensity model for investigating dichotomous treatment variables. Their goal was to investigate the complier average causal effect, which represents the causal effect among treatment compliers when control patients are included in a randomized trial. Su, Steele and Shrier (2020) considered the average treatment effect on recurrent events in an entire population. They applied a semiparametric multiplicative rate model, an inverse probability weighted Nelson-Aalen estimator, and a doubly robust estimator for estimation purposes. The semiparametric multiplicative rate model imposes constraints, similar to the Cox proportional hazards assumption, on the structural framework across all time points. By contrast, Su, Platt and Plante (2022) used pseudo-observations to model recurrent events. An advantage of this approach is that the estimation remains unbiased as proportional assumption is violated. Although these methods are effective for estimating causal effects, they lack the capacity to assess mediation.

The second related research framework focuses on causal mediation analysis in the context of survival outcomes, where the outcome of interest is the time to a single event. Lange and Hansen (2011) suggested using the counterfactual hazard difference to measure natural direct and indirect effects in mediation analysis with survival outcomes. VanderWeele (2011) introduced measurements of direct and indirect effects on mean survival time for the accelerated failure time model and on the hazard ratio for the proportional hazards model when exposure–mediator interactions are being investigated. Tchetgen Tchetgen (2011) proposed multiply robust estimators for direct and indirect effects based on the Cox proportional hazards model and Aalen additive model. For recurrent event data, these methods can be applied to analyze mediation effects on the time to first recurrence. However, as previously noted, subsequent recurrent events following the first one also hold valuable information that should not be disregarded (de Vries et al., 2021; EMA, 2020; Rauch et al., 2018; Zhong and Cook, 2021).

1.3 *Contributions of the present study*

Methodologies for causal inference have been developed in previous studies exploring causation in recurrent event data; however, a methodology for conducting causal mediation analysis in the context of recurrent event data has yet to be fully established. Existing approaches for causal mediation analysis enable handling cases where the outcome variable is survival time, but these approaches do not consider recurrent event data. To address this research gap, the present study extended existing methods, which consider the difference in the number of events as a measure of causal effects, and proposed a novel methodology for the causal mediation analysis of recurrent event data. In this study, natural direct and indirect effects for recurrent event responses were systematically defined on the basis of counterfactual counting processes. Proposed effects were identified from empirical data through conventional assumptions (Pearl, 2001) in causal mediation analysis. Given the empirical expressions of the proposed effects, we developed semiparametric estimators of these effects. Such estimators possess appealing multiple robustness properties for estimation. Having multiple robustness necessitates only a subset of the working models to be correctly specified to lack bias. Comprehensive simulation studies were conducted to investigate the finite sample properties of the proposed robust estimators. Finally, the proposed method was applied to the motivating example (assessment of the mediating effects of renal function on cardiovascular disease recurrence in patients taking two diabetes medications) for illustration purposes.

The remainder of this study is organized as follows. Section 2 defines the notations and proposes the causal effects of interest. The required assumptions are provided, and the empirical expressions are derived through identification. Section 3 presents two methods for estimation. Section 4 presents the results of simulation studies, and in Section 5, data analysis results are shown. Finally, a discussion is provided in Section 6.

## 2. Mediation analysis with outcomes of a counting process outcome

### 2.1 *Notations and definitions*

Consider a dataset containing data from $n$ individuals. Let $T_1, \ldots, T_{K_i}$ denote the occurrence times of recurrent events for person $i$ $(i = 1, \ldots, n)$; $K_i$ is the total number of occurrences throughout person $i$'s lifetime. We introduce counting process notations, which enable the description of event occurrences in detail and have major roles in survival analysis and event history analysis (Andersen et al., 2012). The recurrent events can be represented using the counting process $\tilde{N}_i(t) = \sum_{j=1}^{K_i} I(T_j \leq t)$, which indicates the number of events up to time point $t$. We consider follow-up time $\tau$; right-censoring, with censoring time $T_i^C$; and the at-risk process $Y_i(t) = I(t \leq \min(\tau, T_i^C))$, where $I(\cdot)$ is an indicator function. The observed counting process is then $N_i(t) = \int_0^t Y_i(u) d\tilde{N}_i(u)$. Let $A_i$ represent the exposure or treatment, $M_i$ be the mediator of interest, and $\boldsymbol{C}_i$ be the set of measured baseline confounders. For simplicity, we suppress the subscript $i$ in the following context, provided that doing so does not cause confusion. The underlying structure of our mediation problem is illustrated by the causal directed acyclic graph in Figure 1.

In the present study, $A$ and $M$ are time-fixed variables measured prior to the time origin. To define the causal problem, we adopt a counterfactual framework (Neyman, 1923; Rubin, 1974), which is the dominant conceptual model in the literature on causal inference. Let $M(a)$ be the (counterfactual) mediator value if the exposure has been set to $a$; let $\tilde{N}(t; a, m)$ be the (counterfactual) outcome process if the exposure has been set to $a$ and the mediator has been set to $m$; and let $\tilde{N}(t; a, M(a^*))$ be the (counterfactual) outcome process if the exposure has been set to $a$ and the mediator has been set to the level it would have been had the exposure been set to $a^*$, for any $a$ and $a^*$. On the basis of these counterfactual outcomes, the effects corresponding to our causal problem are specified in the following section.

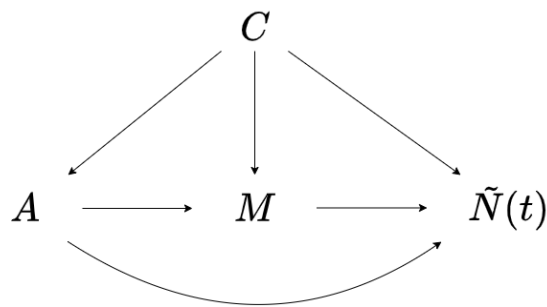

**Figure 1.** Causal directed acyclic graph for mediation analysis with a recurrent outcome.

## 2.2 Direct and indirect effects

We assumed that the exposure variable was binary ($A = 0$ or $1$). To assess mediation, we defined the natural direct effect (NDE) and natural indirect effect (NIE) with recurrent event data as follows:

$$\Delta_{\text{NDE}}(t) = E(\widetilde{N}(t; 1, M(0))) - E(\widetilde{N}(t; 0, M(0))) = \phi(t; 1,0) - \phi(t; 0,0)$$

$$\Delta_{\text{NIE}}(t) = (\widetilde{N}(t; 1, M(1))) - (\widetilde{N}(t; 1, M(0))) = \phi(t; 1,1) - \phi(t; 1,0)$$

Here, $\phi(t; a, a^*) \equiv E(\widetilde{N}(t; a, M(a^*)))$ is referred to as the mediation parameter in this study. The proposed definitions of the mediation parameter, NDE, and NIE are analogous to those provided by Pearl (2001). From the perspective of causal inference, NIE is used to evaluate the influence of exposure on the outcome process via its effect on the mediator. By contrast, NDE captures the effect of exposure on the outcome process that does not operate through the mediator. In a mediation analysis, researchers typically impose a composition assumption (VanderWeele and Vansteelandt, 2009) $\widetilde{N}(t; a) = \widetilde{N}(t; a, M(a))$, which means that for all $a$, intervening to set the exposure to $a$ results in the same counterfactual outcome process as if the intervention that involves setting the exposure to $a$ and setting the mediator to the level it would have been had the exposure been set to $a$. Given the composition assumption, NDE and NIE can be rewritten as $\phi(t; 1,0) - E(\widetilde{N}(t; 0))$ and $E(\widetilde{N}(t; 1)) - \phi(t; 1,0)$, respectively. Accordingly, the average causal effect with recurrent events, the sum of $\Delta_{\text{NDE}}(t)$ and $\Delta_{\text{NIE}}(t)$, is $E(\widetilde{N}(t; 1)) - E(\widetilde{N}(t; 0))$, which definition is identical to the causal effect proposed in previous studies (Su et al., 2020; Su et al., 2022).

## 2.3 Assumptions and identification

In principle, counterfactual variables are inherently unobservable, rendering the causal effects (NDE and NIE) defined earlier impossible to directly assess through observation. Therefore, a procedure for identifying NDE and NIE from empirical data is typically required in the context of causal analysis. Because NDE and NIE are defined as the difference between two mediation parameters, we focused on identification of the mediation parameter. To identify the mediation parameter $\phi(t; a, a^*)$ in this study, we made the following assumptions:

**(A1)** No unmeasured confounders between the exposure and outcome process.

$$\widetilde{N}(t; a, m) \perp\!\!\!\perp A \mid \boldsymbol{C}$$

**(A2)** No unmeasured confounders between the mediator and outcome process.

$$\widetilde{N}(t; a, m) \perp\!\!\!\perp M \mid A, \boldsymbol{C}$$

**(A3)** No unmeasured confounders between the exposure and mediator.

$$M(a) \perp\!\!\!\perp A \mid \boldsymbol{C}$$

**(A4)** Confounders between mediator and outcome process are not affected by exposure.

$$\widetilde{N}(t; a, m) \perp\!\!\!\perp M(a^*) \mid \boldsymbol{C}$$

**(A5)** The counterfactual mediator that intervened by setting exposure to $a$ is equal to the value of observed mediator when an individual actually receives exposure $a$.

$$M(a) = M \mid A = a$$

**(A6)** The counterfactual process that intervened by setting exposure to $a$ and mediator to $m$ is equal to the value of observed outcome process when an individual actually receives exposure $a$ and mediator $m$.

$$\widetilde{N}(t; a, m) = \widetilde{N}(t) \mid A = a, M = m$$

These assumptions can be viewed as an extension of conventional mediation assumptions (Pearl, 2001) for recurrent event data analysis. Specifically, assumptions (A1)–(A4) are known as exchangeability assumptions, and assumptions (A5) and (A6) are consistency assumptions. The presence of unmeasured confounders introduces a backdoor path, producing additional associations between variables and rendering causal interpretations of the estimated effect invalid (Hernán and Robins, 2018). As such, assumptions (A1)–(A3) indicate that $A$-$\widetilde{N}(t)$ confounders, $M$-$\widetilde{N}(t)$ confounders, and $A$-$M$ confounders must all be observed and collected in set $\boldsymbol{C}$. Assumption (A4), also known as the cross-world exchangeability assumption, requires that confounders between the mediator and outcome process are not affected by exposure. In the motivating example, renal function is investigated for its mediating role in the causal pathway from diabetes medication use to cardiovascular risk. Renal function was measured at a time as close to baseline as possible to avoid compromising cross-world exchangeability (VanderWeele and Vansteelandt, 2009). Related issues concerning these assumptions are discussed at the end of this paper. Given assumptions (A1)–(A6), the mediation parameter $\phi(t; a, a^*)$ is identified as $Q(t; a, a^*)$, a function of the distribution of observed data, where

$$Q(t; a, a^*) = \int_c \int_m E(\tilde{N}(t)|A = a, M = m, \boldsymbol{C} = \boldsymbol{c}) \mathrm{d}F_M(m|A = a^*, \boldsymbol{C} = \boldsymbol{c}) \, \mathrm{d}F_{\boldsymbol{C}}(\boldsymbol{c}).$$

$Q(t; a, a^*)$ is the empirical expression of $\phi(t; a, a^*)$ and is referred to as the recurrent event outcome mediation formula, where $F_M(m|A, \boldsymbol{C})$ and $F_{\boldsymbol{C}}(\boldsymbol{c})$ represent the conditional cumulative distribution function of the mediator and the joint cumulative distribution function of confounders, respectively. We emulated the seminal works on the mediation formula by Pearl (2001) and Robins and Greenland (1992) and further adapted the methods in their works to the context of recurrent event data analysis. A detailed description of the identification procedure is provided in Appendix 1. Consequently, NDE and NIE can be identified as follows:

$$\Delta_{NDE}(t) = Q(t; 1,0) - Q(t; 0,0)$$
$$= \int_c \int_m \Big( E(\tilde{N}(t)|A = 1, M = m, \boldsymbol{C} = \boldsymbol{c})$$
$$- E(\tilde{N}(t)|A = 0, M = m, \boldsymbol{C} = \boldsymbol{c}) \Big) \mathrm{d}F_M(m|A = 0, \boldsymbol{C} = \boldsymbol{c}) \mathrm{d}F_{\boldsymbol{C}}(\boldsymbol{c})$$

and

$$\Delta_{NIE}(t) = Q(t; 1,1) - Q(t; 1,0)$$
$$= \int_c \Big( \int_m E(\tilde{N}(t)|A = 1, M = m, \boldsymbol{C} = \boldsymbol{c}) \mathrm{d}F_M(m|A = 1, \boldsymbol{C} = \boldsymbol{c})$$
$$- \int_m E(\tilde{N}(t)|A = 1, M = m, \boldsymbol{C} = \boldsymbol{c}) \mathrm{d}F_M(m|A = 0, \boldsymbol{C} = \boldsymbol{c}) \Big) \mathrm{d}F_{\boldsymbol{C}}(\boldsymbol{c}).$$

## 3. Estimation using a proportional mean model

## 3.1 Regression-based estimation

The empirical formulas derived in the previous section enabled us to assess NIE and NDE through observations. These empirical formulas are linear functions of $Q(t; a, a^*)$. We then focused on developing the estimation procedure for $Q(t; a, a^*)$. We first considered a regression-based (RB) estimator. Typically, the models for the mean function of the recurrent events $E(\tilde{N}(t)|A, M, C)$ and the cumulative distribution function of the mediator $F_M(M|A, C)$ should be prespecified in a parametric or nonparametric manner. The specification of the parametric model for the mediator is flexible in the proposed estimation method. For illustration purposes, we assume that the mediator variable meets the normality assumption. The extension of this approach to a binary mediator is given in Appendix 2.

In modelling recurrent event data, if we assume that the covariates are $X = (A, M, C^T)^T$, the model can be specified as a Cox-type proportional mean function:

$$E(\tilde{N}(t)|A, M, C) = \Lambda_0(t) \exp(\boldsymbol{\beta_0^T X}),$$

where $\boldsymbol{\beta_0} = (\beta_{0A}, \beta_{0M}, \boldsymbol{\beta_{0C}^T})^T$ is a (*p*+2)-dimensional vector of regression parameters with respect to $X$, $\Lambda_0(t)$ is an unspecified baseline mean function, and *p* represents the number of confounders collected in $C$. The model states that an increase in the number of events over time $t$ is exponentially influenced by the multiplicative effects of covariates. During estimation, a conventional assumption known as conditionally independent censoring should be imposed to censored observations.

**(CA.1)** The occurrence of recurrent events is not associated with the censoring mechanism conditional on the exposure $A$, mediator $M$, and confounders $C$.

$$Y(u) \perp\!\!\!\perp d\tilde{N}(u)|A, M, C$$

Under conditionally independent censoring, the coefficient $\boldsymbol{\beta_0}$ in the proportional mean model can be estimated by solving the partial likelihood score equation:

$$U(\boldsymbol{\beta}) = \sum_{i=1}^{n} \int_0^\infty \left\{ \boldsymbol{X}_i - \frac{\sum_{j=1}^n Y_j(u) \boldsymbol{X}_i \exp(\boldsymbol{\beta}^{\mathrm{T}} \boldsymbol{X}_i)}{\sum_{j=1}^n Y_j(u) \exp(\boldsymbol{\beta}^{\mathrm{T}} \boldsymbol{X}_i)} \right\} \mathrm{d} N_i(u),$$

and the solution is denoted as $\widehat{\boldsymbol{\beta}}$. The baseline mean function $\Lambda_0(t)$ can be consistently estimated by the Breslow estimator:

$$\widehat{\Lambda}_0(t; \widehat{\boldsymbol{\beta}}) = \int_0^t \frac{\sum_{j=1}^n \mathrm{d} N_j(u)}{\sum_{j=1}^n Y_j(u) \exp(\widehat{\boldsymbol{\beta}}^{\mathrm{T}} \boldsymbol{X})}.$$

Upon model fitting, $Q(t; a, a^*)$ is estimated by the RB estimator

$$\widehat{Q}^{RB}(t; a, a^*) = \int_{\boldsymbol{c}} \int_m \widehat{\Lambda}_0(t; \widehat{\boldsymbol{\beta}}) \exp(\hat{\beta}_A a + \hat{\beta}_M m + \widehat{\boldsymbol{\beta}}_{\boldsymbol{C}}^{\mathrm{T}} \boldsymbol{c}) \, \mathrm{d}\widehat{F}_M(m | A = a^*, \boldsymbol{C} = \boldsymbol{c}) \, \mathrm{d}\widehat{F}_{\boldsymbol{C}}(\boldsymbol{c}),$$

where $\widehat{F}_M$ and $\widehat{F}_{\boldsymbol{C}}$ are the estimation of $F_M$ and $F_{\boldsymbol{C}}$, respectively. As previously indicated, $\widehat{F}_M$ can be derived through parametric or nonparametric methods, whereas the joint cumulative distribution function of confounders is usually estimated in a nonparametric method. In the motivating example, the mediator (renal function) is assumed to follow a normal distribution with mean $\theta_{00} + \theta_{0A} A + \boldsymbol{\theta}_{0\boldsymbol{C}}^{\mathrm{T}} \boldsymbol{C}$ and variance $\sigma_{0M}^2$. The maximum likelihood method is applied to the estimation of $\boldsymbol{\theta}_0 = (\theta_{00}, \theta_{0A}, \boldsymbol{\theta}_{0\boldsymbol{C}}^{\mathrm{T}}, \sigma_{0M}^2)$, and $\widehat{\boldsymbol{\theta}} = (\hat{\theta}_0, \hat{\theta}_A, \widehat{\boldsymbol{\theta}}_{\boldsymbol{C}}, \hat{\sigma}_M^2)$ denotes the corresponding estimator. Given the normal distribution for $M$, $\widehat{Q}^{RB}(t; a, a^*)$ can be simplified as

$$\widehat{Q}^{RB}(t; a, a^*) = \hat{\eta}(t) \times \exp(\hat{\beta}_A a + \hat{\beta}_M \hat{\theta}_A a^*),$$

where

$$\hat{\eta}(t) = \widehat{\Lambda}_0(t; \widehat{\boldsymbol{\beta}}) \cdot \int_{\boldsymbol{c}} \exp\left( (\hat{\theta}_0 + \widehat{\boldsymbol{\theta}}_{\boldsymbol{C}}^{T} \boldsymbol{C}) \hat{\beta}_M + \frac{1}{2} \hat{\beta}_M^2 \hat{\sigma}_M^2 + \widehat{\boldsymbol{\beta}}_{\boldsymbol{C}}^{T} \boldsymbol{C} \right) \mathrm{d}\widehat{F}_{\boldsymbol{C}}(\boldsymbol{c}),$$

which does not depend on $a$ and $a^*$. Consequently, the estimators of NDE and NIE have the following forms:

$$\widehat{\Delta}_{NDE}(t) = \hat{\eta}(t) \times \{ \exp(\hat{\beta}_A) - 1 \},$$

$$\widehat{\Delta}_{NIE}(t) = \hat{\eta}(t) \times \{ \exp(\hat{\beta}_A + \hat{\beta}_M \hat{\theta}_A) - \exp(\hat{\beta}_A) \}.$$

The asymptotic properties of $\widehat{\Delta}_{NDE}(t)$ and $\widehat{\Delta}_{NIE}(t)$ are determined by the property of $\widehat{Q}^{RB}(t; a, a^*)$. Because the estimators $\widehat{\Lambda}_0(t; \widehat{\boldsymbol{\beta}})$ and $\widehat{\boldsymbol{\beta}}$ converge almost surely to $\Lambda_0(t)$ and $\boldsymbol{\beta}_0$, respectively (Lin et al., 2000), and because $\hat{F}_M$ and $\hat{F}_C$ are consistent, the continuity of $Q(t; a, a^*)$ with respect to unknown parameters permits that $\hat{Q}^{RB}(t; a, a^*)$ preserves consistency (i.e., the estimates are asymptotically unbiased). In terms of statistical inference, Lin et al. (2000) derived the asymptotic variance of $\hat{\boldsymbol{\beta}}$ and $\hat{\Lambda}_0(t; \hat{\boldsymbol{\beta}})$ for the proportional mean model by using the empirical process theory, because the martingale central limit theorem was not applicable. On the basis of their result, the asymptotic variance of $\hat{Q}^{RB}(t; a, a^*)$ can theoretically be derived using the δ-method. To mitigate computational complexity, we used nonparametric bootstrapping to approximate the actual variance.

RB estimation is a straightforward strategy for NDE and NIE estimation, and as discussed herein, the proposed regression-based estimators of NDE and NIE are consistent. These estimators, however, are severely biased if any of the working models are misspecified. To address this problem, we proposed an alternative estimation approach that is robust against model misspecification.

## 3.2 *Triply robust estimation*

Building upon the work by Tchetgen Tchetgen and Shpitser (2012) on semiparametric efficient estimation for causal mediation analysis with time-independent outcomes, we proposed a triply robust (TR) estimator for $Q(t; a, a^*)$ as follows:

$$\hat{Q}^{TR}(t; a, a^*) = \int_0^t \mathbb{P}_n \left\{ \hat{Z}_i(u; a) \frac{\hat{f}_M(M_i|a^*, \boldsymbol{C}_i)}{\hat{f}_M(M_i|a, \boldsymbol{C}_i)} [d\tilde{N}_i(u) - \hat{E}(d\tilde{N}(u)|a, M_i, \boldsymbol{C}_i)] \right.$$
$$\left. + \hat{Z}_i(u; a^*)[\hat{E}(d\tilde{N}(u)|a, M_i, \boldsymbol{C}_i) - d\hat{Q}(u; a, a^*|\boldsymbol{C}_i)] + d\hat{Q}(u; a, a^*|\boldsymbol{C}_i) \right\},$$

where $\mathbb{P}_n\{\cdot\} = n^{-1} \sum_{i=1}^n [\cdot]_i$,

$$\hat{Z}_i(u; a) = \frac{I(A_i = a)\hat{w}_i Y_i(u)}{\frac{1}{n} \sum_{j=1}^n I(A_j = a)\hat{w}_j Y_j(u)},$$

$$d\hat{Q}(u; a, a^*|\boldsymbol{C}) = \int_m \hat{E}(d\tilde{N}(u)|A = a, M = m, \boldsymbol{C}) d\hat{F}_M(m|A = a^*, \boldsymbol{C}), \text{ and}$$

$$\widehat{w}_i = \frac{A_i}{\widehat{E}(A|\boldsymbol{C}_i)} + \frac{1-A_i}{1-\widehat{E}(A|\boldsymbol{C}_i)}.$$

By setting $a = a^*$, the proposed TR estimator, $\widehat{Q}^{TR}(t; a, a)$, is simplifies to

$$\int_0^t \mathbb{P}_n\{\widehat{Z}_i(u;a)\mathrm{d}\widetilde{N}_i(u)\}\} + \int_0^t \mathbb{P}_n\left\{\left(1-\widehat{Z}_i(u;a)\right)\mathrm{d}\widehat{Q}(u;a,a^*|\boldsymbol{C}_i)\right\}.$$

This is equivalent to the doubly robust estimator of the causal effect for recurrent event data presented by Su et al. (2020). The unbiasedness of the proposed TR estimator is demonstrated in a subsequent section. The TR estimator combines three estimators, namely propensity score weighting (PSW) and inverse probability weighting (IPW) estimators and the RB estimator discussed in Section 3.1. The PSW estimator is constructed as follows:

$$\widehat{Q}^{PSW}(t; a, a^*) = \int_0^t \mathbb{P}_n\{\widehat{Z}_i(u;a^*)\widehat{E}(\mathrm{d}\widetilde{N}(u)|a, M_i, \boldsymbol{C}_i)\}.$$

The unbiasedness of the PSW estimator relies on the proper specification of both the recurrent event outcome and exposure models. The IPW method is commonly used for constructing estimators for causal mediation analysis; our IPW estimator is presented as follows:

$$\widehat{Q}^{IPW}(t; a, a^*) = \int_0^t \mathbb{P}_n\left\{\widehat{Z}_i(u;a)\frac{\widehat{f}_M(M_i|a^*,\boldsymbol{C}_i)}{\widehat{f}_M(M_i|a,\boldsymbol{C}_i)}\mathrm{d}\widetilde{N}_i(u)\right\}.$$

The correct specification of the mediator and exposure models ensures the unbiasedness of the IPW estimator. In contrast to the assumptions made in PSW and IPW estimators, the RB estimator $\widehat{Q}^{RB}(t; a, a^*)$ is unbiased under the assumption that the recurrent event outcome and mediator models are correctly specified. Each of these estimators (PSW, IPW, and RB) has its own set of model assumptions. Let $\mathbb{M}_{A,\widetilde{N}}$, $\mathbb{M}_{A,M}$, and $\mathbb{M}_{M,\widetilde{N}}$ represent the sets of model assumptions as follows:

(i) $\mathbb{M}_{A,\widetilde{N}}$, models for $A$ and $\widetilde{N}$ are correctly specified;

(ii) $\mathbb{M}_{A,M}$, models for $A$ and $M$ are correctly specified;

(iii) $\mathbb{M}_{M,\widetilde{N}}$, models for $M$ and $\widetilde{N}$ are correctly specified.

Furthermore, $\mathbb{M}_I$ and $\mathbb{M}_U$ represent the intersection and union, respectively, of the sets $\mathbb{M}_{A,\widetilde{N}}$, $\mathbb{M}_{A,M}$, and $\mathbb{M}_{M,\widetilde{N}}$. Under regular conditions, PSW, IPW, and RB estimators are consistent and asymptotically normal (CAN) in $\mathbb{M}_{A,\tilde{N}}$, $\mathbb{M}_{A,M}$, and $\mathbb{M}_{M,\tilde{N}}$, respectively. However, in practical scenarios where the number of confounders is large, whether the models can be properly specified is typically uncertain. By contrast, the TR estimator guarantees valid inferences about NDE and NIE as long as at least one of the sets of model assumptions ($\mathbb{M}_{A,\tilde{N}}$, $\mathbb{M}_{A,M}$, and $\mathbb{M}_{M,\tilde{N}}$) is satisfied. The TR estimator has higher tolerance against model misspecification than the naïve RB estimator; in $\mathbb{M}_U$, the TR estimator is CAN. The characteristics of the proposed TR estimator are explicitly outlined and demonstrated in a subsequent theorem.

As previously stated, the RB method assumes independence of censoring in relation to recurrent event outcomes (CA.1 assumption). The TR estimation further includes two assumptions regarding censoring in relation to the mediator and exposure, as follows:

**(CA.2)** The mediator is not associated with the censoring mechanism conditional on the exposure and confounders.

$$Y(u) \perp\!\!\!\perp M | A, \boldsymbol{C}$$

**(CA.3)** Confounders are not associated with the censoring mechanism conditional on the exposure.

$$Y(u) \perp\!\!\!\perp C | A$$

Importantly, assumptions (CA.1), (CA.2), and (CA.3) do not require the censoring mechanism to be completely independent of the exposure or treatment assignment ($A$). The estimators proposed in this article remain valid even when the censoring mechanism is related to $A$. The TR estimation necessitates assumptions (CA.1) and (CA.2); however, (CA.3) may not always be a requirement. More specifically, if the mediator and recurrent event outcome models are both correctly specified, as per the conditions outlined in $\mathbb{M}_I$ or $\mathbb{M}_{M,\tilde{N}}$, then the TR estimator can accurately infer NDE and NIE without relying on assumption (CA.3). This implies that a trade-off exists between the plausibility of assumption (CA.3) and the confidence in the correct specification of the exposure model. A summary of the assumptions necessary for the RB estimator and TR estimator in four different scenarios is provided in Table 1.

**Table 1.** Comparison of the premised censoring assumption for the regression-based estimator and the triply robust estimator. A circle was used to represent the requirements for the model assumptions.

| Estimator | Censoring Assumption | | |
|---|---|---|---|
| | $Y(u) \perp\!\!\!\perp d\tilde{N}(u)|A,M,C$ | $Y(u) \perp\!\!\!\perp M|A,C$ | $Y(u) \perp\!\!\!\perp C|A$ |
| RB estimator in $\mathbb{M}_{M,\tilde{N}}$ | ○ | | |
| TR estimator in $\mathbb{M}_I$ | ○ | ○ | |
| TR estimator in $\mathbb{M}_{M,\tilde{N}}$ | ○ | ○ | |
| TR estimator in $\mathbb{M}_{A,\tilde{N}}$ | ○ | ○ | ○ |
| TR estimator in $\mathbb{M}_{A,M}$ | ○ | ○ | ○ |

*Abbreviations: RB, regression-based; TR, triply robust.*

Prior to demonstrating that the TR estimator in $\mathbb{M}_U$ is CAN under the stated assumptions, the TR estimation requires prespecification of the exposure, mediator, and outcome models. As described in Section 3.1, the mediator and recurrent event outcome are assumed to follow a linear regression model and a Cox-type mean model, respectively in the real data application. For exposure, we used a logistic regression model with mean $E(A|C) = \text{expit}(\alpha_0^T(1,C)^T) = \text{expit}(\alpha_{00} + \alpha_{0C}^T C)$ in both the simulation study and real data application. The parameters in the logistic regression model were estimated using the maximum likelihood approach. The TR estimator can be reformulated as follows:

$$\hat{Q}^{TR}\left(t; a, a^*; \hat{\alpha}, \hat{\theta}, \hat{\beta}, \hat{\Lambda}_0(t; \hat{\beta})\right)$$
$$= \int_0^t \mathbb{P}_n \left\{ \hat{Z}_i(u; a; \hat{\alpha}) \frac{\hat{f}_M(M_i|a^*, C_i; \hat{\theta})}{\hat{f}_M(M_i|a, C_i; \hat{\theta})} \left[ d\tilde{N}_i(u) - \hat{E}\left(d\tilde{N}(u) \Big| a, M_i, C_i; \hat{\beta}, \hat{\Lambda}_0(t; \hat{\beta})\right) \right] \right.$$
$$+ \hat{Z}_i(u; a^*; \hat{\alpha}) \left[ \hat{E}\left(d\tilde{N}(u) \Big| a, M_i, C_i; \hat{\beta}, \hat{\Lambda}_0(t; \hat{\beta})\right) \right.$$
$$\left. - d\hat{Q}\left(u; a, a^* \Big| C_i; \hat{\theta}, \hat{\beta}, \hat{\Lambda}_0(t; \hat{\beta})\right) \right] + d\hat{Q}\left(u; a, a^* \Big| C_i; \hat{\theta}, \hat{\beta}, \hat{\Lambda}_0(t; \hat{\beta})\right) \left. \right\}.$$

Under the assumptions of identification and independent censoring, the TR estimator can be shown to be an CAN estimator for $Q(t; a, a^*)$ in $\mathbb{M}_U$. The unbiasedness and asymptotic properties of this estimator are summarized in the following theorem; the proof is available in Appendix 3.

**THEOREM 1: Consistency and asymptotic properties**

*Suppose that identification assumptions (A1)–(A6) and independent censoring assumptions (CA.1)–(CA.3) hold.*

*(a) $\hat{Q}^{TR}\left(t; a, a^*; \hat{\boldsymbol{\alpha}}, \hat{\boldsymbol{\theta}}, \hat{\boldsymbol{\beta}}, \hat{\Lambda}_0(t; \hat{\boldsymbol{\beta}})\right)$ is consistent in $\mathbb{M}_U$.*

*(b) $\sqrt{n}(\hat{Q}^{TR}\left(t; a, a^*;; \hat{\boldsymbol{\alpha}}, \hat{\boldsymbol{\theta}}, \hat{\boldsymbol{\beta}}, \hat{\Lambda}_0(t; \hat{\boldsymbol{\beta}})\right) - Q(t; a, a^*))$ is asymptotically normal with mean zero and variance $E((\Gamma_i(t, a, a^*))^2)$ in $\mathbb{M}_U$, where*
$\Gamma_i(t, a, a^*) = F_1(t; a, a^*; \boldsymbol{\alpha}^*, \boldsymbol{\theta}^*, \boldsymbol{\beta}^*, \Lambda_0^*(u))$

$$\times E\left(\frac{\exp(\boldsymbol{\alpha}^{*T}\boldsymbol{V}_1)\boldsymbol{V}_1\boldsymbol{V}_1^T}{(1+\exp(\boldsymbol{\alpha}^T\boldsymbol{V}_1))^2}\right)^{-1} \boldsymbol{V}_{1,i}\left(A_i - \frac{\exp(\boldsymbol{\alpha}^{*T}\boldsymbol{V}_{1,i})}{1+\exp(\boldsymbol{\alpha}^{*T}\boldsymbol{V}_{1,i})}\right)$$
$$+ F_2(t; a, a^*; \boldsymbol{\alpha}^*, \boldsymbol{\theta}^*, \boldsymbol{\beta}^*, \Lambda_0^*(u))\mathbb{A}_n^M(\boldsymbol{\theta}^*)^{-1}\mathbb{U}_i^M(\boldsymbol{\theta}^*)$$
$$+ \Psi_1(\boldsymbol{\alpha}^*, \boldsymbol{\theta}^*, \boldsymbol{\beta}^*, \Lambda_0^*(t))\mathbb{A}_n^{\tilde{N}}(\boldsymbol{\beta}^*)^{-1}\mathbb{U}_i^{\tilde{N}}(\boldsymbol{\beta}^*, t)$$
$$+ \int_0^t \Psi_2(u; \boldsymbol{\alpha}^*, \boldsymbol{\theta}^*, \boldsymbol{\beta}^*) d\Xi_i(u; \boldsymbol{\beta}^*, \Lambda_0^*(u)) + F_{3,i}(t; a, a^*; \boldsymbol{\alpha}^*, \boldsymbol{\theta}^*, \boldsymbol{\beta}^*, \Lambda_0^*(u))$$
$$+ o_p(1),$$

where $\boldsymbol{\alpha}^*, \boldsymbol{\theta}^*, \boldsymbol{\beta}^*$, and $\Lambda_0^*(u)$ are the actual values to which the estimator converges under the given model [i.e., $\hat{\boldsymbol{\alpha}} \xrightarrow{p} \boldsymbol{\alpha}^*, \hat{\boldsymbol{\theta}} \xrightarrow{p} \boldsymbol{\theta}^*, \hat{\boldsymbol{\beta}} \xrightarrow{p} \boldsymbol{\beta}^*$, and $\hat{\Lambda}_0(t; \hat{\boldsymbol{\beta}}) \xrightarrow{p} \Lambda_0^*(t)$ ], and $\boldsymbol{\alpha}^* = \boldsymbol{\alpha}_0, \boldsymbol{\theta}^* = \boldsymbol{\theta}_0, \boldsymbol{\beta}^* = \boldsymbol{\beta}_0$, and $\Lambda_0^*(t) = \Lambda_0(t)$ if the respective models are correctly specified. The definitions of $F_1, F_2, F_{3,i}, \boldsymbol{V}_1, \mathbb{A}_n^M, \mathbb{A}_n^{\tilde{N}}, \mathbb{U}_i^M, \mathbb{U}_i^{\tilde{N}}, \Psi_1, \Psi_2$, and $\Xi_i$ are detailed in Appendix 3. Their placement in the supporting information rather than in the main text was deemed necessary to retain the readability of this article.

According to Theorem 1, $\hat{Q}^{TR}(t; 1,0) - \hat{Q}^{TR}(t; 0,0)$ and $\hat{Q}^{TR}(t; 1,1) - \hat{Q}^{TR}(t; 1,0)$ are CAN for NDE and NIE, respectively. The asymptotic variances of these estimators can be calculated as $E\left((\Gamma_i(t, 1,0) - \Gamma_i(t, 0,0))^2\right)$ and $E\left((\Gamma_i(t, 1,1) - \Gamma_i(t, 1,0))^2\right)$. However, the empirical estimators of the asymptotic variances do not have multiply robust properties. These variance estimators may be inconsistent under $\mathbb{M}_{A,\tilde{N}}, \mathbb{M}_{A,M}$, or $\mathbb{M}_{M,\tilde{N}}$, even if the estimator $\hat{Q}^{TR}(t; a, a^*)$ remains consistent. Therefore, nonparametric bootstrap methods or similar resampling techniques for variance estimation are recommended in practical applications.

## 4. Numerical studies

In this section, we present four simulations to evaluate the performance of the proposed method. The settings for each simulation are provided, and the results of each experiment are presented in different subsections. Unless otherwise noted, the simulation settings were as follows: for $i = 1, \ldots, n$, the baseline confounders $C_{1,i}$ and $C_{2,i}$ were respectively distributed in a standard normal distribution with mean 0 and variance 1, and a Bernoulli distribution with a probability of success 0.6. The exposure $A_i$ and mediator $M_i$ followed a Bernoulli distribution and normal distributions, respectively, under the following mechanisms:

$$A_i \sim \text{Bernoulli}\left(\text{expit}(1 + C_{1,i} - 2C_{2,i})\right), \text{and}$$

$$M_i \sim \text{Normal}(\mu = 3 - A_i - C_{1,i} + 1.5C_{2,i}, \sigma^2 = 2).$$

The recurrent outcome model was expressed as follows:

$$\Lambda_i(t) = 0.05t \times \exp(A_i + 0.1M_i + 0.2C_{1,i} - 0.22C_{2,i}).$$

All parameters were chosen to resemble our motivating example. The average number of occurrences by the end of the follow-up period ($\tau = 24$ months) was 2.042. The results were displayed at four time points: 4.8, 9.6, 14.4, and 19.2 months, corresponding to the 20th, 40th, 60th, and 80th percentiles of the follow-up period.

*Experiment 1: Robustness of the TR estimator under different scenarios*

The first simulation experiment aimed to validate the robustness of the proposed TR estimator under scenarios $\mathbb{M}_I$, $\mathbb{M}_{A,\widetilde{N}}$, $\mathbb{M}_{A,M}$, and $\mathbb{M}_{M,\widetilde{N}}$ with a sample size of 1000 by comparison with the RB estimator. To examine the performance of the proposed method when the models were misspecified, we introduced measurement errors into the corresponding models. For example, if the model for the mediator is misspecified, it will result in $\boldsymbol{\theta}^* = \boldsymbol{\theta_0} + \boldsymbol{\varepsilon}$, where $\boldsymbol{\varepsilon}$ is a multivariate normal variable with a nonzero mean, leading to bias. The actual implementation details are presented in Appendix 4. The results of the empirical bias and the empirical standard error for NDE and NIE estimations are presented in Table 2. As anticipated, the TR estimator is empirically unbiased regardless of which working model is misspecified, whereas the RB estimator only performs well under $\mathbb{M}_I$ and $\mathbb{M}_{M,\widetilde{N}}$.

**Table 2.** Empirical bias and empirical standard error for the estimated natural direct effect and natural indirect effect for the regression-based estimator and the triply robust estimator under $\mathbb{M}_I$, $\mathbb{M}_{A,\widetilde{N}}$, $\mathbb{M}_{A,M}$, and $\mathbb{M}_{M,\widetilde{N}}$. The four time points considered are 20th, 40th, 60th, and 80th percentiles of the follow-up period.

|  |  | NDE (20%) | | NDE (40%) | | NDE (60%) | | NDE (80%) | |
|---|---|---|---|---|---|---|---|---|---|
|  |  | Bias ($\times 10^{-3}$) | ESE | Bias ($\times 10^{-3}$) | ESE | Bias ($\times 10^{-3}$) | ESE | Bias ($\times 10^{-3}$) | ESE |
| $\mathbb{M}_I$ | TR | -0.674 | 0.076 | 2.454 | 0.117 | 7.472 | 0.152 | 10.767 | 0.180 |
|  | RB | 3.386 | 0.040 | 7.271 | 0.073 | 10.427 | 0.106 | 12.871 | 0.137 |
| $\mathbb{M}_{A,\widetilde{N}}$ | TR | -0.726 | 0.088 | 2.624 | 0.135 | 8.825 | 0.172 | 12.020 | 0.204 |
|  | RB | 36.931 | 0.043 | 74.396 | 0.080 | 111.085 | 0.117 | 147.038 | 0.153 |
| $\mathbb{M}_{A,M}$ | TR | 0.229 | 0.077 | 4.151 | 0.122 | 9.810 | 0.161 | 13.763 | 0.194 |
|  | RB | 83.590 | 0.055 | 167.794 | 0.105 | 251.103 | 0.154 | 333.618 | 0.202 |
| $\mathbb{M}_{M,\widetilde{N}}$ | TR | -0.049 | 0.067 | 3.786 | 0.103 | 8.399 | 0.135 | 11.522 | 0.163 |
|  | RB | 3.386 | 0.040 | 7.271 | 0.073 | 10.427 | 0.106 | 12.871 | 0.137 |
|  |  | NIE (20%) | | NIE (40%) | | NIE (60%) | | NIE (80%) | |
|  |  | Bias ($\times 10^{-3}$) | ESE | Bias ($\times 10^{-3}$) | ESE | Bias ($\times 10^{-3}$) | ESE | Bias ($\times 10^{-3}$) | ESE |
| $\mathbb{M}_I$ | TR | -0.181 | 0.036 | 0.854 | 0.057 | -0.459 | 0.079 | 0.365 | 0.098 |
|  | RB | -0.382 | 0.016 | -0.851 | 0.033 | -1.206 | 0.049 | -1.49 | 0.065 |
| $\mathbb{M}_{A,\widetilde{N}}$ | TR | -0.069 | 0.052 | 0.829 | 0.078 | -1.624 | 0.103 | -0.657 | 0.124 |
|  | RB | -22.03 | 0.021 | -44.165 | 0.041 | -66.165 | 0.062 | -88.083 | 0.083 |
| $\mathbb{M}_{A,M}$ | TR | -1.121 | 0.042 | -0.939 | 0.071 | -2.900 | 0.104 | -2.734 | 0.135 |
|  | RB | -60.735 | 0.030 | -121.638 | 0.060 | -182.329 | 0.089 | -242.885 | 0.119 |
| $\mathbb{M}_{M,\widetilde{N}}$ | TR | -0.219 | 0.033 | 0.142 | 0.051 | -0.714 | 0.071 | 0.107 | 0.088 |
|  | RB | -0.382 | 0.016 | -0.851 | 0.033 | -1.206 | 0.049 | -1.49 | 0.065 |

*Abbreviations: ESE, empirical standard error; RB, regression-based; TR, triply robust; NDE, natural direct effect; NIE, natural indirect effect.*

*Experiment 2: Performance of the TR estimator with a low occurrence rate*

The second simulation experiment assessed the performance of the TR estimator when the occurrence rate was varied. This was achieved by introducing a scalar parameter ($\zeta$) to control the occurrence rate as follows:

$$E\left(\tilde{N}_i(t)\right) = \zeta \times \Lambda_i(t).$$

$\zeta$ was set to three values: 1, 2, and 3, with $\zeta = 1$ representing the standard setting. The average occurrences during the 24-month follow-up period for $\zeta = 1$, 2, and 3 were 2.042, 4.084, and 6.126, respectively. The results were displayed for different sample sizes, including $n = 200$, 500, and 1,000. As shown in Table 3, as the occurrence rate increased, the variance of the estimation increased. However, the accuracy of the TR estimator remained satisfactory even with a small sample size of 200.

**Table 3.** Empirical bias and empirical standard error for the estimated natural direct effect and natural indirect effect determined using the triply robust estimator under $\mathbb{M}_U$. The cross-tabulations show the variation in performance of the triply robust method at different sample sizes and occurrence rates. The four time points considered are 20th, 40th, 60th, and 80th percentiles of the follow-up period.

| Bias ($10^{-3}$) [ESE] | | NDE | | | NIE | | |
|---|---|---|---|---|---|---|---|
| Time point | $\zeta$ | n = 200 | n = 500 | n = 1000 | n = 200 | n = 500 | n = 1000 |
| 20% | 1 | 1.498 | -0.950 | -0.674 | -3.852 | -0.081 | -0.181 |
| | | [0.1858] | [0.1123] | [0.0758] | [0.0866] | [0.0537] | [0.0365] |
| | 2 | 0.681 | -3.901 | -2.098 | 1.706 | 0.746 | 1.953 |
| | | [0.2451] | [0.1564] | [0.1093] | [0.1233] | [0.0806] | [0.0528] |
| | 3 | 3.809 | 4.589 | 1.572 | -2.323 | -0.482 | 0.737 |
| | | [0.3108] | [0.1883] | [0.1340] | [0.1740] | [0.1032] | [0.0698] |
| 40% | 1 | -0.101 | 2.317 | 2.454 | -7.648 | 1.949 | 0.854 |
| | | [0.2566] | [0.1662] | [0.1166] | [0.1316] | [0.0842] | [0.0568] |
| | 2 | -0.244 | -10.081 | -0.115 | -5.775 | 4.190 | 5.122 |
| | | [0.3654] | [0.2212] | [0.1560] | [0.2048] | [0.1344] | [0.0900] |
|  | 3 | 14.567 | -0.609 | 5.205 | -8.997 | -0.094 | 2.410 |
|  |  | [0.4602] | [0.2732] | [0.2043] | [0.2838] | [0.1755] | [0.1236] |
| 60% | 1 | -1.053 | 3.022 | 7.472 | -8.442 | 3.798 | -0.459 |
|  |  | [0.3261] | [0.2081] | [0.1516] | [0.1728] | [0.1115] | [0.0791] |
|  | 2 | 3.275 | -5.255 | 1.019 | -1.174 | 3.725 | 5.389 |
|  |  | [0.4428] | [0.2913] | [0.1978] | [0.2708] | [0.1822] | [0.1255] |
|  | 3 | 32.791 | 2.929 | 11.244 | -7.871 | 3.718 | 3.846 |
|  |  | [0.6062] | [0.3467] | [0.2606] | [0.3872] | [0.2431] | [0.1753] |
| 80% | 1 | -6.208 | -6.277 | 10.767 | -6.157 | 6.158 | 0.365 |
|  |  | [0.3962] | [0.2509] | [0.1801] | [0.2181] | [0.1406] | [0.0977] |
|  | 2 | 6.934 | -2.694 | 3.666 | -3.008 | 5.065 | 5.637 |
|  |  | [0.5285] | [0.3608] | [0.2424] | [0.3457] | [0.2366] | [0.1629] |
|  | 3 | 23.361 | 7.774 | 16.212 | -5.468 | 7.137 | 3.936 |
|  |  | [0.7174] | [0.4314] | [0.3142] | [0.4946] | [0.3224] | [0.2292] |

*Abbreviations: ESE, empirical standard error; NDE, natural direct effect; NIE, natural indirect effect.*

*Experiment 3: Behavior of the TR estimator under early censoring*

The third simulation experiment sought to evaluate the performance of the TR estimator under excessive early censoring of individuals, a common scenario in practice. A mean proportion of censoring time (PC) was defined to measure the extent of censoring:

$$\mathrm{PC} = E\left(\frac{\tau - T_i^C}{\tau} I(T_i^C < \tau)\right),$$

where τ was 24 in this experiment. The focus was on the effect of the PC on the performance of the TR estimator. The results were presented for a sample size of 1000. We considered four PC values: 0%, 15%, 30%, and 50%. PC values were coerced to 15%, 30%, and 50% by having $T_i^C$ follow a uniform distribution over the interval $(0, x)$ where $x$ was 80, 40, and 24, respectively, and to 0% by having $T_i^C = \tau$. The results of this experiment are summarized in Table 4. A moderate increase in PC did not affect the accuracy of the proposed estimator, but the empirical standard error may have slightly increased due to the completely random censoring time $T_i^C$.

**Table 4.** Empirical bias and empirical standard error for the estimated natural direct effect and natural indirect effect using the triply robust estimator under $\mathbb{M}_U$ with a different mean proportion of censored time. The four time points considered are 20th, 40th, 60th, and 80th percentiles of the follow-up period.

| PC | NDE (20%) Bias ($10^{-3}$) | ESE | NDE (40%) Bias ($10^{-3}$) | ESE | NDE (60%) Bias ($10^{-3}$) | ESE | NDE (80%) Bias ($10^{-3}$) | ESE |
|---|---|---|---|---|---|---|---|---|
| 0%  | -0.707 | 0.074 | 1.77  | 0.109 | -0.707 | 0.074 | 1.77  | 0.109 |
| 15% | 2.78   | 0.079 | 4.45  | 0.111 | 2.78   | 0.079 | 4.45  | 0.111 |
| 30% | 5.06   | 0.079 | 0.747 | 0.115 | 5.06   | 0.079 | 0.747 | 0.115 |
| 50% | 1.27   | 0.075 | 1.58  | 0.117 | 1.27   | 0.075 | 1.58  | 0.117 |

| PC | NIE (20%) Bias ($10^{-3}$) | ESE | NIE (40%) Bias ($10^{-3}$) | ESE | NIE (60%) Bias ($10^{-3}$) | ESE | NIE (80%) Bias ($10^{-3}$) | ESE |
|---|---|---|---|---|---|---|---|---|
| 0%  | 0.865  | 0.035 | 1.30  | 0.057 | 0.865  | 0.035 | 1.30  | 0.057 |
| 15% | -1.96  | 0.036 | -1.58 | 0.053 | -1.96  | 0.036 | -1.58 | 0.053 |
| 30% | -2.22  | 0.036 | -1.99 | 0.057 | -2.22  | 0.036 | -1.99 | 0.057 |
| 50% | 0.960  | 0.037 | 1.11  | 0.058 | 0.960  | 0.037 | 1.11  | 0.058 |

*Abbreviations: ESE, empirical standard error; PC, mean proportion of censored time; NDE, natural direct effect; NIE, natural indirect effect.*

*Experiment 4: Effect of the violation of independent censoring assumptions*

The fourth and final simulation experiment investigated how the violation of each independent censoring assumption affected the performance of the TR estimator under various sets of working models. Three scenarios where the censoring time depended on the exposure, confounders, or mediator were considered in this simulation, leading to a failure of assumption (CA.2) when it was dependent on the mediator or assumption (CA.3) when it was dependent on the confounders. Each scenario was evaluated under the assumptions of $\mathbb{M}_I$, $\mathbb{M}_{A,\widetilde{N}}$, $\mathbb{M}_{A,M}$, and $\mathbb{M}_{M,\widetilde{N}}$, separately. The sample size was 3000. The boxplots of empirical biases under different scenarios are presented in Figure 2. As shown in Figure 2(A), the TR estimator was not influenced by the exposure's involvement in the censoring mechanism, which confirms the theoretical conclusion that the censoring mechanism can depend on the exposure or treatment assignment. By contrast, when the censoring mechanism was related to the mediator, the TR estimator failed to estimate NDE and NIE accurately (Figure 2(B)). Similarly, the proposed method produced biased estimates under $\mathbb{M}_{A,\widetilde{N}}$ and $\mathbb{M}_{A,M}$ when the censoring mechanism was a function of the confounders (Figure 2(C)). However, we observed that the estimator remained empirically unbiased under $\mathbb{M}_I$ and $\mathbb{M}_{M,\widetilde{N}}$, implying that assumption (CA.3) can be disregarded if the mediator and outcome models are both correctly specified. The results obtained are consistent with the theoretical property of the TR estimator.

In conclusion, the four simulation experiments presented in this section offer valuable insights into the performance of the proposed TR estimator under various scenarios, occurrence rates, censoring levels, and violations of independent censoring assumptions. The results confirm the robustness and accuracy of the TR estimator, underscoring its potential utility in practice.

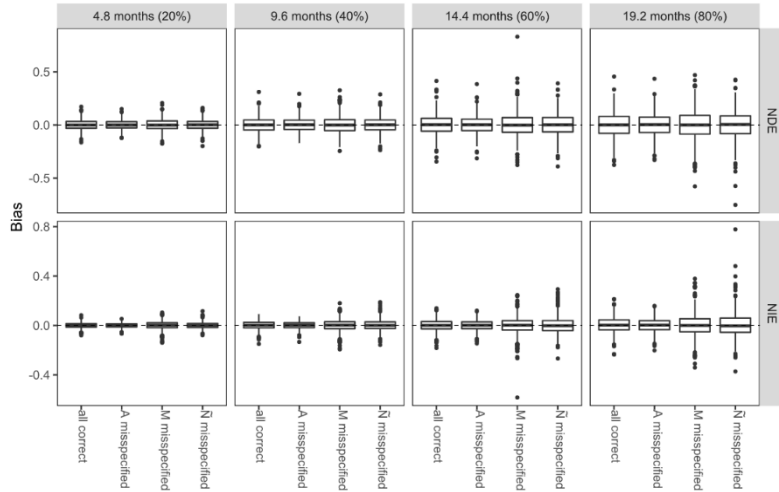

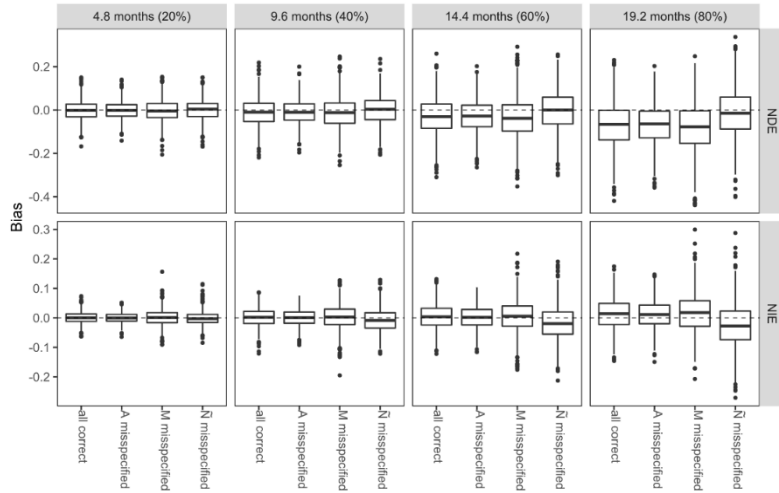

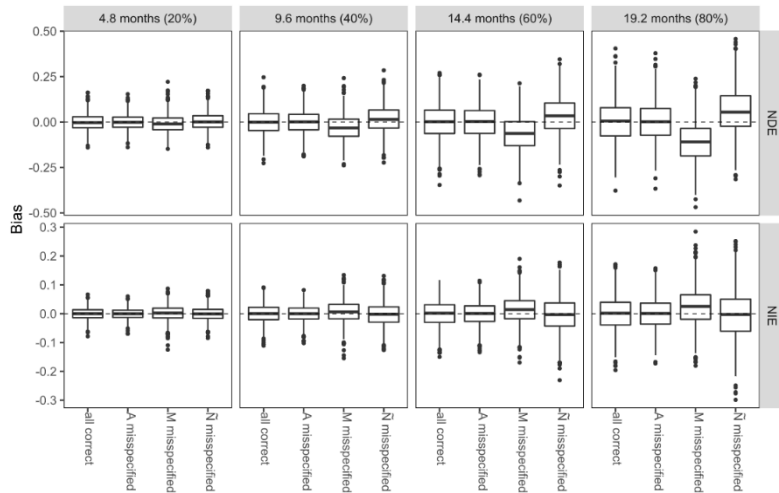

**Figure 2.** Boxplot of an empirical bias for a triply robust estimator (A) when the censoring mechanism is a function of exposure, (B) when the censoring mechanism is a function of mediator (assumption CA.2 is not valid), and (C) when the censoring mechanism is a function of confounders (assumption CA.3 is not valid).

# 5. Real data analysis

To achieve optimal glucose levels, most patients with type 2 diabetes require treatment beyond metformin monotherapy (Association, 2017; McCoy et al., 2021). Second-line options include DPP-4 inhibitors and SGLT2 inhibitors. DPP-4 inhibitors enhance insulin secretion by inhibiting the breakdown of the incretin glucagon-like peptide-1, which consequently lowers blood glucose levels (Drucker, 2003; Liu et al., 2012). SGLT2 inhibitors attenuate blood glucose levels by blocking glucose reabsorption in renal tubules (Vallon, 2015). Cardiovascular disease is a primary cause of mortality among patients with diabetes mellitus (Sowers et al., 2001; Strain and Paldánius, 2018). Therefore, evaluating the cardiovascular implications of SGLT2 and DPP-4 inhibitors is crucial for making informed treatment decisions for patients at high cardiovascular risk, ensuring that physicians and patients have a comprehensive understanding of the associated benefits and risks.

Moreover, investigating the underlying mechanisms of action for oral diabetes medications and their effects on cardiovascular disease risk is essential. Diabetes is a significant risk factor for chronic renal failure, with 20%–40% of adult patients with diabetes developing diabetic kidney disease (ElSayed et al., 2022). Both SGLT2 and DPP-4 inhibitors effectively control blood glucose levels and reduce the risk of chronic renal failure (Cherney et al., 2014; Heerspink et al., 2017; Kanasaki et al., 2014; Kojima et al., 2013; Neal, Perkovic and Matthews, 2017; Ojima et al., 2015; Sarnak et al., 2003; Sharkovska et al., 2014; Vavrinec et al., 2014; Wanner et al., 2016; Zelniker and Braunwald, 2018). Furthermore, a reduced glomerular filtration rate due to renal failure contributes to an increased incidence of cardiovascular disease and mortality (Fox et al., 2012; Go et al., 2004; Levey et al., 2007; Matsushita et al., 2010; Sarnak et al., 2003; Zelniker and Braunwald, 2018). Accordingly, renal function may be a mediator of the effects of these drugs on cardiovascular disease, elucidating how these drugs influence cardiovascular disease through renal function is essential.

To this end, a causal mediation analysis was conducted to assess whether differences in estimated glomerular filtration rate (eGFR) can explain the causal effects of DPP-4 inhibitors versus SGLT2 inhibitors on cardiovascular disease recurrence. The analysis involved using data from a retrospective cohort consisting of 843 patients with type 2 diabetes mellitus who had previously had cardiovascular disease. Among these patients, 218 were in the DPP-4 inhibitor group and 625 were in the SGLT2 inhibitor group. Follow-up concluded on December 31, 2019. The total follow-up period for the DPP-4 inhibitor and SGLT2 inhibitor groups were 529 and 1364 person-years, respectively, with an average follow-up period of 820 days.

Over 3 years of stable dosing, only a small number of individuals experienced a recurrence of cardiovascular disease. Among these people, 16 individuals in the DPP-4 inhibitor group experienced a total of 34 recurrences of cardiovascular disease; 65 individuals in the SGLT2 inhibitor group experienced a total of 105 recurrences. The remaining participants did not report any incidence of cardiovascular disease. The overall recurrence rate was 0.0734 times per person-year, with recurrence rates being 0.0643 times per person-year in the DPP-4 inhibitor group and 0.077 times per person-year in the SGLT2 inhibitor group. eGFR was measured at or before 24 weeks of stable drug use, and the index date for the recurrence of cardiovascular disease was set at 24 weeks of stable dosing.

To satisfy the assumptions of no unmeasured confounding, this analysis accounted for two types of confounders: pretreatment factors, including sex, age, and body mass index for $A$-$M$, $A$-$\widetilde{N}$, and $M$-$\widetilde{N}$ confounding; and a posttreatment factor—cardiovascular disease history during the first 24 weeks of stable dosing—for $M$-$\widetilde{N}$ confounding. The consideration of this posttreatment factor was due to the potential delayed effect of the drugs on cardiovascular function. The causal directed acyclic graph in Figure 3 served as a basis for applying the proposed recurrent causal mediation analysis to investigate how eGFR mediates the effect of diabetes drugs (1 = DPP-4 inhibitor; 0 = SGLT2 inhibitor) on cardiovascular disease. The proposed mediation method requires three prespecified models for the exposure drug, for the mediator eGFR, and for the recurrence of cardiovascular disease. These models are specified as follows:

- Logistic model for the exposure drug:

$$E(\text{drug}|\text{sex, age, bmi, cvd\_history}) = \text{expit}(\alpha_0 + \alpha_1 \text{sex} + \alpha_2 \text{age} + \alpha_3 \text{bmi} + \alpha_4 \text{cvd\_history})$$

- Linear model for the mediator eGFR:

$$E(\text{eGFR}|\text{drug, sex, age, and bmi, cvd\_history})$$
$$= \theta_0 + \theta_1 \text{drug} + \theta_2 \text{sex} + \theta_3 \text{age} + \theta_4 \text{bmi} + \theta_5 \text{cvd\_history}$$

- Proportional mean model for the recurrence of cardiovascular disease:

$$E(\tilde{N}(t)|\text{drug, eGFR, sex, age, bmi, cvd\_history})$$
$$= \Lambda_0(t) \exp(\beta_1 \text{drug} + \beta_2 \text{eGFR} + \beta_3 \text{sex} + \beta_4 \text{age} + \beta_5 \text{bmi} + \beta_6 \text{cvd\_history})$$

The estimated values of these parameters and $\Lambda_0(t)$ are provided in Appendix 5.

We subsequently employed the proposed TR estimators to evaluate NDE and NIE; the results are displayed in Figure 4. The estimates of NDE and NIE along with their corresponding 95% confidence intervals at the 1-year, 2-year, and 3-year marks of stable drug use are presented in Table 5. The 95% confidence intervals were also calculated through bootstrapping. The primary result in Figure 4 indicates that the estimates of NDE and NIE are opposite in direction (NIE is positive and NDE is negative). This implies that the use of DPP-4 inhibitors increases the incidence of cardiovascular disease by affecting eGFR—an indicator of renal glomerular function, compared with SGLT2 inhibitors. However, when not mediated by eGFR, DPP-4 inhibitors perform better in reducing the recurrence of cardiovascular disease. The 95% confidence intervals indicate that the estimated NIE was not significant at all time points and that the estimated NDE was only significant at certain time points.

As presented in Table 5, the NIE estimate at the 3-year mark of stable drug administration was 0.042. This finding suggests that, compared with those prescribed SGLT2 inhibitors, patients with diabetes and a history of cardiovascular disease who are prescribed DPP-4 inhibitors may experience an additional 0.042 recurrences of cardiovascular events per person-year on average, through the drug's effects on renal glomerular function. Conversely, the estimate of NDE at the 3-year mark of stable drug use is −0.102, indicating that the number of cardiovascular disease recurrences not mediated by renal glomerular function was reduced, on average, by 0.102 per person-years for patients with diabetes and a history of cardiovascular disease who are assigned to receive DPP-4 inhibitors; the comparison group comprised patients given SGLT2 inhibitors.

The findings suggest that SGLT2 inhibitors outperform DPP-4 inhibitors in controlling the recurrence of cardiovascular disease through the mechanism mediated by renal glomerular function, but DPP-4 inhibitors may offer protection against heart disease through a mechanism unrelated to glomerular function.

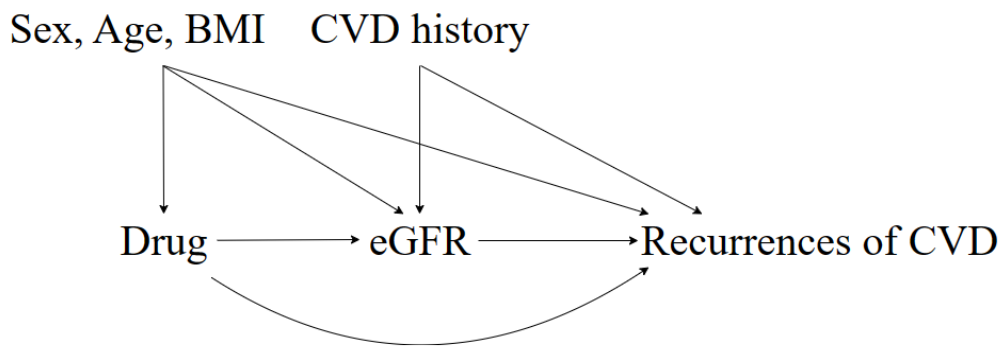

**Figure 3.** Causal directed acyclic graph for a real data analysis with the exposure being drugs, the mediator being estimated glomerular filtration rate (eGFR), and the recurrent events of interest being cardiovascular disease (CVD).

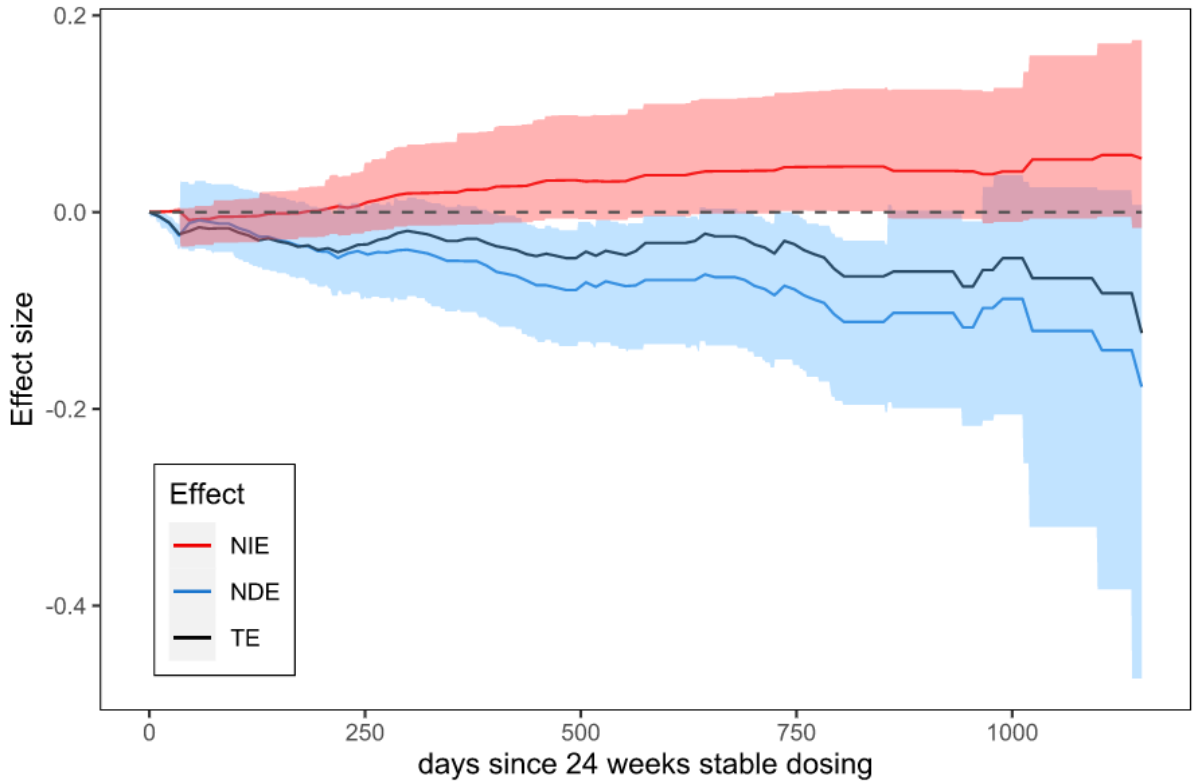

**Figure 4.** Mediating effect of diabetes drugs on the recurrence of cardiovascular disease, mediated by estimated glomerular filtration rate.

**Table 5.** Estimated value of total effect, natural indirect effect, and natural direct effect and their 95% confidence intervals at 1-, 2-, and 3-year marks since the start of stable dosing.

| Est. (95 C.I.) | 1 year | 2 years | 3 years |
|---|---|---|---|
| TE  | -0.038 (-0.069, -0.006) | -0.040 (-0.115, 0.046) | -0.060 (-0.171, 0.057) |
| NIE | 0.001 (-0.027, 0.026) | 0.035 (-0.008, 0.102) | 0.042 (-0.007, 0.121) |
| NDE | -0.040 (-0.076, 0.005) | -0.074 (-0.139, -0.011) | -0.102 (-0.200, 0.000) |

*Abbreviations: TE, total effect; NDE, natural direct effect; NIE, natural indirect effect.*

# 6. Discussion

This study introduced a novel method for causal mediation analysis with recurrent event outcomes. The method initially developed an RB estimator to estimate relevant estimands, and the proportional mean model was used as the outcome model. This model extends the Cox-type model for analyzing recurrent event data. A weighted-based approach was then incorporated into the RB estimator to propose a TR estimator for mediation analysis with recurrent event outcomes. The TR method requires only two out of three models for the exposure, mediator, and outcome to be correctly specified, whereas the RB estimator necessitates that both mediator and outcome models be correctly specified. This implies that if the proportional mean model does not accurately represent the underlying mechanism, the TR estimator remains unbiased, relying on the other two models of the exposure and mediator. Notably, identifying which model is misspecified is unnecessary, because any of the models can be allowed to fail. This study presents a large-sample theorem demonstrating the asymptotic distribution of the TR method and its uniform consistency.

Although the TR estimator provides robustness by allowing for a one-time misspecification, it also imposes more censoring assumptions than does the RB estimator. For the TR estimator to outperform the RB estimator, all three censoring assumptions (CA.1) (CA.2), and (CA.3) must to be satisfied. Although both methods require assumption (CA.1), the performance of the two estimators is similar if (CA.3) is not met, even if (CA.1) and (CA.2) are both satisfied. Specifically, they are biased under the sets $\mathbb{M}_{A,\widetilde{N}}$ and $\mathbb{M}_{A,M}$ and unbiased under $\mathbb{M}_I$ and $\mathbb{M}_{M,\widetilde{N}}$. In this case, the TR estimator fails if assumption (CA.2) is not met but the RB estimator does not, as long as (CA.1) holds. In summary, we must be careful with regard to whether the censoring assumptions could be true. If we can be confident that the three censoring assumptions can be satisfied, then the TR estimator's robustness will aid in achieving better estimation.

The proposed method has several appealing features, but it also has some limitations that warrant attention. In particular, when a recurrent event occurs between the measurement times of $A$ and $M$, denoted as $L$, then $L$ is considered a recanting witness if it meets three conditions simultaneously: (i) $L$ is affected by $A$, (ii) $M$ is affected by $L$, and (iii) $L$ exerts influence on the outcome process $\widetilde{N}(t)$. This scenario is illustrated in Figure 5. If a recanting witness is present, it would violate assumption (A4) and introduce a bias in estimation.

Therefore, verification of whether $L$ satisfies these three conditions is crucial. Because $L$ and $\widetilde{N}(t)$ provide the information of the same type of event, conditions (ii) and (iii) are typically satisfied, which implies $L$ is a potential confounder between $M$ and $\widetilde{N}(t)$. The only situation in which the presence of $L$ would not violate assumption (A4) is if $A$ had a delayed effect such that the recurrent event was not affected by exposure until the measurement time of the mediator. This situation arises if the time span is sufficiently short. To avoid violating (A4) in practice, we can either measure A and M at around the same time or simultaneously at baseline (VanderWeele and Vansteelandt, 2009).

Another limitation of the proposed methods is that they are applicable only to time-fixed variables for both the exposure and mediator. However, in the context of a longitudinal study, understanding how the mediator influences the effect of $A$ on $\widetilde{N}(t)$ over time is crucial. Consequently, a potential avenue for future research is to develop methods for recurrent event outcome with time-varying mediators. In conclusion, this study presents a novel method for causal mediation analysis with recurrent event outcomes that provides a more robust approach than RB estimators. Despite its limitations, the proposed method addresses an important gap in the literature and offers a valuable tool for researchers examining the causal mechanisms underlying recurrent event data. By carefully considering the assumptions and limitations of the TR estimator, researchers can make informed decisions about which approach is most appropriate for their study. Further advancements in this research area, such as the development of methods for time-varying mediators, will continue to expand our understanding of the complex relationships between exposure, mediators, and recurrent event outcomes.

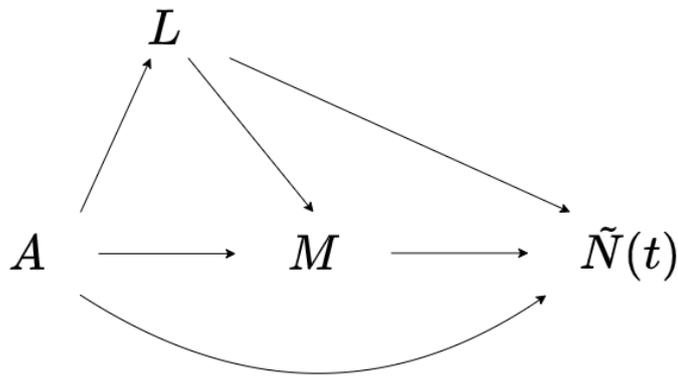

**Figure 5.** Causal directed acyclic graph for mediation analysis with a recurrent outcome and the presence of an exposure-induced mediator–outcome confounder.

## Acknowledgement

This manuscript was edited by Wallace Academic Editing. This work was supported by the National Science and Technology Council of Taiwan [Grant No. NSTC 111-2118-M-006-010-MY3].

# Appendix

## Appendix 1: Procedure of identification

To identify the mediation parameter, we made the following assumptions:

*Assumptions*

**(A1)** No unmeasured confounders between the exposure and outcome process.
$$\tilde{N}(t; a, m) \perp\!\!\!\perp A \mid \boldsymbol{C}$$
**(A2)** No unmeasured confounders between the mediator and outcome process.
$$\tilde{N}(t; a, m) \perp\!\!\!\perp M \mid A, \boldsymbol{C}$$
**(A3)** No unmeasured confounders between the exposure and mediator.
$$M(a) \perp\!\!\!\perp A \mid \boldsymbol{C}$$
**(A4)** Confounders between the mediator and outcome process are not affected by exposure.
$$\tilde{N}(t; a, m) \perp\!\!\!\perp M(a^*) \mid \boldsymbol{C}$$
**(A5)** The counterfactual mediator intervened by setting exposure to $a$ is equal to the observed mediator when the individual actually receives exposure $a$.
$$M(a) = M \mid A = a$$
**(A6)** The counterfactual process intervened by setting the exposure to $a$ and mediator to $m$ is equal to the observed outcome process when the individual actually receives exposure $a$ and mediator $m$.
$$\tilde{N}(t; a, m) = \tilde{N}(t) \mid A = a, M = m$$

The mediation parameter $\phi(t; a, a^*)$ can be identified by $Q(t; a, a^*)$

$$\phi(t; a, a^*) = E\left(\tilde{N}(t; a, M(a^*))\right)$$

$$= \int_{\boldsymbol{c}} E\left(\tilde{N}(t; a, M(a^*)) \mid \boldsymbol{C} = \boldsymbol{c}\right) dF_{\boldsymbol{C}}(\boldsymbol{c})$$

$$= \int_{\boldsymbol{c}} \int_m E\left(\tilde{N}(t; a, M(a^*)) \mid M(a^*) = m, \boldsymbol{C} = \boldsymbol{c}\right) dF_{M(a^*)}(m \mid \boldsymbol{C} = \boldsymbol{c}) \, dF_{\boldsymbol{C}}(\boldsymbol{c})$$

$$\stackrel{A4}{=} \int_{\boldsymbol{c}} \int_m E\left(\tilde{N}(t; a, m) \mid \boldsymbol{C} = \boldsymbol{c}\right) dF_{M(a^*)}(m \mid \boldsymbol{C} = \boldsymbol{c}) \, dF_{\boldsymbol{C}}(\boldsymbol{c})$$

$$\stackrel{A1}{=} \int_{\boldsymbol{c}} \int_m E\left(\tilde{N}(t; a, m) \mid A = a, \boldsymbol{C} = \boldsymbol{c}\right) dF_{M(a^*)}(m \mid \boldsymbol{C} = \boldsymbol{c}) \, dF_{\boldsymbol{C}}(\boldsymbol{c})$$

$$\stackrel{A2}{=} \int_{\boldsymbol{c}} \int_m E\left(\tilde{N}(t; a, m) \mid A = a, M = m, \boldsymbol{C} = \boldsymbol{c}\right) dF_{M(a^*)}(m \mid \boldsymbol{C} = \boldsymbol{c}) \, dF_{\boldsymbol{C}}(\boldsymbol{c})$$

$$\stackrel{A3}{=} \int_{\boldsymbol{c}} \int_m E\left(\tilde{N}(t; a, m) \mid A = a, M = m, \boldsymbol{C} = \boldsymbol{c}\right) dF_{M(a^*)}(m \mid A = a^*, \boldsymbol{C} = \boldsymbol{c}) \, dF_{\boldsymbol{C}}(\boldsymbol{c})$$

$$\stackrel{A5,A6}{=} \int_{\boldsymbol{c}} \int_m E\left(\tilde{N}(t) \mid A = a, M = m, \boldsymbol{C} = \boldsymbol{c}\right) dF_M(m \mid A = a^*, \boldsymbol{C} = \boldsymbol{c}) \, dF_{\boldsymbol{C}}(\boldsymbol{c})$$

$$\equiv Q(t; a, a^*)$$

# Appendix 2: Results for the binary mediator

For the binary mediator, we fit a logistic regression model:

$$\log\left(\frac{\Pr(M=1|a,\boldsymbol{c})}{1-\Pr(M=1|a,\boldsymbol{c})}\right) = \theta_0 + \theta_A a + \theta_{\boldsymbol{C}}^T \boldsymbol{c}$$

If the proportional mean model is used, the identified mediator parameter $Q(t;.,.)$ will be

$$Q(t; a, a^*) = \int_{\boldsymbol{c}} \mu_0(t) \exp(\beta_A a + \beta_{\boldsymbol{C}}^T \boldsymbol{c}) E\left(e^{\beta_M M} | a^*, \boldsymbol{c}\right) dF_{\boldsymbol{C}}(\boldsymbol{c})$$

$$= \mu_0(t) \int_{\boldsymbol{c}} \exp(\beta_A a + \beta_{\boldsymbol{C}}^T \boldsymbol{c}) \left( \frac{1}{1 + \exp(\theta_0 + \theta_A a^* + \theta_{\boldsymbol{C}}^T \boldsymbol{c})} \right.$$

$$\left. + \frac{\exp(\theta_0 + \theta_A a^* + \theta_{\boldsymbol{C}}^T \boldsymbol{c})}{1 + \exp(\theta_0 + \theta_A a^* + \theta_{\boldsymbol{C}}^T \boldsymbol{c})} e^{\beta_M} \right) dF_{\boldsymbol{C}}(\boldsymbol{c})$$

And the regression-based estimators for NDE and NIE will be

$$\widehat{\Delta}_{NDE}(t) = \hat{\mu}_0(t) \int_{\boldsymbol{c}} \left(\exp(\hat{\beta}_A + \hat{\beta}_{\boldsymbol{C}}^T \boldsymbol{c}) - \exp(\hat{\beta}_{\boldsymbol{C}}^T \boldsymbol{c})\right) \left( \frac{1}{1 + \exp(\hat{\theta}_0 + \hat{\theta}_{\boldsymbol{C}}^T \boldsymbol{c})} \right.$$

$$\left. + \frac{\exp(\hat{\theta}_0 + \hat{\theta}_{\boldsymbol{C}}^T \boldsymbol{c})}{1 + \exp(\hat{\theta}_0 + \hat{\theta}_{\boldsymbol{C}}^T \boldsymbol{c})} e^{\hat{\beta}_M} \right) d\hat{F}_{\boldsymbol{C}}(\boldsymbol{c})$$

$$\widehat{\Delta}_{NIE}(t) = \hat{\mu}_0(t) \int_{\boldsymbol{c}} \left( \exp(\hat{\beta}_A + \hat{\beta}_{\boldsymbol{C}}^T \boldsymbol{c}) \left( \left( \frac{1}{1 + \exp(\hat{\theta}_0 + \hat{\theta}_A + \hat{\theta}_{\boldsymbol{C}}^T \boldsymbol{c})} \right. \right. \right.$$

$$\left. + \frac{\exp(\hat{\theta}_0 + \hat{\theta}_A + \hat{\theta}_{\boldsymbol{C}}^T \boldsymbol{c})}{1 + \exp(\hat{\theta}_0 + \hat{\theta}_A + \hat{\theta}_{\boldsymbol{C}}^T \boldsymbol{c})} e^{\hat{\beta}_M} \right)$$

$$\left. \left. - \left( \frac{1}{1 + \exp(\hat{\theta}_0 + \hat{\theta}_{\boldsymbol{C}}^T \boldsymbol{c})} + \frac{\exp(\hat{\theta}_0 + \hat{\theta}_{\boldsymbol{C}}^T \boldsymbol{c})}{1 + \exp(\hat{\theta}_0 + \hat{\theta}_{\boldsymbol{C}}^T \boldsymbol{c})} e^{\hat{\beta}_M} \right) \right) \right) d\hat{F}_{\boldsymbol{C}}(\boldsymbol{c})$$

# Appendix 3: Consistency and asymptotic normality

## Appendix 3.1 Consistency

We show that $d\widehat{Q}^{TR}\left(u; a, a^*; \widehat{\boldsymbol{\alpha}}, \widehat{\boldsymbol{\theta}}, \widehat{\boldsymbol{\beta}}, \widehat{\Lambda}_0(t; \widehat{\boldsymbol{\beta}})\right)$ in $\mathbb{M}_U$ converge in probability to $dQ(u; a, a^*)$ uniformly in $u \in (0, \tau)$; thus, $\widehat{Q}^{TR}\left(t; a, a^*; \widehat{\boldsymbol{\alpha}}, \widehat{\boldsymbol{\theta}}, \widehat{\boldsymbol{\beta}}, \widehat{\Lambda}_0(t; \widehat{\boldsymbol{\beta}})\right) = \int_0^t d\widehat{Q}^{TR}\left(u; a, a^*; \widehat{\boldsymbol{\alpha}}, \widehat{\boldsymbol{\theta}}, \widehat{\boldsymbol{\beta}}, \widehat{\Lambda}_0(t; \widehat{\boldsymbol{\beta}})\right)$ converge in probability to $Q(u; a, a^*)$ uniformly.

Without loss of generality, we set $a = 1, a^* = 0$. We do not indicate which model is correctly specified. Under the given model, let $\hat{\boldsymbol{\alpha}} \xrightarrow{p} \boldsymbol{\alpha}^*, \hat{\boldsymbol{\theta}} \xrightarrow{p} \boldsymbol{\theta}^*, \hat{\boldsymbol{\beta}} \xrightarrow{p} \boldsymbol{\beta}^*, \hat{\Lambda}_0(t; \hat{\boldsymbol{\beta}}) \xrightarrow{p} \Lambda_0^*(t)$ and $\boldsymbol{\alpha}^* = \boldsymbol{\alpha_0}, \boldsymbol{\theta}^* = \boldsymbol{\theta_0}, \boldsymbol{\beta}^* = \boldsymbol{\beta_0}, \Lambda_0^*(t) = \Lambda_0(t)$ when the respective models are correctly specified. The required censoring assumptions are listed as follows:

**(CA.1)** The occurrence of recurrent events is not associated with the censoring mechanism conditional on the exposure $A$, mediator $M$, and confounders $\boldsymbol{C}$.

$$Y(u) \perp\!\!\!\perp d\tilde{N}(u) | A, M, \boldsymbol{C}$$

**(CA.2)** The mediator is not associated with the censoring mechanism conditional on the exposure $A$ and confounders $\boldsymbol{C}$.

$$Y(u) \perp\!\!\!\perp M | A, \boldsymbol{C}$$

**(CA.3)** Confounders are not associated with the censoring mechanism conditional on the exposure $A$.

$$Y(u) \perp\!\!\!\perp \boldsymbol{C} | A$$

According to the weak law of large numbers,

$$d\hat{Q}^{TR}\left(u; 1,0; \hat{\boldsymbol{\alpha}}, \hat{\boldsymbol{\theta}}, \hat{\boldsymbol{\beta}}, \hat{\Lambda}_0(t;\hat{\boldsymbol{\beta}})\right)$$

$$= \frac{1}{n} \sum_{i=1}^{n} \left\{ \frac{I(A_i = 1)\hat{w}_i(\hat{\boldsymbol{\alpha}}) Y_i(u)}{\frac{1}{n}\sum_{j=1}^{n} I(A_j = 1)\hat{w}_j(\hat{\boldsymbol{\alpha}}) Y_j(u)} \frac{\hat{f}_M(M_i | A = 0, \boldsymbol{C}_i; \hat{\boldsymbol{\theta}})}{\hat{f}_M(M_i | A = 1, \boldsymbol{C}_i; \hat{\boldsymbol{\theta}})} \right.$$

$$\times \left[ d\tilde{N}_i(u) - \hat{E}\left(d\tilde{N}(u) \Big| A = 1, M_i, \boldsymbol{C}_i; \hat{\boldsymbol{\beta}}, \hat{\Lambda}_0(t; \hat{\boldsymbol{\beta}})\right) \right]$$

$$+ \frac{I(A_i = 0)\hat{w}_i(\hat{\boldsymbol{\alpha}}) Y_i(u)}{\frac{1}{n}\sum_{j=1}^{n} I(A_i = 0)\hat{w}_j(\hat{\boldsymbol{\alpha}}) Y_j(u)} \left[ \hat{E}\left(d\tilde{N}(u) \Big| A = 1, M_i, \boldsymbol{C}_i; \hat{\boldsymbol{\beta}}, \hat{\Lambda}_0(t; \hat{\boldsymbol{\beta}})\right) \right.$$

$$\left. - d\hat{Q}\left(u; 1,0 \Big| \boldsymbol{C}_i; \hat{\boldsymbol{\theta}}, \hat{\boldsymbol{\beta}}, \hat{\Lambda}_0(t; \hat{\boldsymbol{\beta}})\right) \right] + d\hat{Q}\left(u; 1,0 \Big| \boldsymbol{C}_i; \hat{\boldsymbol{\theta}}, \hat{\boldsymbol{\beta}}, \hat{\Lambda}_0(t; \hat{\boldsymbol{\beta}})\right) \right\}$$

$$\xrightarrow{p} E\left\{\frac{I(A_1=1)\widehat{w}_1(\boldsymbol{\alpha}^*)Y_1(u)}{\pi(u,1,\boldsymbol{\alpha}^*)}\frac{\hat{f}_M(M_1|A=0,\boldsymbol{C}_1;\boldsymbol{\theta}^*)}{\hat{f}_M(M_1|A=1,\boldsymbol{C}_1;\boldsymbol{\theta}^*)}\right.$$

$$\times\left[d\widetilde{N}_1(u)-\hat{E}\left(d\widetilde{N}(u)\big|A=1,M_1,\boldsymbol{C}_1;\boldsymbol{\beta}^*,\Lambda_0^*(t)\right)\right]$$

$$+\frac{I(A_1=0)\widehat{w}_1(\boldsymbol{\alpha}^*)Y_1(u)}{\pi(u,0,\boldsymbol{\alpha}^*)}\left[\hat{E}\left(d\widetilde{N}(u)\big|A=1,M_1,\boldsymbol{C}_1;\boldsymbol{\beta}^*,\Lambda_0^*(t)\right)\right.$$

$$\left.\left.-d\hat{Q}(u;1,0|\boldsymbol{C}_1;\boldsymbol{\theta}^*,\boldsymbol{\beta}^*,\Lambda_0^*(t))\right]+d\hat{Q}(u;1,0|\boldsymbol{C}_1;\boldsymbol{\theta}^*,\boldsymbol{\beta}^*,\Lambda_0^*(t))\right\}$$

Where $\pi(u,a;\boldsymbol{\alpha}^*) = E(I(A=a)\widehat{w}_1(\boldsymbol{\alpha}^*)Y(u)) = E\{E(I(A=a)\widehat{w}_1(\boldsymbol{\alpha}^*)Y(u)|\boldsymbol{C})\} = E\{E[I(A=a)Y(u)|\boldsymbol{C}]*\widehat{w}_1(\boldsymbol{\alpha}^*)\} = E\left\{E[I(A=a)Y(u)|A=a,\boldsymbol{C}]*\frac{1}{\hat{P}(A=a|\boldsymbol{C};\boldsymbol{\alpha}^*)}*P(A=a|\boldsymbol{C})\right\}$

If model A is correctly specified, we have $\boldsymbol{\alpha}^* = \boldsymbol{\alpha}_0$; then,
$$\pi(u,a;\boldsymbol{\alpha}_0) = E(E(Y(u)|A=a,\boldsymbol{C})) = E(Y(u)|A=a)$$

***Case 1: In*** $\mathbb{M}_{M,\widetilde{N}}$

$$d\hat{Q}^{TR}\left(u;1,0;\hat{\boldsymbol{\alpha}},\hat{\boldsymbol{\theta}},\hat{\boldsymbol{\beta}},\hat{\Lambda}_0(t;\hat{\boldsymbol{\beta}})\right)$$

$$\xrightarrow{p} E\left\{\frac{I(A_1=1)\widehat{w}_1(\boldsymbol{\alpha}^*)Y_1(u)}{\pi(u,1;\boldsymbol{\alpha}^*)}\frac{\hat{f}_M(M_1|A=0,\boldsymbol{C}_1;\boldsymbol{\theta}_0)}{\hat{f}_M(M_1|A=1,\boldsymbol{C}_1;\boldsymbol{\theta}_0)}\right.$$

$$\times\left[d\widetilde{N}_1(u)-\hat{E}\left(d\widetilde{N}(u)\big|A=1,M_1,\boldsymbol{C}_1;\boldsymbol{\beta}_0,\Lambda_0(t)\right)\right]$$

$$+\frac{I(A_1=0)\widehat{w}_1(\boldsymbol{\alpha}^*)Y_1(u)}{\pi(u,0;\boldsymbol{\alpha}^*)}\left[\hat{E}\left(d\widetilde{N}(u)\big|A=1,M_1,\boldsymbol{C}_1;\boldsymbol{\beta}_0,\Lambda_0(t)\right)\right.$$

$$\left.\left.-d\hat{Q}(u;1,0|\boldsymbol{C}_1;\boldsymbol{\theta}_0,\boldsymbol{\beta}_0,\Lambda_0(t))\right]+d\hat{Q}(u;1,0|\boldsymbol{C}_1;\boldsymbol{\theta}_0,\boldsymbol{\beta}_0,\Lambda_0(t))\right\}$$

$$= E\left\{E\left\{\frac{\widehat{w}_1(\boldsymbol{\alpha}^*)Y_1(u)}{\pi(u,1;\boldsymbol{\alpha}^*)}\frac{\hat{f}_M(M_1|A=0,\boldsymbol{C}_1;\boldsymbol{\theta}_0)}{\hat{f}_M(M_1|A=1,\boldsymbol{C}_1;\boldsymbol{\theta}_0)}\times\left[d\widetilde{N}_1(u)-\hat{E}\left(d\widetilde{N}(u)\big|A=1,M_1,\boldsymbol{C}_1;\boldsymbol{\beta}_0,\Lambda_0(t)\right)\right]\bigg|A_1=1,M_1,Y_1(u),\boldsymbol{C}_1\right\}\right.$$

$$*P(A_1=1|M_1,Y_1(u),\boldsymbol{C}_1)\bigg\}$$

$$+E\left\{\frac{I(A_1=0)\widehat{w}_1(\boldsymbol{\alpha}^*)Y_1(u)}{\pi(u,0;\boldsymbol{\alpha}^*)}\left[\hat{E}\left(d\widetilde{N}(u)\big|A=1,M_1,\boldsymbol{C}_1;\boldsymbol{\beta}_0,\Lambda_0(t)\right)-d\hat{Q}(u;1,0|\boldsymbol{C}_1;\boldsymbol{\theta}_0,\boldsymbol{\beta}_0,\Lambda_0(t))\right]\right\}$$

$$+E\{d\hat{Q}(u;1,0|\boldsymbol{C}_1;\boldsymbol{\theta}_0,\boldsymbol{\beta}_0,\Lambda_0(t))\}$$

$$= E\left\{\frac{\widehat{w}_1(\boldsymbol{\alpha}^*)Y_1(u)}{\pi(u,1;\boldsymbol{\alpha}^*)}\frac{\hat{f}_M(M_1|A=0,\boldsymbol{C}_1;\boldsymbol{\theta}_0)}{\hat{f}_M(M_1|A=1,\boldsymbol{C}_1;\boldsymbol{\theta}_0)}\times\left[E(d\widetilde{N}(u)|A=1,M_1,Y_1(u),\boldsymbol{C}_1)-\right.\right.$$

$$\left.\hat{E}\left(d\widetilde{N}(u)\big|A=1,M_1,\boldsymbol{C}_1;\boldsymbol{\beta}_0,\Lambda_0(t)\right)\right]\times P(A=1|M_1,Y_1(u),\boldsymbol{C}_1)\bigg\}+$$

$$\mathrm{E}\left\{\frac{I(A_1=0)\widehat{w}_1(\boldsymbol{\alpha}^*)Y_1(u)}{\pi(u,0;\boldsymbol{\alpha}^*)}\left[\hat{E}\left(\mathrm{d}\widetilde{N}(u)\Big|A=1,M_1,\boldsymbol{C}_1;\boldsymbol{\beta_0},\Lambda_0(t)\right)-\mathrm{d}\hat{Q}(u;1,0|\boldsymbol{C}_1;\boldsymbol{\theta_0},\boldsymbol{\beta_0},\Lambda_0(t))\right]\right\}+$$

$$E\{\mathrm{d}\hat{Q}(u;1,0|\boldsymbol{C}_1;\boldsymbol{\theta_0},\boldsymbol{\beta_0},\Lambda_0(t))\} \qquad (3.1.1)$$

$$= E\left\{\frac{\widehat{w}_1(\boldsymbol{\alpha}^*)Y_1(u)}{\pi(u,1;\boldsymbol{\alpha}^*)}\frac{\hat{f}_M(M_1|A=0,\boldsymbol{C}_1;\boldsymbol{\theta_0})}{\hat{f}_M(M_1|A=1,\boldsymbol{C}_1;\boldsymbol{\theta_0})}\Big[E(\mathrm{d}\widetilde{N}(u)|A=1,M_1,\boldsymbol{C}_1)\right.$$

$$\left.-\hat{E}\left(\mathrm{d}\widetilde{N}(u)\Big|A=1,M_1,\boldsymbol{C}_1;\boldsymbol{\beta_0},\Lambda_0(t)\right)\Big]\times P(A=1|M_1,Y_1(u),\boldsymbol{C}_1)\right\}$$

$$+\mathrm{E}\left\{\frac{I(A_1=0)\widehat{w}_1(\boldsymbol{\alpha}^*)Y_1(u)}{\pi(u,0;\boldsymbol{\alpha}^*)}\Big[\hat{E}\left(\mathrm{d}\widetilde{N}(u)\Big|A=1,M_1,\boldsymbol{C}_1;\boldsymbol{\beta_0},\Lambda_0(t)\right)\right.$$

$$\left.-\mathrm{d}\hat{Q}(u;1,0|\boldsymbol{C}_1;\boldsymbol{\theta_0},\boldsymbol{\beta_0},\Lambda_0(t))\Big]\right\}$$

$$+E\{\mathrm{d}\hat{Q}(u;1,0|\boldsymbol{C}_1;\boldsymbol{\theta_0},\boldsymbol{\beta_0},\Lambda_0(t))\}\ (\text{By CA.1})$$

$$= E\left\{\frac{I(A_1=0)\widehat{w}_1(\boldsymbol{\alpha}^*)Y_1(u)}{\pi(u,0;\boldsymbol{\alpha}^*)}\left[\hat{E}\left(\mathrm{d}\widetilde{N}(u)\Big|A=1,M_1,\boldsymbol{C}_1;\boldsymbol{\beta_0},\Lambda_0(t)\right)-\mathrm{d}\hat{Q}(u;1,0|\boldsymbol{C}_1;\boldsymbol{\theta_0},\boldsymbol{\beta_0},\Lambda_0(t))\right]\right\}$$

$$+E\{\mathrm{d}\hat{Q}(u;1,0|\boldsymbol{C}_1;\boldsymbol{\theta_0},\boldsymbol{\beta_0},\Lambda_0(t))\}\ (\text{If model }\widetilde{N}\text{ is correctly specified})$$

$$= E\left\{E\left\{\frac{\widehat{w}_1(\boldsymbol{\alpha}^*)Y_1(u)}{\pi(u,0;\boldsymbol{\alpha}^*)}\left[\hat{E}\left(\mathrm{d}\widetilde{N}(u)\Big|A=1,M_1,\boldsymbol{C}_1;\boldsymbol{\beta_0},\Lambda_0(t)\right)-\mathrm{d}\hat{Q}\left(u;1,0|\boldsymbol{C}_1;\boldsymbol{\theta_0},\boldsymbol{\beta_0},\Lambda_0(t)\right)\right]\Big|A_1=0,\boldsymbol{C}_1\right\}\right.$$

$$\left.*P(A=0|\boldsymbol{C}_1)\right\}+E\left\{\mathrm{d}\hat{Q}\left(u;1,0|\boldsymbol{C}_1;\boldsymbol{\theta_0},\boldsymbol{\beta_0},\Lambda_0(t)\right)\right\}$$

$$= E\left\{\frac{E(Y_1(u)|A_1=0,\boldsymbol{C}_1)}{\hat{P}(A_1=0|\boldsymbol{C}_1;\boldsymbol{\alpha}^*)\pi(u,0;\boldsymbol{\alpha}^*)}E\left\{\left[\hat{E}\left(\mathrm{d}\widetilde{N}(u)\Big|A=1,M_1,\boldsymbol{C}_1;\boldsymbol{\beta_0},\Lambda_0(t)\right)-\mathrm{d}\hat{Q}(u;1,0|\boldsymbol{C}_1;\boldsymbol{\theta_0},\boldsymbol{\beta_0},\Lambda_0(t))\right]\Big|A_1=0,\boldsymbol{C}_1\right\}\right.$$

$$\left.*P(A=0|\boldsymbol{C}_1)\right\}+E\{\mathrm{d}\hat{Q}(u;1,0|\boldsymbol{C}_1;\boldsymbol{\theta_0},\boldsymbol{\beta_0},\Lambda_0(t))\}\ (\text{By CA.2})$$

$$= E\left\{\frac{E(Y(u)|A=0,\boldsymbol{C}_1)}{\hat{P}(A_1=0|\boldsymbol{C}_1;\boldsymbol{\alpha}^*)\pi(u,0;\boldsymbol{\alpha}^*)}\left[E\left(\hat{E}\left(\mathrm{d}\widetilde{N}(u)\Big|A=1,M_1,\boldsymbol{C}_1;\boldsymbol{\beta_0},\Lambda_0(t)\right)\Big|A=0,\boldsymbol{C}_1\right)\right.\right.$$

$$\left.\left.-\mathrm{d}\hat{Q}(u;1,0|\boldsymbol{C}_1;\boldsymbol{\theta_0},\boldsymbol{\beta_0},\Lambda_0(t))\right]*P(A=0|\boldsymbol{C}_1)\right\}+E\{\mathrm{d}\hat{Q}(u;1,0|\boldsymbol{C}_1;\boldsymbol{\theta_0},\boldsymbol{\beta_0},\Lambda_0(t))\}$$

$$= E\{\mathrm{d}\hat{Q}(u;1,0|\boldsymbol{C}_1;\boldsymbol{\theta_0},\boldsymbol{\beta_0},\Lambda_0(t))\}\ (\text{If model M is correctly specified})$$

$$= \mathrm{d}Q(u;1,0)\ \ (\text{If models M and }\widetilde{N}\text{ are correctly specified})$$

*Case 2: In* $\mathbb{M}_{A,\widetilde{N}}$

$$\mathrm{d}\hat{Q}^{TR}\left(u;1,0;\widehat{\boldsymbol{\alpha}},\widehat{\boldsymbol{\theta}},\widehat{\boldsymbol{\beta}},\widehat{\Lambda}_0(t;\widehat{\boldsymbol{\beta}})\right)$$

$$\xrightarrow{p} E\left\{\frac{\widehat{w}_1(\boldsymbol{\alpha}_0)Y_1(u)}{\pi(u,1;\boldsymbol{\alpha}_0)}\frac{\hat{f}_M(M_1|A=0,\boldsymbol{C}_1;\boldsymbol{\theta}^*)}{\hat{f}_M(M_1|A=1,\boldsymbol{C}_1;\boldsymbol{\theta}^*)}\times\Big[E(\mathrm{d}\widetilde{N}(u)|A=1,M_1,Y_1(u),\boldsymbol{C}_1)\ -\right.$$

$$\hat{E}\left(\mathrm{d}\widetilde{N}(u)\Big|A=1,M_1,\boldsymbol{C}_1;\boldsymbol{\beta_0},\Lambda_0(t)\right)\Big]\times P(A=1|M_1,Y_1(u),\boldsymbol{C}_1)\right\}+$$

$$\mathrm{E}\left\{\frac{I(A_1=0)\widehat{w}_1(\boldsymbol{\alpha}_0)Y_1(u)}{\pi(u,0;\boldsymbol{\alpha}_0)}\left[\hat{E}\left(\mathrm{d}\widetilde{N}(u)\Big|A=1,M_1,\boldsymbol{C}_1;\boldsymbol{\beta_0},\Lambda_0(t)\right)-\mathrm{d}\hat{Q}(u;1,0|\boldsymbol{C}_1;\boldsymbol{\theta}^*,\boldsymbol{\beta_0},\Lambda_0(t))\right]\right\}+$$

$E\{d\hat{Q}(u; 1,0|C_1; \boldsymbol{\theta}^*, \boldsymbol{\beta_0}, \Lambda_0(t))\}$ (By 3.1.1)

$$= E\left\{\frac{\widehat{w}_1(\boldsymbol{\alpha_0})Y_1(u)}{\pi(u,1;\boldsymbol{\alpha_0})}\frac{\hat{f}_M(M_1|A=0,C_1;\boldsymbol{\theta}^*)}{\hat{f}_M(M_1|A=1,C_1;\boldsymbol{\theta}^*)}\left[E(d\tilde{N}(u)|A=1,M_1,C_1) - \hat{E}\left(d\tilde{N}(u)\big|A=1,M_1,C_1;\boldsymbol{\beta_0},\Lambda_0(t)\right)\right]\right.$$
$$\left. \times P(A=1|M_1,Y_1(u),C_1)\right\}$$
$$+ E\left\{\frac{I(A_1=0)\widehat{w}_1(\boldsymbol{\alpha_0})Y_1(u)}{\pi(u,0;\boldsymbol{\alpha_0})}\left[\hat{E}\left(d\tilde{N}(u)\big|A=1,M_1,C_1;\boldsymbol{\beta_0},\Lambda_0(t)\right)\right.\right.$$
$$\left.\left. - d\hat{Q}(u;1,0|C_1;\boldsymbol{\theta}^*,\boldsymbol{\beta_0},\Lambda_0(t))\right]\right\} + E\{d\hat{Q}(u;1,0|C_1;\boldsymbol{\theta}^*,\boldsymbol{\beta_0},\Lambda_0(t))\} \text{ (By CA.1)}$$

$$= E\left\{\frac{I(A_1=0)\widehat{w}_1(\boldsymbol{\alpha_0})Y_1(u)}{\pi(u,0;\boldsymbol{\alpha_0})}\left[\hat{E}\left(d\tilde{N}(u)\big|A=1,M_1,C_1;\boldsymbol{\beta_0},\Lambda_0(t)\right) - d\hat{Q}(u;1,0|C_1;\boldsymbol{\theta}^*,\boldsymbol{\beta_0},\Lambda_0(t))\right]\right\} +$$

$E\{d\hat{Q}(u;1,0|C_1;\boldsymbol{\theta}^*,\boldsymbol{\beta_0},\Lambda_0(t))\}$ (If model $\tilde{N}$ is correctly specified)

$$= E\left\{\frac{I(A_1=0)\widehat{w}_1(\boldsymbol{\alpha_0})Y_1(u)}{\pi(u,0;\boldsymbol{\alpha_0})}\hat{E}\left(d\tilde{N}(u)\big|A=1,M_1,C_1;\boldsymbol{\beta_0},\Lambda_0(t)\right)\right\}$$
$$+ E\left\{\frac{I(A_1=0)\widehat{w}_1(\boldsymbol{\alpha_0})Y_1(u) - \pi(u,0;\boldsymbol{\alpha_0})}{\pi(u,0;\boldsymbol{\alpha_0})}d\hat{Q}(u;1,0|C_1;\boldsymbol{\theta}^*,\boldsymbol{\beta_0},\Lambda_0(t))\right\}$$

$$= E\left\{\frac{I(A_1=0)\widehat{w}_1(\boldsymbol{\alpha_0})Y_1(u)}{\pi(u,0;\boldsymbol{\alpha_0})}\hat{E}\left(d\tilde{N}(u)\big|A=1,M_1,C_1;\boldsymbol{\beta_0},\Lambda_0(t)\right)\right\}$$
$$+ E\left\{E\left(\frac{\widehat{w}_1(\boldsymbol{\alpha_0})Y_1(u) - \pi(u,0;\boldsymbol{\alpha_0})}{\pi(u,0;\boldsymbol{\alpha_0})}d\hat{Q}(u;1,0|C_1;\boldsymbol{\theta}^*,\boldsymbol{\beta_0},\Lambda_0(t))\bigg|A_1=0,C_1\right)P(A_1=0|C_1)\right\}$$

$$= E\left\{\frac{I(A_1=0)\widehat{w}_1(\boldsymbol{\alpha_0})Y_1(u)}{\pi(u,0;\boldsymbol{\alpha_0})}\hat{E}\left(d\tilde{N}(u)\big|A=1,M_1,C_1;\boldsymbol{\beta_0},\Lambda_0(t)\right)\right\}$$
$$+ E\left\{\frac{P(Y_1(u)|A_1=0,C_1) - \pi(u,0;\boldsymbol{\alpha_0})}{\hat{P}(A=0|C_1;\boldsymbol{\alpha_0})\pi(u,0;\boldsymbol{\alpha_0})}d\hat{Q}(u;1,0|C_1;\boldsymbol{\theta}^*,\boldsymbol{\beta_0},\Lambda_0(t))P(A=0|C_1)\right\}$$

$$= E\left\{\frac{I(A_1=0)\widehat{w}_1(\boldsymbol{\alpha_0})Y_1(u)}{\pi(u,0;\boldsymbol{\alpha_0})}\hat{E}\left(d\tilde{N}(u)\big|A=1,M_1,C_1;\boldsymbol{\beta_0},\Lambda_0(t)\right)\right\}$$
$$+ E\left\{\frac{P(Y_1(u)|A_1=0,C_1) - E(Y(u)|A=0)}{\hat{P}(A=0|C_1;\boldsymbol{\alpha_0})\pi(u,0;\boldsymbol{\alpha_0})}d\hat{Q}(u;1,0|C_1;\boldsymbol{\theta}^*,\boldsymbol{\beta_0},\Lambda_0(t))P(A=0|C_1)\right\}$$

(If model $A$ is correctly specified)

$$= E\left\{\frac{I(A_1=0)\widehat{w}_1(\boldsymbol{\alpha_0})Y_1(u)}{\pi(u,0;\boldsymbol{\alpha_0})}\hat{E}\left(d\tilde{N}(u)\big|A=1,M_1,C_1;\boldsymbol{\beta_0},\Lambda_0(t)\right)\right\} \text{ (By CA.3)}$$

$$= E\left\{E\left(\frac{Y_1(u)}{\pi(u,0;\boldsymbol{\alpha_0})}\hat{E}\left(d\tilde{N}(u)\big|A=1,M_1,C_1;\boldsymbol{\beta_0},\Lambda_0(t)\right)\bigg|A_1=0,C_1\right)\frac{P(A_1=0|C_1)}{\hat{P}(A=0|C_1)}\right\}$$

$$= E\left\{E\left(\frac{Y_1(u)}{\pi(u,0;\boldsymbol{\alpha_0})}\hat{E}\left(d\tilde{N}(u)\big|A=1,M_1,C_1;\boldsymbol{\beta_0},\Lambda_0(t)\right)\bigg|A_1=0,C_1\right)\right\}$$

(As $A$ model is correctly specified)

$$= E\left\{E\left(\hat{E}\left(d\tilde{N}(u)\big|A=1,M_1,C_1;\boldsymbol{\beta_0},\Lambda_0(t)\right)\bigg|A_1=0,C_1\right)\frac{E(Y_1(u)|A_1=0,C_1)}{\pi(u,0;\boldsymbol{\alpha_0})}\right\} \text{ (By CA.2)}$$

$$= \mathrm{E}\left\{E\left(\hat{E}\left(\mathrm{d}\tilde{N}(u)\big|A=1, M_1, \boldsymbol{C}_1; \boldsymbol{\beta_0}, \Lambda_0(t)\right)\bigg|A_1=0, \boldsymbol{C}_1\right)\frac{E(Y_1(u)|A_1=0)}{\pi(u,0;\boldsymbol{\alpha_0})}\right\} \text{ (By CA.3)}$$

$$= \mathrm{E}\left\{E\left(\hat{E}\left(\mathrm{d}\tilde{N}(u)\big|A=1, M_1, \boldsymbol{C}_1; \boldsymbol{\beta_0}, \Lambda_0(t)\right)\bigg|A=0, \boldsymbol{C}_1\right)\right\} \text{ (If model } A \text{ is correctly specified)}$$

$$= \mathrm{d}Q(u; 1,0) \text{ (If model } \tilde{N} \text{ is correctly specified)}$$

*Case 3: In* $\mathbb{M}_{A,M}$

$$\mathrm{d}\hat{Q}^{TR}\left(u; 1,0; \hat{\boldsymbol{\alpha}}, \hat{\boldsymbol{\theta}}, \hat{\boldsymbol{\beta}}, \hat{\Lambda}_0(t; \hat{\boldsymbol{\beta}})\right)$$

$$\xrightarrow{p} E\left\{\frac{I(A_1=1)\hat{w}_1(\boldsymbol{\alpha_0})Y_1(u)}{\pi(u,1;\boldsymbol{\alpha_0})}\frac{\hat{f}_M(M_1|A=0,\boldsymbol{C}_1;\boldsymbol{\theta_0})}{\hat{f}_M(M_1|A=1,\boldsymbol{C}_1;\boldsymbol{\theta_0})}\right.$$
$$\times \left[\mathrm{d}\tilde{N}_1(u) - \hat{E}\left(\mathrm{d}\tilde{N}(u)\big|A=1, M_1, \boldsymbol{C}_1; \boldsymbol{\beta}^*, \Lambda_0^*(t)\right)\right]$$
$$+ \frac{I(A_1=0)\hat{w}_1(\boldsymbol{\alpha_0})Y_1(u)}{\pi(u,0;\boldsymbol{\alpha_0})}\left[\hat{E}\left(\mathrm{d}\tilde{N}(u)\big|A=1, M_1, \boldsymbol{C}_1; \boldsymbol{\beta}^*, \Lambda_0^*(t)\right)\right.$$
$$\left.\left.- \mathrm{d}\hat{Q}\left(u; 1,0\big|\boldsymbol{C}_1; \boldsymbol{\theta_0}, \boldsymbol{\beta}^*, \Lambda_0^*(t)\right)\right] + \mathrm{d}\hat{Q}\left(u; 1,0\big|\boldsymbol{C}_1; \boldsymbol{\theta_0}, \boldsymbol{\beta}^*, \Lambda_0^*(t)\right)\right\}$$

$$= E\left\{\frac{I(A_1=1)\hat{w}_1(\boldsymbol{\alpha_0})Y_1(u)}{\pi(u,1;\boldsymbol{\alpha_0})}\frac{\hat{f}_M(M_1|A=0,\boldsymbol{C}_1;\boldsymbol{\theta_0})}{\hat{f}_M(M_1|A=1,\boldsymbol{C}_1;\boldsymbol{\theta_0})}\right.$$
$$\left.\times\left[\mathrm{d}\tilde{N}_1(u) - \hat{E}\left(\mathrm{d}\tilde{N}(u)\big|A=1, M_1, \boldsymbol{C}_1; \boldsymbol{\beta}^*, \Lambda_0^*(t)\right)\right]\right\}$$
$$+ E\left\{\frac{I(A_1=0)\hat{w}_1(\boldsymbol{\alpha_0})Y_1(u)}{\pi(u,0;\boldsymbol{\alpha_0})}\hat{E}\left(\mathrm{d}\tilde{N}(u)\big|A=1, M_1, \boldsymbol{C}_1; \boldsymbol{\beta}^*, \Lambda_0^*(t)\right)\right\}$$
$$+ E\left\{\left(\frac{I(A_1=0)\hat{w}_1(\boldsymbol{\alpha_0})Y_1(u)-\pi(u,0;\boldsymbol{\alpha_0})}{\pi(u,0;\boldsymbol{\alpha_0})}\right)\mathrm{d}\hat{Q}\left(u; 1,0\big|\boldsymbol{C}_1; \boldsymbol{\theta_0}, \boldsymbol{\beta}^*, \Lambda_0^*(t)\right)\right\}$$

$$= E\left\{\frac{I(A_1=1)\hat{w}_1(\boldsymbol{\alpha_0})Y_1(u)}{\pi(u,1;\boldsymbol{\alpha_0})}\frac{\hat{f}_M(M_1|A=0,\boldsymbol{C}_1;\boldsymbol{\theta_0})}{\hat{f}_M(M_1|A=1,\boldsymbol{C}_1;\boldsymbol{\theta_0})}\right.$$
$$\left.\times\left[\mathrm{d}\tilde{N}_1(u) - \hat{E}\left(\mathrm{d}\tilde{N}(u)\big|A=1, M_1, \boldsymbol{C}_1; \boldsymbol{\beta}^*, \Lambda_0^*(t)\right)\right]\right\}$$
$$+ E\left\{\frac{I(A_1=0)\hat{w}_1(\boldsymbol{\alpha_0})Y_1(u)}{\pi(u,0;\boldsymbol{\alpha_0})}\hat{E}\left(\mathrm{d}\tilde{N}(u)\big|A=1, M_1, \boldsymbol{C}_1; \boldsymbol{\beta}^*, \Lambda_0^*(t)\right)\right\}$$
$$+ E\left\{E\left(\frac{\hat{w}_1(\boldsymbol{\alpha_0})Y_1(u)-\pi(u,0;\boldsymbol{\alpha_0})}{\pi(u,0;\boldsymbol{\alpha_0})}\mathrm{d}\hat{Q}\left(u; 1,0\big|\boldsymbol{C}_1; \boldsymbol{\theta_0}, \boldsymbol{\beta}^*, \Lambda_0^*(t)\right)\bigg|A_1=0,\boldsymbol{C}_1\right)P(A_1=0|\boldsymbol{C}_1)\right\}$$

$$= E\left\{\frac{I(A_1=1)\widehat{w}_1(\boldsymbol{\alpha_0})Y_1(u)}{\pi(u,1;\boldsymbol{\alpha_0})}\frac{\hat{f}_M(M_1|A=0,\boldsymbol{C_1};\boldsymbol{\theta_0})}{\hat{f}_M(M_1|A=1,\boldsymbol{C_1};\boldsymbol{\theta_0})}\right.$$

$$\left.\times\left[d\widetilde{N}_1(u)-\hat{E}\left(d\widetilde{N}(u)\big|A=1,M_1,\boldsymbol{C_1};\boldsymbol{\beta}^*,\Lambda_0^*(t)\right)\right]\right\}$$

$$+E\left\{\frac{I(A_1=0)\widehat{w}_1(\boldsymbol{\alpha_0})Y_1(u)}{\pi(u,0;\boldsymbol{\alpha_0})}\hat{E}\left(d\widetilde{N}(u)\big|A=1,M_1,\boldsymbol{C_1};\boldsymbol{\beta}^*,\Lambda_0^*(t)\right)\right\}$$

$$+E\left\{\frac{P(Y_1(u)|A_1=0,\boldsymbol{C_1})-\pi(u,0;\boldsymbol{\alpha_0})}{\hat{P}(A=0|\boldsymbol{C_1};\boldsymbol{\alpha_0})\pi(u,0;\boldsymbol{\alpha_0})}d\hat{Q}(u;1,0|\boldsymbol{C_1};\boldsymbol{\theta_0},\boldsymbol{\beta}^*,\Lambda_0^*(t))P(A_1=0|\boldsymbol{C_1})\right\}\text{ (By CA.2)}$$

$$= E\left\{\frac{I(A_1=1)\widehat{w}_1(\boldsymbol{\alpha_0})Y_1(u)}{\pi(u,1;\boldsymbol{\alpha_0})}\frac{\hat{f}_M(M_1|A=0,\boldsymbol{C_1};\boldsymbol{\theta_0})}{\hat{f}_M(M_1|A=1,\boldsymbol{C_1};\boldsymbol{\theta_0})}\right.$$

$$\left.\times\left[d\widetilde{N}_1(u)-\hat{E}\left(d\widetilde{N}(u)\big|A=1,M_1,\boldsymbol{C_1};\boldsymbol{\beta}^*,\Lambda_0^*(t)\right)\right]\right\}$$

$$+E\left\{\frac{I(A_1=0)\widehat{w}_1(\boldsymbol{\alpha_0})Y_1(u)}{\pi(u,0;\boldsymbol{\alpha_0})}\hat{E}\left(d\widetilde{N}(u)\big|A=1,M_1,\boldsymbol{C_1};\boldsymbol{\beta}^*,\Lambda_0^*(t)\right)\right\}$$

$$+E\left\{\frac{P(Y_1(u)|A_1=0,\boldsymbol{C_1})-E(Y(u)|A=0)}{\hat{P}(A=0|\boldsymbol{C_1};\boldsymbol{\alpha_0})\pi(u,0;\boldsymbol{\alpha_0})}d\hat{Q}(u;1,0|\boldsymbol{C_1};\boldsymbol{\theta_0},\boldsymbol{\beta}^*,\Lambda_0^*(t))P(A_1=0|\boldsymbol{C_1})\right\}$$

(If model $A$ is correctly specified)

$$= E\left\{\frac{I(A_1=1)\widehat{w}_1(\boldsymbol{\alpha_0})Y_1(u)}{\pi(u,1;\boldsymbol{\alpha_0})}\frac{\hat{f}_M(M_1|A=0,\boldsymbol{C_1};\boldsymbol{\theta_0})}{\hat{f}_M(M_1|A=1,\boldsymbol{C_1};\boldsymbol{\theta_0})}\right.$$

$$\left.\times\left[d\widetilde{N}_1(u)-\hat{E}\left(d\widetilde{N}(u)\big|A=1,M_1,\boldsymbol{C_1};\boldsymbol{\beta}^*,\Lambda_0^*(t)\right)\right]\right\}$$

$$+E\left\{\frac{I(A_1=0)\widehat{w}_1(\boldsymbol{\alpha_0})Y_1(u)}{\pi(u,0;\boldsymbol{\alpha_0})}\hat{E}\left(d\widetilde{N}(u)\big|A=1,M_1,\boldsymbol{C_1};\boldsymbol{\beta}^*,\Lambda_0^*(t)\right)\right\}\text{ (By CA.3)}$$

$$= E\left\{\frac{I(A_1=1)\widehat{w}_1(\boldsymbol{\alpha_0})Y_1(u)}{\pi(u,1;\boldsymbol{\alpha_0})}\frac{\hat{f}_M(M_1|A=0,\boldsymbol{C_1};\boldsymbol{\theta_0})}{\hat{f}_M(M_1|A=1,\boldsymbol{C_1};\boldsymbol{\theta_0})}\right.$$

$$\left.\times\left[d\widetilde{N}_1(u)-\hat{E}\left(d\widetilde{N}(u)\big|A=1,M_1,\boldsymbol{C_1};\boldsymbol{\beta}^*,\Lambda_0^*(t)\right)\right]\right\}$$

$$+E\left\{E\left(\frac{Y_1(u)}{\pi(u,0;\boldsymbol{\alpha_0})}\hat{E}\left(d\widetilde{N}(u)\big|A=1,M_1,\boldsymbol{C_1};\boldsymbol{\beta}^*,\Lambda_0^*(t)\right)\bigg|A_1=0,\boldsymbol{C_1}\right)\frac{P(A_1=0|\boldsymbol{C_1})}{\hat{P}(A=0|\boldsymbol{C_1})}\right\}$$

$$= E\left\{\frac{I(A_1=1)\widehat{w}_1(\boldsymbol{\alpha_0})Y_1(u)}{\pi(u,1;\boldsymbol{\alpha_0})}\frac{\hat{f}_M(M_1|A=0,\boldsymbol{C_1};\boldsymbol{\theta_0})}{\hat{f}_M(M_1|A=1,\boldsymbol{C_1};\boldsymbol{\theta_0})}\right.$$

$$\left.\times\left[d\widetilde{N}_1(u)-\hat{E}\left(d\widetilde{N}(u)\big|A=1,M_1,\boldsymbol{C_1};\boldsymbol{\beta}^*,\Lambda_0^*(t)\right)\right]\right\}$$

$$+E\left\{E\left(\frac{Y_1(u)}{\pi(u,0;\boldsymbol{\alpha_0})}\hat{E}\left(d\widetilde{N}(u)\big|A=1,M_1,\boldsymbol{C_1};\boldsymbol{\beta}^*,\Lambda_0^*(t)\right)\bigg|A_1=0,\boldsymbol{C_1}\right)\right\}$$

(If model $A$ is correctly specified)

$$= E\left\{\frac{I(A_1 = 1)\widehat{w}_1(\boldsymbol{\alpha_0})Y_1(u)}{\pi(u, 1; \boldsymbol{\alpha_0})} \frac{\hat{f}_M(M_1|A = 0, \boldsymbol{C_1}; \boldsymbol{\theta_0})}{\hat{f}_M(M_1|A = 1, \boldsymbol{C_1}; \boldsymbol{\theta_0})}\right.$$

$$\left.\times \left[d\widetilde{N}_1(u) - \hat{E}\left(d\widetilde{N}(u)\Big|A = 1, M_1, \boldsymbol{C_1}; \boldsymbol{\beta^*}, \Lambda_0^*(t)\right)\right]\right\}$$

$$+ E\left\{E\left(\hat{E}\left(d\widetilde{N}(u)\Big|A = 1, M_1, \boldsymbol{C_1}; \boldsymbol{\beta^*}, \Lambda_0^*(t)\right)\Big|A_1 = 0, \boldsymbol{C_1}\right)\frac{E(Y_1(u)|A_1 = 0, \boldsymbol{C_1})}{\pi(u, 0; \boldsymbol{\alpha_0})}\right\} \text{ (By CA. 2)}$$

$$= E\left\{\frac{I(A_1 = 1)\widehat{w}_1(\boldsymbol{\alpha_0})Y_1(u)}{\pi(u, 1; \boldsymbol{\alpha_0})} \frac{\hat{f}_M(M_1|A = 0, \boldsymbol{C_1}; \boldsymbol{\theta_0})}{\hat{f}_M(M_1|A = 1, \boldsymbol{C_1}; \boldsymbol{\theta_0})}\right.$$

$$\left.\times \left[d\widetilde{N}_1(u) - \hat{E}\left(d\widetilde{N}(u)\Big|A = 1, M_1, \boldsymbol{C_1}; \boldsymbol{\beta^*}, \Lambda_0^*(t)\right)\right]\right\}$$

$$+ E\left\{E\left(\hat{E}\left(d\widetilde{N}(u)\Big|A = 1, M_1, \boldsymbol{C_1}; \boldsymbol{\beta^*}, \Lambda_0^*(t)\right)\Big|A_1 = 0, \boldsymbol{C_1}\right)\frac{E(Y_1(u)|A_1 = 0)}{\pi(u, 0; \boldsymbol{\alpha_0})}\right\} \text{ (By CA. 3)}$$

$$= E\left\{\frac{I(A_1 = 1)\widehat{w}_1(\boldsymbol{\alpha_0})Y_1(u)}{\pi(u, 1; \boldsymbol{\alpha_0})} \frac{\hat{f}_M(M_1|A = 0, \boldsymbol{C_1}; \boldsymbol{\theta_0})}{\hat{f}_M(M_1|A = 1, \boldsymbol{C_1}; \boldsymbol{\theta_0})}\right.$$

$$\left.\times \left[d\widetilde{N}_1(u) - \hat{E}\left(d\widetilde{N}(u)\Big|A = 1, M_1, \boldsymbol{C_1}; \boldsymbol{\beta^*}, \Lambda_0^*(t)\right)\right]\right\}$$

$$+ E\left\{E\left(\hat{E}\left(d\widetilde{N}(u)\Big|A = 1, M_1, \boldsymbol{C_1}; \boldsymbol{\beta^*}, \Lambda_0^*(t)\right)\Big|A = 0, \boldsymbol{C_1}\right)\right\} \text{ (If model A is correctly specified)}$$

$$= E\left\{\frac{I(A_1 = 1)\widehat{w}_1(\boldsymbol{\alpha_0})Y_1(u)}{\pi(u, 1; \boldsymbol{\alpha_0})} \frac{\hat{f}_M(M_1|A = 0, \boldsymbol{C_1}; \boldsymbol{\theta_0})}{\hat{f}_M(M_1|A = 1, \boldsymbol{C_1}; \boldsymbol{\theta_0})}\right.$$

$$\left.\times \left[d\widetilde{N}_1(u) - \hat{E}\left(d\widetilde{N}(u)\Big|A = 1, M_1, \boldsymbol{C_1}; \boldsymbol{\beta^*}, \Lambda_0^*(t)\right)\right]\right\}$$

$$+ E\left\{E\left(\hat{E}\left(d\widetilde{N}(u)\Big|A = 1, M_1, \boldsymbol{C_1}; \boldsymbol{\beta^*}, \Lambda_0^*(t)\right)\Big|A = 0, \boldsymbol{C_1}\right)\right\}$$

$$= E\left\{\frac{I(A_1 = 1)\widehat{w}_1(\boldsymbol{\alpha_0})Y_1(u)}{\pi(u, 1; \boldsymbol{\alpha_0})} \frac{\hat{f}_M(M_1|A = 0, \boldsymbol{C_1}; \boldsymbol{\theta_0})}{\hat{f}_M(M_1|A = 1, \boldsymbol{C_1}; \boldsymbol{\theta_0})} d\widetilde{N}_1(u)\right\}$$

$$- E\left\{\frac{I(A_1 = 1)\widehat{w}_1(\boldsymbol{\alpha_0})Y_1(u)}{\pi(u, 1; \boldsymbol{\alpha_0})} \frac{\hat{f}_M(M_1|A = 0, \boldsymbol{C_1}; \boldsymbol{\theta_0})}{\hat{f}_M(M_1|A = 1, \boldsymbol{C_1}; \boldsymbol{\theta_0})} \hat{E}\left(d\widetilde{N}(u)\Big|A = 1, M_1, \boldsymbol{C_1}; \boldsymbol{\beta^*}, \Lambda_0^*(t)\right)\right\}$$

$$+ E\left\{E\left(\hat{E}\left(d\widetilde{N}(u)\Big|A = 1, M_1, \boldsymbol{C_1}; \boldsymbol{\beta^*}, \Lambda_0^*(t)\right)\Big|A = 0, \boldsymbol{C_1}\right)\right\}$$

$$= E\left\{\frac{I(A_1 = 1)\widehat{w}_1(\boldsymbol{\alpha_0})Y_1(u)}{\pi(u, 1; \boldsymbol{\alpha_0})} \frac{\hat{f}_M(M_1|A = 0, \boldsymbol{C_1}; \boldsymbol{\theta_0})}{\hat{f}_M(M_1|A = 1, \boldsymbol{C_1}; \boldsymbol{\theta_0})} d\widetilde{N}_1(u)\right\}$$

$$- E\left\{\frac{E(Y_1(u)|A_1 = 1, \boldsymbol{C_1})}{\hat{P}(A_1 = 1|\boldsymbol{C_1}; \boldsymbol{\alpha_0})\pi(u, 1; \boldsymbol{\alpha_0})}\right.$$

$$\left.\times E\left(\frac{\hat{f}_M(M_1|A = 0, \boldsymbol{C_1}; \boldsymbol{\theta_0})}{\hat{f}_M(M_1|A = 1, \boldsymbol{C_1}; \boldsymbol{\theta_0})} \hat{E}\left(d\widetilde{N}(u)\Big|A = 1, M_1, \boldsymbol{C_1}; \boldsymbol{\beta^*}, \Lambda_0^*(t)\right)\Big|A_1 = 1, \boldsymbol{C_1}\right) P(A_1 = 1|\boldsymbol{C_1})\right\}$$

$$+ E\left\{E\left(\hat{E}\left(d\widetilde{N}(u)\Big|A = 1, M_1, \boldsymbol{C_1}; \boldsymbol{\beta^*}, \Lambda_0^*(t)\right)\Big|A = 0, \boldsymbol{C_1}\right)\right\} \text{ (By CA. 2)}$$

$$= E\left\{\frac{I(A_1=1)\hat{w}_1(\boldsymbol{\alpha_0})Y_1(u)}{\pi(u,1;\boldsymbol{\alpha_0})}\frac{\hat{f}_M(M_1|A=0,\boldsymbol{C}_1;\boldsymbol{\theta_0})}{\hat{f}_M(M_1|A=1,\boldsymbol{C}_1;\boldsymbol{\theta_0})}d\tilde{N}_1(u)\right\}$$

$$- E\left\{\frac{E(Y_1(u)|A_1=1,\boldsymbol{C}_1)}{\pi(u,1;\boldsymbol{\alpha_0})}\right.$$

$$\times E\left(\frac{\hat{f}_M(M_1|A=0,\boldsymbol{C}_1;\boldsymbol{\theta_0})}{\hat{f}_M(M_1|A=1,\boldsymbol{C}_1;\boldsymbol{\theta_0})}\hat{E}\left(d\tilde{N}(u)\Big|A=1,M_1,\boldsymbol{C}_1;\boldsymbol{\beta}^*,\Lambda_0^*(t)\right)\Big|A_1=1,\boldsymbol{C}_1\right)\right\}$$

$$+ E\left\{E\left(\hat{E}\left(d\tilde{N}(u)\Big|A=1,M_1,\boldsymbol{C}_1;\boldsymbol{\beta}^*,\Lambda_0^*(t)\right)\Big|A=0,\boldsymbol{C}_1\right)\right\} \text{ (If model } A \text{ is correctly specified)}$$

$$= E\left\{\frac{I(A_1=1)\hat{w}_1(\boldsymbol{\alpha_0})Y_1(u)}{\pi(u,1;\boldsymbol{\alpha_0})}\frac{\hat{f}_M(M_1|A=0,\boldsymbol{C}_1;\boldsymbol{\theta_0})}{\hat{f}_M(M_1|A=1,\boldsymbol{C}_1;\boldsymbol{\theta_0})}d\tilde{N}_1(u)\right\}$$

$$- E\left\{\frac{E(Y_1(u)|A_1=1)}{\pi(u,1;\boldsymbol{\alpha_0})}\right.$$

$$\times E\left(\frac{\hat{f}_M(M_1|A=0,\boldsymbol{C}_1;\boldsymbol{\theta_0})}{\hat{f}_M(M_1|A=1,\boldsymbol{C}_1;\boldsymbol{\theta_0})}\hat{E}\left(d\tilde{N}(u)\Big|A=1,M_1,\boldsymbol{C}_1;\boldsymbol{\beta}^*,\Lambda_0^*(t)\right)\Big|A_1=1,\boldsymbol{C}_1\right)\right\}$$

$$+ E\left\{E\left(\hat{E}\left(d\tilde{N}(u)\Big|A=1,M_1,\boldsymbol{C}_1;\boldsymbol{\beta}^*,\Lambda_0^*(t)\right)\Big|A=0,\boldsymbol{C}_1\right)\right\} \text{ (By CA.3)}$$

$$= E\left\{\frac{I(A_1=1)\hat{w}_1(\boldsymbol{\alpha_0})Y_1(u)}{\pi(u,1;\boldsymbol{\alpha_0})}\frac{\hat{f}_M(M_1|A=0,\boldsymbol{C}_1;\boldsymbol{\theta_0})}{\hat{f}_M(M_1|A=1,\boldsymbol{C}_1;\boldsymbol{\theta_0})}d\tilde{N}_1(u)\right\}$$

$$- E\left\{E\left(\frac{\hat{f}_M(M_1|A=0,\boldsymbol{C}_1;\boldsymbol{\theta_0})}{\hat{f}_M(M_1|A=1,\boldsymbol{C}_1;\boldsymbol{\theta_0})}\hat{E}\left(d\tilde{N}(u)\Big|A=1,M_1,\boldsymbol{C}_1;\boldsymbol{\beta}^*,\Lambda_0^*(t)\right)\Big|A_1=1,\boldsymbol{C}_1\right)\right\}$$

$$+ E\left\{E\left(\hat{E}\left(d\tilde{N}(u)\Big|A=1,M_1,\boldsymbol{C}_1;\boldsymbol{\beta}^*,\Lambda_0^*(t)\right)\Big|A=0,\boldsymbol{C}_1\right)\right\} \text{ (If model } A \text{ is correctly specified)}$$

$$= E\left\{\frac{I(A_1=1)\hat{w}_1(\boldsymbol{\alpha_0})Y_1(u)}{\pi(u,1;\boldsymbol{\alpha_0})}\frac{\hat{f}_M(M_1|A=0,\boldsymbol{C}_1;\boldsymbol{\theta_0})}{\hat{f}_M(M_1|A=1,\boldsymbol{C}_1;\boldsymbol{\theta_0})}d\tilde{N}_1(u)\right\}$$

$$- E\left\{\int_m \frac{\hat{f}_M(m|A=0,\boldsymbol{C}_1;\boldsymbol{\theta_0})}{\hat{f}_M(m|A=1,\boldsymbol{C}_1;\boldsymbol{\theta_0})}\hat{E}\left(d\tilde{N}(u)\Big|A=1,m,\boldsymbol{C}_1;\boldsymbol{\beta}^*,\Lambda_0^*(t)\right)f_M(m|A=1,\boldsymbol{C}_1)dm\right\}$$

$$+ E\left\{E\left(\hat{E}\left(d\tilde{N}(u)\Big|A=1,M_1,\boldsymbol{C}_1;\boldsymbol{\beta}^*,\Lambda_0^*(t)\right)\Big|A=0,\boldsymbol{C}_1\right)\right\}$$

$$= E\left\{\frac{I(A_1=1)\hat{w}_1(\boldsymbol{\alpha_0})Y_1(u)}{\pi(u,1;\boldsymbol{\alpha_0})}\frac{\hat{f}_M(M_1|A=0,\boldsymbol{C}_1;\boldsymbol{\theta_0})}{\hat{f}_M(M_1|A=1,\boldsymbol{C}_1;\boldsymbol{\theta_0})}d\tilde{N}_1(u)\right\}$$

$$- E\left\{\int_m \hat{f}_M(m|A=0,\boldsymbol{C}_1;\boldsymbol{\theta_0})\hat{E}\left(d\tilde{N}(u)\Big|A=1,m,\boldsymbol{C}_1;\boldsymbol{\beta}^*,\Lambda_0^*(t)\right)dm\right\}$$

$$+ E\left\{E\left(\hat{E}\left(d\tilde{N}(u)\Big|A=1,M_1,\boldsymbol{C}_1;\boldsymbol{\beta}^*,\Lambda_0^*(t)\right)\Big|A=0,\boldsymbol{C}_1\right)\right\} \text{ (If model } M \text{ is correctly specified)}$$

$$= E\left\{\frac{I(A_1=1)\hat{w}_1(\boldsymbol{\alpha_0})Y_1(u)}{\pi(u,1;\boldsymbol{\alpha_0})}\frac{\hat{f}_M(M_1|A=0,\boldsymbol{C}_1;\boldsymbol{\theta_0})}{\hat{f}_M(M_1|A=1,\boldsymbol{C}_1;\boldsymbol{\theta_0})}d\tilde{N}_1(u)\right\} \text{ (If model } M \text{ is correctly specified)}$$

$$= E\left\{E\left(\frac{Y_1(u)}{\hat{P}(A_1=1|\boldsymbol{C}_1;\boldsymbol{\alpha_0})\pi(u,1;\boldsymbol{\alpha_0})}\frac{\hat{f}_M(M_1|A=0,\boldsymbol{C}_1;\boldsymbol{\theta_0})}{\hat{f}_M(M_1|A=1,\boldsymbol{C}_1;\boldsymbol{\theta_0})}d\tilde{N}_1(u)\Big|A_1=1,\boldsymbol{C}_1\right)P(A_1=1|\boldsymbol{C}_1)\right\}$$

$$= E\left\{E\left(\frac{Y_1(u)}{\pi(u,1;\boldsymbol{\alpha_0})}\frac{\hat{f}_M(M_1|A=0,\boldsymbol{C}_1;\boldsymbol{\theta_0})}{\hat{f}_M(M_1|A=1,\boldsymbol{C}_1;\boldsymbol{\theta_0})}d\tilde{N}_1(u)\bigg|A_1=1,\boldsymbol{C}_1\right)\right\} \text{ (If model } A \text{ is correctly specified)}$$

$$= E\left\{E\left(E\left(\frac{Y_1(u)}{\pi(u,1;\boldsymbol{\alpha_0})}\frac{\hat{f}_M(M_1|A=0,\boldsymbol{C}_1;\boldsymbol{\theta_0})}{\hat{f}_M(M_1|A=1,\boldsymbol{C}_1;\boldsymbol{\theta_0})}d\tilde{N}_1(u)\bigg|A_1=1,M_1,\boldsymbol{C}_1\right)\bigg|A_1=1,\boldsymbol{C}_1\right)\right\}$$

$$= E\left\{E\left(\frac{E(Y_1(u)|A_1=1,M_1,\boldsymbol{C}_1)}{\pi(u,1;\boldsymbol{\alpha_0})}\frac{\hat{f}_M(M_1|A=0,\boldsymbol{C}_1;\boldsymbol{\theta_0})}{\hat{f}_M(M_1|A=1,\boldsymbol{C}_1;\boldsymbol{\theta_0})}E(d\tilde{N}_1(u)|A_1=1,M_1,\boldsymbol{C}_1)\bigg|A_1=1,\boldsymbol{C}_1\right)\right\}$$

(By CA. 1)

$$= E\left\{E\left(\frac{E(Y_1(u)|A_1=1,\boldsymbol{C}_1)}{\pi(u,1;\boldsymbol{\alpha_0})}\frac{\hat{f}_M(M_1|A=0,\boldsymbol{C}_1;\boldsymbol{\theta_0})}{\hat{f}_M(M_1|A=1,\boldsymbol{C}_1;\boldsymbol{\theta_0})}E(d\tilde{N}_1(u)|A_1=1,M_1,\boldsymbol{C}_1)\bigg|A_1=1,\boldsymbol{C}_1\right)\right\} \text{ (By CA. 2)}$$

$$= E\left\{E\left(\frac{\hat{f}_M(M_1|A=0,\boldsymbol{C}_1;\boldsymbol{\theta_0})}{\hat{f}_M(M_1|A=1,\boldsymbol{C}_1;\boldsymbol{\theta_0})}E(d\tilde{N}_1(u)|A_1=1,M_1,\boldsymbol{C}_1)\bigg|A_1=1,\boldsymbol{C}_1\right)\right\}$$

(If model $A$ is correctly specified)

$$= E\left\{\int_m\frac{\hat{f}_M(M_1|A=0,\boldsymbol{C}_1;\boldsymbol{\theta_0})}{\hat{f}_M(M_1|A=1,\boldsymbol{C}_1;\boldsymbol{\theta_0})}E(d\tilde{N}_1(u)|A_1=1,M_1,\boldsymbol{C}_1)f_M(M_1|A=1,\boldsymbol{C}_1)\right\}$$

$$= E\left\{\int_m \hat{f}_M(M_1|A=0,\boldsymbol{C}_1;\boldsymbol{\theta_0})E(d\tilde{N}_1(u)|A_1=1,M_1,\boldsymbol{C}_1)\right\} \text{ (If model } M \text{ is correctly specified)}$$

$$= E\{E(E(d\tilde{N}_1(u)|A_1=1,M_1,\boldsymbol{C}_1)|A_1=1,\boldsymbol{C}_1)\} \text{ (If model } M \text{ is correctly specified)}$$

$$= dQ(u;1,0)$$

## Appendix 3.2 Asymptotic distribution

We show that $\sqrt{n}\left(\hat{Q}^{TR}\left(t;a,a^*;\hat{\boldsymbol{\alpha}},\hat{\boldsymbol{\theta}},\hat{\boldsymbol{\beta}},\hat{\Lambda}_0(t;\hat{\boldsymbol{\beta}})\right)-Q(t;a,a^*)\right)$ is normally asymptotic with mean 0. The asymptotic variance will also be presented. Because the target can be decomposed into four elements,

$$\sqrt{n}\left(\hat{Q}^{TR}\left(t;a,a^*;\hat{\boldsymbol{\alpha}},\hat{\boldsymbol{\theta}},\hat{\boldsymbol{\beta}},\hat{\Lambda}_0(t;\hat{\boldsymbol{\beta}})\right)-Q(t;a,a^*)\right)$$

$$=\sqrt{n}\left(\hat{Q}^{TR}\left(t;a,a^*;\hat{\boldsymbol{\alpha}},\hat{\boldsymbol{\theta}},\hat{\boldsymbol{\beta}},\hat{\Lambda}_0(t;\hat{\boldsymbol{\beta}})\right)-\hat{Q}^{TR}\left(t;a,a^*;\boldsymbol{\alpha}^*,\hat{\boldsymbol{\theta}},\hat{\boldsymbol{\beta}},\hat{\Lambda}_0(t;\hat{\boldsymbol{\beta}})\right)\right)$$

$$+\sqrt{n}\left(\hat{Q}^{TR}\left(t;a,a^*;\boldsymbol{\alpha}^*,\hat{\boldsymbol{\theta}},\hat{\boldsymbol{\beta}},\hat{\Lambda}_0(t;\hat{\boldsymbol{\beta}})\right)-\hat{Q}^{TR}\left(t;a,a^*;\boldsymbol{\alpha}^*,\boldsymbol{\theta}^*,\hat{\boldsymbol{\beta}},\hat{\Lambda}_0(t;\hat{\boldsymbol{\beta}})\right)\right)$$

$$+\sqrt{n}\left(\hat{Q}^{TR}\left(t;a,a^*;\boldsymbol{\alpha}^*,\boldsymbol{\theta}^*,\hat{\boldsymbol{\beta}},\hat{\Lambda}_0(t;\hat{\boldsymbol{\beta}})\right)-\hat{Q}^{TR}(t;a,a^*;\boldsymbol{\alpha}^*,\boldsymbol{\theta}^*,\boldsymbol{\beta}^*,\Lambda_0^*(t))\right)$$

$$+\sqrt{n}\left(\hat{Q}^{TR}(t;a,a^*;\boldsymbol{\alpha}^*,\boldsymbol{\theta}^*,\boldsymbol{\beta}^*,\Lambda_0^*(t;\boldsymbol{\beta}^*))-Q(t;a,a^*)\right)$$

Using Taylor expansion, the first element is obtained by

$$\sqrt{n}\left(\hat{Q}^{TR}\left(t;a,a^*;\hat{\boldsymbol{\alpha}},\hat{\boldsymbol{\theta}},\hat{\boldsymbol{\beta}},\hat{\Lambda}_0(t;\hat{\boldsymbol{\beta}})\right)-\hat{Q}^{TR}\left(t;a,a^*;\boldsymbol{\alpha}^*,\hat{\boldsymbol{\theta}},\hat{\boldsymbol{\beta}},\hat{\Lambda}_0(t;\hat{\boldsymbol{\beta}})\right)\right)$$

$$= \int_0^t \mathbb{P}_n \left\{ \frac{\partial}{\partial \boldsymbol{\alpha}} \hat{Z}_i(u; a; \boldsymbol{\alpha}) \frac{\hat{f}(M_i|a^*, \boldsymbol{C}_i; \boldsymbol{\theta})}{\hat{f}(M_i|a, \boldsymbol{C}_i; \boldsymbol{\theta})} [d\tilde{N}_i(u) - \hat{E}(d\tilde{N}_i(u)|a, M_i, \boldsymbol{C}_i; \boldsymbol{\beta})] \right.$$
$$+ \frac{\partial}{\partial \boldsymbol{\alpha}} \hat{Z}_i(u; a^*; \boldsymbol{\alpha}) [\hat{E}(d\tilde{N}_i(u)|a, M_i, \boldsymbol{C}_i; \boldsymbol{\beta})$$
$$\left. - d\hat{Q}(u; a, a^*|\boldsymbol{C}_i; \boldsymbol{\theta}, \boldsymbol{\beta})] \right\} \bigg|_{\boldsymbol{\alpha}=\boldsymbol{\alpha}^*, \boldsymbol{\theta}=\boldsymbol{\theta}^*, \boldsymbol{\beta}=\boldsymbol{\beta}^*, \Lambda_0(u;\boldsymbol{\beta})=\Lambda_0^*(u)} \sqrt{n}(\hat{\boldsymbol{\alpha}} - \boldsymbol{\alpha}^*) + o_p(1)$$

as $\hat{\boldsymbol{\alpha}} \xrightarrow{a.s.} \boldsymbol{\alpha}^*$ where

$$\frac{\partial}{\partial \boldsymbol{\alpha}} \hat{Z}_i(u; a; \boldsymbol{\alpha}) = \frac{\partial}{\partial \boldsymbol{\alpha}} \frac{I(A_i = a)\hat{w}_i(\boldsymbol{\alpha})Y_i(u)}{\frac{1}{n}\sum_{j=1}^n I(A_j = a)\hat{w}_j(\boldsymbol{\alpha})Y_j(u)}$$

$$= \frac{[I(A_i = a)Y_i(u)] \left[ \left(\frac{1}{n}\sum_{j=1}^n I(A_j = a)\hat{w}_j(\boldsymbol{\alpha})Y_j(u)\right) \frac{\partial}{\partial \boldsymbol{\alpha}} \hat{w}_i(\boldsymbol{\alpha}) - \hat{w}_i(\boldsymbol{\alpha}) \left(\frac{1}{n}\sum_{j=1}^n I(A_j = a) \frac{\partial}{\partial \boldsymbol{\alpha}} \hat{w}_j(\boldsymbol{\alpha})Y_j(u)\right) \right]}{\left(\frac{1}{n}\sum_{j=1}^n I(A_j = a)\hat{w}_j(\boldsymbol{\alpha})Y_j(u)\right)^2}$$

and where

$$\frac{\partial}{\partial \boldsymbol{\alpha}} \hat{w}_i(\boldsymbol{\alpha}) = \frac{\partial}{\partial \boldsymbol{\alpha}} \left( \frac{A_i}{\hat{E}(A_i|\boldsymbol{C}_i; \boldsymbol{\alpha})} + \frac{1 - A_i}{1 - \hat{E}(A_i|\boldsymbol{C}_i; \boldsymbol{\alpha})} \right)$$

$$= \left( \frac{1 - A_i}{\left(1 - \hat{E}(A_i|\boldsymbol{C}_i; \boldsymbol{\alpha})\right)^2} - \frac{A_i}{\left(\hat{E}(A_i|\boldsymbol{C}_i; \boldsymbol{\alpha})\right)^2} \right) \frac{\partial}{\partial \boldsymbol{\alpha}} \hat{E}(A_i|\boldsymbol{C}_i; \boldsymbol{\alpha})$$

$$= \left( \frac{1 - A_i}{\left(1 - \hat{E}(A_i|\boldsymbol{C}_i; \boldsymbol{\alpha})\right)^2} - \frac{A_i}{\left(\hat{E}(A_i|\boldsymbol{C}_i; \boldsymbol{\alpha})\right)^2} \right) \frac{\boldsymbol{V}_{1,i} \exp(\boldsymbol{\alpha}^T \boldsymbol{V}_{1,i})}{\left(1 + \exp(\boldsymbol{\alpha}^T \boldsymbol{V}_{1,i})\right)^2}$$

Using Equation (B-8) in Su, Steele and Shrier (2020),

$$\sqrt{n}(\hat{\boldsymbol{\alpha}} - \boldsymbol{\alpha}^*) = E\left[ \frac{\exp(\boldsymbol{\alpha}^{*T}\boldsymbol{V}_1)\boldsymbol{V}_1\boldsymbol{V}_1'}{(1 + \exp(\boldsymbol{\alpha}^T\boldsymbol{V}_1))^2} \right]^{-1} n^{-\frac{1}{2}} \sum_{i=1}^n \boldsymbol{V}_{1,i} \left( A_i - \frac{\exp(\boldsymbol{\alpha}^{*T}\boldsymbol{V}_{1,i})}{1 + \exp(\boldsymbol{\alpha}^{*T}\boldsymbol{V}_{1,i})} \right) + o_p(1)$$

where

$$\boldsymbol{V}_{1,i} = \left(1, \boldsymbol{C}_i^T\right)^T$$

$$\boldsymbol{V}_1 = (\boldsymbol{V}_{1,1}, \boldsymbol{V}_{1,2}, \ldots, \boldsymbol{V}_{1,n})$$

hence,

$$\sqrt{n}\left(\hat{Q}^{TR}\left(t; a, a^*; \hat{\boldsymbol{\alpha}}, \hat{\boldsymbol{\theta}}, \hat{\boldsymbol{\beta}}, \hat{\Lambda}_0(t; \hat{\boldsymbol{\beta}})\right) - \hat{Q}^{TR}\left(t; a, a^*; \boldsymbol{\alpha}^*, \hat{\boldsymbol{\theta}}, \hat{\boldsymbol{\beta}}, \hat{\Lambda}_0(t; \hat{\boldsymbol{\beta}})\right)\right)$$

$$= \int_0^t \mathbb{P}_n \left\{ \frac{\partial}{\partial \boldsymbol{\alpha}} \hat{Z}_i(u; a; \boldsymbol{\alpha}) \exp\left( \frac{1}{\sigma_M^2} \left[ (\theta_A(M_i - \theta_0 - \boldsymbol{\theta}_C^T \boldsymbol{C}))(a^* - a) - \frac{1}{2}\theta_A^2(a^{*2} - a^2) \right] \right) \right.$$
$$\times [d\tilde{N}_i(u) - d\Lambda_0(u; \beta) \exp(\beta_A a + \beta_M M_i + \boldsymbol{\beta}_C^T \boldsymbol{C}_i)]$$

$$
\begin{aligned}
&+ \frac{\partial}{\partial \boldsymbol{\alpha}} \hat{Z}_i(u; a^*; \boldsymbol{\alpha})\big[\mathrm{d}\Lambda_0(u;\beta)\exp(\beta_A a + \beta_M M_i + \boldsymbol{\beta}_C^{\mathrm{T}} \boldsymbol{C}_i) \\
&\qquad - \mathrm{d}\hat{Q}(u; a, a^*|\boldsymbol{C}_i; \boldsymbol{\theta}, \boldsymbol{\beta})\big]\Big\}\Big|_{\boldsymbol{\alpha}=\boldsymbol{\alpha}^*, \boldsymbol{\theta}=\boldsymbol{\theta}^*, \boldsymbol{\beta}=\boldsymbol{\beta}^*, \Lambda_0(u;\beta)=\Lambda_0^*(u)} \sqrt{n}(\hat{\boldsymbol{\alpha}} - \boldsymbol{\alpha}^*) + o_p(1) \\
&= F_1\big(t; a, a^*; \boldsymbol{\alpha}^*, \boldsymbol{\theta}^*, \boldsymbol{\beta}^*, \Lambda_0^*(u)\big) \\
&\qquad \times E\left[\frac{\exp(\boldsymbol{\alpha}^{*\mathrm{T}}\boldsymbol{V}_1)\boldsymbol{V}_1\boldsymbol{V}_1^{\mathrm{T}}}{(1+\exp(\boldsymbol{\alpha}^{\mathrm{T}}\boldsymbol{V}_1))^2}\right]^{-1} n^{-1/2} \sum_{i=1}^n \boldsymbol{V}_{1,i}\left(A_i - \frac{\exp(\boldsymbol{\alpha}^{*\mathrm{T}}\boldsymbol{V}_{1,i})}{1+\exp(\boldsymbol{\alpha}^{*\mathrm{T}}\boldsymbol{V}_{1,i})}\right) + o_p(1)
\end{aligned}
$$

(3.2.1)

where

$$
\begin{aligned}
&F_1\big(t; a, a^*; \boldsymbol{\alpha}^*, \boldsymbol{\theta}^*, \boldsymbol{\beta}^*, \Lambda_0^*(u)\big) \\
&= \int_0^t \frac{E\left[I(A_1 = a)Y_1(u)\left(\pi(u,a;\boldsymbol{\alpha}^*)B(\boldsymbol{\alpha}^*) - \widehat{w}_1(\boldsymbol{\alpha}^*)E\big(B(\boldsymbol{\alpha}^*)I(A_1=a)Y_1(u)\big)\right)\right]}{\big(\pi(u,a;\boldsymbol{\alpha}^*)\big)^2} \\
&\quad \times E\big[\Phi_1^*(a, a^*; \boldsymbol{\theta}^*)\big(\mathrm{d}\widetilde{N}_1(u) - \mathrm{d}\Lambda_0^*(u)\exp(\beta_A^* a + \beta_M^* M_1 + \boldsymbol{\beta}_C^{*\mathrm{T}} \boldsymbol{C}_1)\big)\big] \\
&\quad + \frac{E\left[I(A_1 = a^*)Y_1(u)\big[\pi(u,a^*;\boldsymbol{\alpha}^*)B(\boldsymbol{\alpha}^*) - \widehat{w}_1(\boldsymbol{\alpha}^*)E\big(B(\boldsymbol{\alpha}^*)I(A_1=a^*)Y_1(u)\big)\big]\right]}{\big(\pi(u,a^*;\boldsymbol{\alpha}^*)\big)^2} \\
&\quad \times E\bigg[\mathrm{d}\Lambda_0^*(u)\exp(\beta_A^* a + \beta_M^* M_1 + \boldsymbol{\beta}_C^{*\mathrm{T}} \boldsymbol{C}_1) \\
&\qquad - \int_m \mathrm{d}\Lambda_0^*(u)\exp(\beta_A^* a + \beta_M^* m + \boldsymbol{\beta}_C^{*\mathrm{T}} \boldsymbol{C}_1)\Phi_1(a^*; \boldsymbol{\theta}^*)\mathrm{d}m\bigg]\bigg\} \\
&\pi(u, a; \boldsymbol{\alpha}^*) = E\big(I(A_1 = a)\widehat{w}_1(\boldsymbol{\alpha}^*)Y_1(u)\big) \\
&B(\boldsymbol{\alpha}^*) = \frac{\partial}{\partial \boldsymbol{\alpha}} \widehat{w}_1(\boldsymbol{\alpha})\bigg|_{\boldsymbol{\alpha}=\boldsymbol{\alpha}^*} = \left(\frac{1-A_1}{\big(1-\hat{E}(A_1|\boldsymbol{C}_1;\boldsymbol{\alpha})\big)^2} - \frac{A_1}{\big(\hat{E}(A_1|\boldsymbol{C}_1;\boldsymbol{\alpha})\big)^2}\right)\frac{\boldsymbol{V}_{1,1}\exp(\boldsymbol{\alpha}^{\mathrm{T}}\boldsymbol{V}_{1,1})}{(1+\exp(\boldsymbol{\alpha}^{\mathrm{T}}\boldsymbol{V}_{1,1}))^2} \\
&\Phi_i(a^*; \boldsymbol{\theta}^*) = \frac{1}{\sqrt{2\pi\sigma_M^{*2}}} \exp\left(\frac{-(m - \theta_0^* - \theta_A^* a^* - \boldsymbol{\theta}_C^{*\mathrm{T}}\boldsymbol{C}_i)^2}{2\sigma_M^{*2}}\right) \\
&\Phi_i^*(a, a^*; \boldsymbol{\theta}^*) = \frac{\hat{f}(M_i|a^*, \boldsymbol{C}_i; \boldsymbol{\theta}^*)}{\hat{f}(M_i|a, \boldsymbol{C}_i; \boldsymbol{\theta}^*)} \\
&\qquad\qquad\quad = \exp\left(\frac{-1}{\sigma_M^{*2}}\bigg[\Big(\theta_A^*\big(M_i - \theta_0^* - \boldsymbol{\theta}_C^{*\mathrm{T}}\boldsymbol{C}\big)\Big)(a - a^*) - \frac{1}{2}\theta_A^{*2}(a^2 - a^{*2})\bigg]\right)
\end{aligned}
$$

For the second element,

$$
\sqrt{n}\left(\hat{Q}^{TR}\big(t; a, a^*; \boldsymbol{\alpha}^*, \hat{\boldsymbol{\theta}}, \hat{\boldsymbol{\beta}}, \hat{\Lambda}_0(t; \hat{\boldsymbol{\beta}})\big) - \hat{Q}^{TR}\big(t; a, a^*; \boldsymbol{\alpha}^*, \boldsymbol{\theta}^*, \hat{\boldsymbol{\beta}}, \hat{\Lambda}_0(t; \hat{\boldsymbol{\beta}})\big)\right)
$$

$$= \sqrt{n}\left(\hat{Q}^{TR}(t;a,a^*;\boldsymbol{\alpha}^*,\boldsymbol{\theta},\boldsymbol{\beta},\Lambda_0(t;\boldsymbol{\beta}))\right.$$
$$\left.- \hat{Q}^{TR}(t;a,a^*;\boldsymbol{\alpha}^*,\boldsymbol{\theta}^*,\boldsymbol{\beta},\Lambda_0(t;\boldsymbol{\beta}))\right)\Big|_{\boldsymbol{\theta}=\boldsymbol{\theta}^*,\boldsymbol{\beta}=\boldsymbol{\beta}^*,\Lambda_0(t;\boldsymbol{\beta})=\Lambda_0^*(t)}$$
$$+ \frac{\partial}{\partial \boldsymbol{\theta}} \hat{Q}^{TR}(t;a,a^*;\boldsymbol{\alpha}^*,\boldsymbol{\theta},\boldsymbol{\beta},\Lambda_0(t;\boldsymbol{\beta}))\Big|_{\boldsymbol{\theta}=\boldsymbol{\theta}^*,\boldsymbol{\beta}=\boldsymbol{\beta}^*,\Lambda_0(t;\boldsymbol{\beta})=\Lambda_0^*(t)} \sqrt{n}(\hat{\boldsymbol{\theta}} - \boldsymbol{\theta}^*)$$
$$+ o_p(1)$$
$$= \int_0^t \mathbb{P}_n \left\{\hat{Z}_i(u;a;\boldsymbol{\alpha}^*)[d\tilde{N}_i(u) - d\Lambda_0^*(u)\exp(\beta_A^* a + \beta_M^* M_i + \boldsymbol{\beta}_C^{*T}\boldsymbol{C}_i)]\Phi_i^*(a,a^*;\boldsymbol{\theta}^*)\right.$$
$$\times \boldsymbol{G}_{1,i}(a,a^*;\boldsymbol{\theta}^*) + \left(1 - \hat{Z}_i(u;a^*;\boldsymbol{\alpha}^*)\right) \times \boldsymbol{\rho}_i(u;a,a^*;\boldsymbol{\theta}^*,\boldsymbol{\beta}^*,\Lambda_0^*(u))\Big\}$$
$$\times \sqrt{n}(\hat{\boldsymbol{\theta}} - \boldsymbol{\theta}^*) + o_p(1)$$

where

$$\boldsymbol{G}_{1,i}(a,a^*;\boldsymbol{\theta}^*) = \begin{bmatrix} \dfrac{\theta_A(a-a^*)}{\sigma_M^{*2}} \\ \dfrac{\theta_A^*(a^2 - a^{*2}) - (M_i - \theta_0^* - \boldsymbol{\theta}_C^{*T}\boldsymbol{C}_i)(a-a^*)}{\sigma_M^{*2}} \\ \dfrac{\boldsymbol{C}_i \theta_A^*(a-a^*)}{\sigma_M^{*2}} \\ \dfrac{\theta_A^*(M_i - \theta_0^* - \boldsymbol{\theta}_C^{*T}\boldsymbol{C}_i)(a-a^*) - \frac{1}{2}\theta_A^{*2}(a^2-a^{*2})}{\sigma_M^{*4}} \end{bmatrix}$$

$$\boldsymbol{\rho}_i(u;a,a^*;\boldsymbol{\theta}^*,\boldsymbol{\beta}^*,\Lambda_0^*(u)) = \frac{\partial}{\partial \boldsymbol{\theta}}\left(d\hat{Q}(u;a,a^*|\boldsymbol{C}_i;\boldsymbol{\theta},\boldsymbol{\beta})\right)\Big|_{\boldsymbol{\theta}=\boldsymbol{\theta}^*,\boldsymbol{\beta}=\boldsymbol{\beta}^*,\Lambda_0(u;\boldsymbol{\beta})=\Lambda_0^*(u)}$$
$$= \int_m d\Lambda_0^*(u)\exp(\beta_A^* a + \beta_M^* m + \boldsymbol{\beta}_C^{*T}\boldsymbol{C}_i) \times \boldsymbol{G}_{2,i}(a^*;\boldsymbol{\theta}^*) \times \Phi_i(a^*;\boldsymbol{\theta}^*)\,dm$$

$$\boldsymbol{G}_{2,i}(a^*;\boldsymbol{\theta}^*) = \begin{bmatrix} \dfrac{1}{\sigma_M^{*2}}(M_i - \theta_0^* - \theta_A^* a^* - \boldsymbol{\theta}_C^{*T}\boldsymbol{C}_i) \\ \dfrac{a^*}{\sigma_M^{*2}}(M_i - \theta_0^* - \theta_A^* a^* - \boldsymbol{\theta}_C^{*T}\boldsymbol{C}_i) \\ \dfrac{\boldsymbol{C}_i}{\sigma_M^{*2}}(M_i - \theta_0^* - \theta_A^* a^* - \boldsymbol{\theta}_C^{*T}\boldsymbol{C}_i) \\ \dfrac{-1}{2\sigma_M^{*3}} + \dfrac{(M_i - \theta_0^* - \theta_A^* a^* - \boldsymbol{\theta}_C^{*T}\boldsymbol{C}_i)^2}{2\sigma_M^{*4}} \end{bmatrix}$$

Let

$$\mathbb{U}^M(\boldsymbol{\theta}) = \frac{\partial}{\partial \boldsymbol{\theta}} \left( \frac{-n}{2} \log(\sigma_M^2) - \frac{1}{2\sigma_M^2} (\boldsymbol{M} - \boldsymbol{V}_2^T \boldsymbol{\theta}_-)' (\boldsymbol{M} - \boldsymbol{V}_2^T \boldsymbol{\theta}_-) \right)$$

$$= \begin{bmatrix} \frac{1}{\sigma_M^2} \boldsymbol{V}_2 (\boldsymbol{M} - \boldsymbol{V}_2^T \boldsymbol{\theta}_-) \\ \frac{-n}{2\sigma_M^2} + \frac{1}{2\sigma_M^4} (\boldsymbol{M} - \boldsymbol{V}_2^T \boldsymbol{\theta}_-)^T (\boldsymbol{M} - \boldsymbol{V}_2^T \boldsymbol{\theta}_-) \end{bmatrix}$$

be the maximum likelihood estimating equation for the linear model with respect to M. Where

$$\boldsymbol{M} = (M_1, \ldots, M_n)^T$$

$$\boldsymbol{V}_2 = (\boldsymbol{V}_{2,1}, \ldots, \boldsymbol{V}_{2,n}); \quad \boldsymbol{V}_{2,i} = (1, A_i, \boldsymbol{C}_i^T)^T$$

$$\boldsymbol{\theta}_- = (\theta_0, \theta_A, \boldsymbol{\theta}_C^T)^T$$

by the integral form of the mean value theorem (Feng et al., 2013),

$$\mathbb{U}^M(\widehat{\boldsymbol{\theta}}) = \mathbb{U}^M(\boldsymbol{\theta}^*) + \frac{\partial}{\partial \boldsymbol{\theta}^T} \mathbb{U}^M(\boldsymbol{\theta}) (\widehat{\boldsymbol{\theta}} - \boldsymbol{\theta}^*) \Big|_{\boldsymbol{\theta} = \boldsymbol{\theta}^*} + o(\|\widehat{\boldsymbol{\theta}} - \boldsymbol{\theta}^*\|)$$

Because the $\widehat{\boldsymbol{\theta}}$ maximized the log-likelihood such that $\mathbb{U}^M(\widehat{\boldsymbol{\theta}}) = 0$ and $\widehat{\boldsymbol{\theta}} \xrightarrow{a.s.} \boldsymbol{\theta}^*$, which implies that $\sqrt{n}(\widehat{\boldsymbol{\theta}} - \boldsymbol{\theta}^*) = \mathbb{A}_n^M(\boldsymbol{\theta}^*)^{-1} n^{-1/2} \mathbb{U}^M(\boldsymbol{\theta}^*) + o_p(1)$, where $\mathbb{A}_n^M(\boldsymbol{\theta}^*) = -n^{-1} \frac{\partial}{\partial \boldsymbol{\theta}^T} \mathbb{U}^M(\boldsymbol{\theta}) \Big|_{\boldsymbol{\theta} = \boldsymbol{\theta}^*}$. Then

$$\sqrt{n} \left( \widehat{Q}^{TR}(t; a, a^*; \boldsymbol{\alpha}^*, \widehat{\boldsymbol{\theta}}, \widehat{\boldsymbol{\beta}}, \widehat{\Lambda}_0(t; \widehat{\boldsymbol{\beta}})) - \widehat{Q}^{TR}(t; a, a^*; \boldsymbol{\alpha}^*, \boldsymbol{\theta}^*, \widehat{\boldsymbol{\beta}}, \widehat{\Lambda}_0(t; \widehat{\boldsymbol{\beta}})) \right)$$

$$= F_2(t; a, a^*; \boldsymbol{\alpha}^*, \boldsymbol{\theta}^*, \boldsymbol{\beta}^*, \Lambda_0^*(u)) \mathbb{A}_n^M(\boldsymbol{\theta}^*)^{-1} n^{-1/2} \sum_{i=1}^n \mathbb{U}_i^M(\boldsymbol{\theta}^*) + o_p(1) \qquad (3.2.2)$$

where
$F_2(t; a, a^*; \boldsymbol{\alpha}^*, \boldsymbol{\theta}^*, \boldsymbol{\beta}^*, \Lambda_0^*(u))$

$$= \int_0^t E \left( \frac{I(A_1 = a) \widehat{w}_1(\boldsymbol{\alpha}^*) Y_1(u)}{\pi(u, a; \boldsymbol{\alpha}^*)} [d\widetilde{N}_1(u) - d\Lambda_0^*(u) \exp(\beta_A^* a + \beta_M^* M_1 + \boldsymbol{\beta}_C^{*T} \boldsymbol{C}_1)] \times \Phi_1^*(a, a^*; \boldsymbol{\theta}^*) G_{1,1}(a, a^*; \boldsymbol{\theta}^*) \right.$$

$$\left. + \left(1 - \frac{I(A_1 = a) \widehat{w}_1(\boldsymbol{\alpha}^*) Y_1(u)}{\pi(u, a^*; \boldsymbol{\alpha}^*)}\right) \rho_1(u; a, a^*; \boldsymbol{\theta}^*, \boldsymbol{\beta}^*, \Lambda_0^*(u)) \right)$$

$$\mathbb{U}_i^M(\boldsymbol{\theta}^*) = \begin{bmatrix} \frac{1}{\sigma_M^{*2}} \boldsymbol{V}_{2,i} (M_i - \boldsymbol{V}_{2,i}^T \boldsymbol{\theta}_-^*) \\ \frac{-1}{2\sigma_M^{*2}} + \frac{1}{2\sigma_M^{*4}} (M_i - \boldsymbol{V}_{2,i}^T \boldsymbol{\theta}_-^*)^2 \end{bmatrix}$$

For the third element,

$$\sqrt{n}\left(\hat{Q}^{TR}\left(t;a,a^*;\boldsymbol{\alpha}^*,\boldsymbol{\theta}^*,\widehat{\boldsymbol{\beta}},\widehat{\Lambda}_0(t;\widehat{\boldsymbol{\beta}})\right) - \hat{Q}^{TR}\left(t;a,a^*;\boldsymbol{\alpha}^*,\boldsymbol{\theta}^*,\boldsymbol{\beta}^*,\Lambda_0^*(t)\right)\right)$$

$$= \sqrt{n}\left(\hat{Q}^{TR}(t;a,a^*;\boldsymbol{\alpha}^*,\boldsymbol{\theta}^*,\boldsymbol{\beta},\Lambda_0(t;\boldsymbol{\beta}))\right.$$

$$\left. - \hat{Q}^{TR}(t;a,a^*;\boldsymbol{\alpha}^*,\boldsymbol{\theta}^*,\boldsymbol{\beta}^*,\Lambda_0^*(t;\boldsymbol{\beta}^*)))\right|_{\boldsymbol{\beta}=\boldsymbol{\beta}^*,\Lambda_0(t;\boldsymbol{\beta})=\Lambda_0^*(t)}$$

$$+ \frac{\partial}{\partial \boldsymbol{\beta}}\hat{Q}^{TR}(t;a,a^*;\boldsymbol{\alpha}^*,\boldsymbol{\theta}^*,\boldsymbol{\beta},\Lambda_0(t;\boldsymbol{\beta}))\bigg|_{\boldsymbol{\beta}=\boldsymbol{\beta}^*,\Lambda_0(t;\boldsymbol{\beta})=\Lambda_0^*(t)} \sqrt{n}(\widehat{\boldsymbol{\beta}}-\boldsymbol{\beta}^*)$$

$$+ \int_0^t \frac{\partial}{\partial d\Lambda_0(u;\boldsymbol{\beta})} d\hat{Q}^{TR}(u;a,a^*;\boldsymbol{\alpha}^*,\boldsymbol{\theta}^*,\boldsymbol{\beta},\Lambda_0(u;\boldsymbol{\beta}))\bigg|_{\boldsymbol{\beta}=\boldsymbol{\beta}^*,\Lambda_0(u;\boldsymbol{\beta})=\Lambda_0^*(u)}$$

$$\times \sqrt{n}d\left(\widehat{\Lambda}_0(u;\widehat{\boldsymbol{\beta}}) - \Lambda_0^*(u;\boldsymbol{\beta}^*)\right) + o_p(1)$$

$$= \int_0^t \mathbb{P}_n\left\{\left(\hat{Z}_i(u;a^*;\boldsymbol{\alpha}^*) - \hat{Z}_i(u;a;\boldsymbol{\alpha}^*)\frac{\hat{f}(M_i|a^*,C_i;\boldsymbol{\theta}^*)}{\hat{f}(M_i|a,C_i;\boldsymbol{\theta}^*)}\right)\right.$$

$$\times \left[\exp\left(\boldsymbol{\beta}^{*T}(a,M_i,C_i^T)^T\right)(a,M_i,C_i^T)^T d\Lambda_0^*(u)\right]$$

$$+ \left(1 - \hat{Z}_i(u;a^*;\boldsymbol{\alpha}^*)\right)$$

$$\left. \times \int_m d\Lambda_0^*(u)\exp\left(\boldsymbol{\beta}^{*T}(a,m,C_i^T)^T\right)(a,m,C_i^T)^T \Phi_i(a^*;\boldsymbol{\theta}^*)dm\right\}\sqrt{n}(\widehat{\boldsymbol{\beta}}-\boldsymbol{\beta}^*)$$

$$+ \int_0^t \mathbb{P}_n\left\{\left(\hat{Z}_i(u;a^*;\boldsymbol{\alpha}^*) - \hat{Z}_i(u;a;\boldsymbol{\alpha}^*)\frac{\hat{f}(M_i|a^*,C_i;\boldsymbol{\theta}^*)}{\hat{f}(M_i|a,C_i;\boldsymbol{\theta}^*)}\right)\right.$$

$$\times \left[\exp\left(\boldsymbol{\beta}^{*T}(a,M_i,C_i^T)^T\right)\right]$$

$$\left. + \left(1 - \hat{Z}_i(u;a^*;\boldsymbol{\alpha}^*)\right) \times \int_m \exp\left(\boldsymbol{\beta}^{*T}(a,m,C_i^T)^T\right)\Phi_i(a^*;\boldsymbol{\theta}^*)dm\right\}$$

$$\times d\left(\sqrt{n}\left(\widehat{\Lambda}_0(u;\widehat{\boldsymbol{\beta}}) - \Lambda_0^*(u)\right)\right) + o_p(1)$$

If the following conditions are met,

(i) $\Pr(Y_i(\tau;a,M(a^*))=1) > 0$ and $\widetilde{N}_i(\tau;a,M(a^*))$ is bounded by a constant for all $i=1,\ldots,n$, $a=0,1$.

(ii) $X_i = (A_i, M_i, C_i^T)^T$ is bounded; that is, there exists a constant $\xi$ such that $\|X_i\| < \xi$

(iii) For $\boldsymbol{\beta}$ in some neighborhood of $\boldsymbol{\beta}_0$, $\exp(\boldsymbol{\beta}^T X_i)$ is bounded away from 0, for all $i=1,\ldots,n$.

(iv) For $\boldsymbol{\theta}$ in some neighborhood of $\boldsymbol{\theta}_0$, there exists constants $\xi_1, \xi_2 > 0$, such that $0 < \xi_1 < \frac{\hat{f}(M_i|a^*,C_i;\boldsymbol{\theta})}{\hat{f}(M_i|a,C_i;\boldsymbol{\theta})} < \xi_2 < \infty$, for all $i=1,\ldots,n$

(v) For $\boldsymbol{\alpha}$ in some neighborhood of $\boldsymbol{\alpha}_0$, there exists constants $\xi_1, \xi_2 > 0$, such that $0 < \xi_1 < \widehat{w}_i(\boldsymbol{\alpha}) < \xi_2 < \infty$, for all $i = 1, \ldots, n$

Then,

$$\sqrt{n}\left(\widehat{Q}^{TR}\left(t; a, a^*; \boldsymbol{\alpha}^*, \boldsymbol{\theta}^*, \widehat{\boldsymbol{\beta}}, \widehat{\Lambda}_0(t; \widehat{\boldsymbol{\beta}})\right) - \widehat{Q}^{TR}\left(t; a, a^*; \boldsymbol{\alpha}^*, \boldsymbol{\theta}^*, \boldsymbol{\beta}^*, \Lambda_0^*(t)\right)\right)$$

$$= \Psi_1(\boldsymbol{\alpha}^*, \boldsymbol{\theta}^*, \boldsymbol{\beta}^*, \Lambda_0^*(t))\sqrt{n}(\widehat{\boldsymbol{\beta}} - \boldsymbol{\beta}^*)$$

$$+ \int_0^t \Psi_2(u; \boldsymbol{\alpha}^*, \boldsymbol{\theta}^*, \boldsymbol{\beta}^*) d\left(\sqrt{n}\left(\widehat{\Lambda}_0(u; \widehat{\boldsymbol{\beta}}) - \Lambda_0^*(u)\right)\right) + o_p(1)$$

where

$\Psi_1(\boldsymbol{\alpha}^*, \boldsymbol{\theta}^*, \boldsymbol{\beta}^*, \Lambda_0^*(t))$

$$= E\left(\int_0^t \left(\frac{I(A_1 = a^*)}{\pi(u, a^*; \boldsymbol{\alpha}^*)} - \frac{I(A_1 = a)}{\pi(u, a; \boldsymbol{\alpha}^*)} \Phi_1^*(a, a^*; \boldsymbol{\theta}^*)\right) \widehat{w}_1(\boldsymbol{\alpha}^*) Y_1(u)\right.$$

$$\times \exp(\boldsymbol{\beta}^{*T}(a, M_1, \boldsymbol{C}_1^T)^T)(a, M_1, \boldsymbol{C}_1^T)^T d\Lambda_0^*(u)$$

$$+ \left(1 - \frac{I(A_1 = a^*)\widehat{w}_1(\boldsymbol{\alpha}^*) Y_1(u)}{\pi(u, a^*; \boldsymbol{\alpha}^*)}\right)$$

$$\left.\times \int_m d\Lambda_0^*(u) \exp(\boldsymbol{\beta}^{*T}(a, m, \boldsymbol{C}_1^T)^T)(a, m, \boldsymbol{C}_1^T)^T \Phi_1(a^*; \boldsymbol{\theta}^*) dm\right)$$

$\Psi_2(u; \boldsymbol{\alpha}^*, \boldsymbol{\theta}^*, \boldsymbol{\beta}^*)$

$$= E\left(\left(\frac{I(A_1 = a^*)}{\pi(u, a^*; \boldsymbol{\alpha}^*)} - \frac{I(A_1 = a)}{\pi(u, a; \boldsymbol{\alpha}^*)} \Phi_1^*(a, a^*; \boldsymbol{\theta}^*)\right) \widehat{w}_1(\boldsymbol{\alpha}^*) Y_1(u)\right.$$

$$\times \exp(\boldsymbol{\beta}^{*T}(a, M_1, \boldsymbol{C}_1^T)^T)$$

$$\left.+ \left(1 - \frac{I(A_1 = a^*)\widehat{w}_1(\boldsymbol{\alpha}^*) Y_1(u)}{\pi(u, a^*; \boldsymbol{\alpha}^*)}\right) \times \int_m \exp(\boldsymbol{\beta}^{*T}(a, m, \boldsymbol{C}_1^T)^T) \Phi_1(a^*; \boldsymbol{\theta}^*) dm\right)$$

Following the conditions and the results in Lin et al. (2000), who derived $\widehat{\boldsymbol{\beta}} \xrightarrow{a.s.} \boldsymbol{\beta}^*$,

$\widehat{\Lambda}_0(t; \widehat{\boldsymbol{\beta}}) \xrightarrow{a.s.} \Lambda_0^*(t)$ uniformly in $t$, then

$$\sqrt{n}(\widehat{\boldsymbol{\beta}} - \boldsymbol{\beta}^*) = \mathbb{A}_n^{\widetilde{N}}(\boldsymbol{\beta}^*)^{-1} n^{-1/2} \sum_{i=1}^n \mathbb{U}_i^{\widetilde{N}}(\boldsymbol{\beta}^*, t) + o_p(1)$$

$$\sqrt{n}\left(\widehat{\Lambda}_0(t; \widehat{\boldsymbol{\beta}}) - \Lambda_0^*(t)\right) = n^{-1/2} \sum_{i=1}^n \Xi_i(t; \boldsymbol{\beta}^*, \Lambda_0^*(t)) + o_p(1)$$

where

$$\Xi_i(t; \boldsymbol{\beta}^*, \Lambda_0^*(t))$$
$$= \int_0^t \frac{\mathrm{d}\mathcal{M}_i(u)}{S^{(0)}(\boldsymbol{\beta}^*, u)}$$
$$- \int_0^t \frac{S^{(1)}(\boldsymbol{\beta}^*, u)\mathrm{d}(\sum_{i=1}^n N_i(u))}{n(S^{(0)}(\boldsymbol{\beta}^*, u))^2} \mathbb{A}_n^{\tilde{N}}(\boldsymbol{\beta}^*)^{-1} \int_0^\tau \left(\boldsymbol{X}_i - \frac{S^{(1)}(\boldsymbol{\beta}^*, u)}{S^{(0)}(\boldsymbol{\beta}^*, u)}\right) \mathrm{d}\mathcal{M}_i(u)$$

$$\mathbb{A}_n^{\tilde{N}}(\boldsymbol{\beta}^*) = -n^{-1} \frac{\partial}{\partial \boldsymbol{\beta}^{\mathrm{T}}} \mathbb{U}^{\tilde{N}}(\boldsymbol{\beta}^*, t)\bigg|_{\boldsymbol{\beta}=\boldsymbol{\beta}^*}$$

$$\mathbb{U}_i^{\tilde{N}}(\boldsymbol{\beta}^*, t) = \int_0^t \left(\boldsymbol{X}_i - \frac{S^{(1)}(\boldsymbol{\beta}^*, u)}{S^{(0)}(\boldsymbol{\beta}^*, u)}\right) \mathrm{d}\mathcal{M}_i(u)$$

$$S^{(0)}(\boldsymbol{\beta}^*, t) = n^{-1} \sum_{i=1}^n Y_i(t) \exp(\boldsymbol{\beta}^{\mathrm{T}} \boldsymbol{X}_i)$$

$$S^{(1)}(\boldsymbol{\beta}^*, t) = n^{-1} \sum_{i=1}^n Y_i(t) \boldsymbol{X}_i \exp(\boldsymbol{\beta}^{\mathrm{T}} \boldsymbol{X}_i)$$

$$\mathcal{M}_i(t) = N_i(t) - \int_0^t Y_i(u) \exp(\boldsymbol{\beta}^{*\mathrm{T}} \boldsymbol{X}_i) \, \mathrm{d}\Lambda_0^*(u)$$

Therefore,

$$\sqrt{n} \left( \hat{Q}^{TR}\left(t; a, a^*; \boldsymbol{\alpha}^*, \boldsymbol{\theta}^*, \hat{\boldsymbol{\beta}}, \hat{\Lambda}_0(t; \hat{\boldsymbol{\beta}})\right) - \hat{Q}^{TR}(t; a, a^*; \boldsymbol{\alpha}^*, \boldsymbol{\theta}^*, \boldsymbol{\beta}^*, \Lambda_0^*(t)) \right) =$$

$$\Psi_1(\boldsymbol{\alpha}^*, \boldsymbol{\theta}^*, \boldsymbol{\beta}^*, \Lambda_0^*(t)) \mathbb{A}_n^{\tilde{N}}(\boldsymbol{\beta}^*)^{-1} n^{-1/2} \sum_{i=1}^n \mathbb{U}_i^{\tilde{N}}(\boldsymbol{\beta}^*, t) +$$

$$\int_0^t \Psi_2(u; \boldsymbol{\alpha}^*, \boldsymbol{\theta}^*, \boldsymbol{\beta}^*) n^{-1/2} \sum_{i=1}^n \mathrm{d}\Xi_i(u; \boldsymbol{\beta}^*, \Lambda_0^*(u)) + o_p(1) \qquad (3.2.3)$$

For the fourth element,

$$\sqrt{n} \left( \hat{Q}^{TR}(t; a, a^*; \boldsymbol{\alpha}^*, \boldsymbol{\theta}^*, \boldsymbol{\beta}^*, \Lambda_0^*(t; \boldsymbol{\beta}^*)) - Q(t; a, a^*) \right)$$

$$= \sqrt{n} \Bigg\{ \int_0^t \mathbb{P}_n \Bigg\{ \hat{Z}_i(u; a; \boldsymbol{\alpha}^*) \frac{\hat{f}(M_i|a^*, \boldsymbol{C}_i; \boldsymbol{\theta}^*)}{\hat{f}(M_i|a, \boldsymbol{C}_i; \boldsymbol{\theta}^*)} [d\tilde{N}_i(u) - \hat{E}(d\tilde{N}_i(u)|a, M_i, \boldsymbol{C}_i; \boldsymbol{\beta}^*)]$$
$$+ \hat{Z}_i(u; a^*; \boldsymbol{\alpha}^*) [\hat{E}(d\tilde{N}_i(u)|a, M_i, \boldsymbol{C}_i; \boldsymbol{\beta}^*) - d\hat{Q}(u; a, a^*|\boldsymbol{C}_i; \boldsymbol{\theta}^*, \boldsymbol{\beta}^*)]$$
$$+ d\hat{Q}(u; a, a^*|\boldsymbol{C}_i; \boldsymbol{\theta}^*, \boldsymbol{\beta}^*) \Bigg\} - Q(t; a, a^*) \Bigg\}$$

$$= \int_0^t n^{1/2} \mathbb{P}_n \left\{ \frac{I(A_i = a)}{\pi(u, a; \boldsymbol{\alpha}^*)} \frac{\Phi_i(a^*; \boldsymbol{\theta}^*)}{\Phi_i(a; \boldsymbol{\theta}^*)} d\widetilde{N}_i(u) \right.$$

$$+ \left( \frac{I(A_i = a^*)}{\pi(u, a^*; \boldsymbol{\alpha}^*)} - \frac{I(A_i = a)}{\pi(u, a; \boldsymbol{\alpha}^*)} \frac{\Phi_i(a^*; \boldsymbol{\theta}^*)}{\Phi_i(a; \boldsymbol{\theta}^*)} \right)$$

$$\times \widehat{w}_i(\boldsymbol{\alpha}^*) Y_i(u) \left[ d\Lambda_0^*(u) \exp(\beta_A^* a + \beta_M^* M_i + \boldsymbol{\beta}_C^{*\mathrm{T}} \boldsymbol{C}_i) \right]$$

$$+ \left( 1 - \frac{I(A_i = a^*) \widehat{w}_i(\boldsymbol{\alpha}^*) Y_i(u)}{\pi(u, a^*; \boldsymbol{\alpha}^*)} \right)$$

$$\left. \times \int_m d\Lambda_0^*(u) \exp(\beta_A^* a + \beta_M^* m + \boldsymbol{\beta}_C^{*\mathrm{T}} \boldsymbol{C}_i) \Phi_i(a^*; \boldsymbol{\theta}^*) dm - dQ(t; a, a^*) \right\}$$

$$+ o_p(1)$$

$$= n^{-1/2} \sum_{i=1}^n F_{3,i}\big(t; a, a^*; \boldsymbol{\alpha}^*, \boldsymbol{\theta}^*, \boldsymbol{\beta}^*, \Lambda_0^*(u)\big) + o_p(1) \qquad (3.2.4)$$

where

$$F_3\big(t; a, a^*; \boldsymbol{\alpha}^*, \boldsymbol{\theta}^*, \boldsymbol{\beta}^*, \Lambda_0^*(u)\big)$$

$$= \frac{I(A_i = a)}{\pi(u, a; \boldsymbol{\alpha}^*)} \frac{\Phi_i(a^*; \boldsymbol{\theta}^*)}{\Phi_i(a; \boldsymbol{\theta}^*)} d\widetilde{N}_i(u)$$

$$+ \left( \frac{I(A_i = a^*)}{\pi(u, a^*; \boldsymbol{\alpha}^*)} - \frac{I(A_i = a)}{\pi(u, a; \boldsymbol{\alpha}^*)} \frac{\Phi_i(a^*; \boldsymbol{\theta}^*)}{\Phi_i(a; \boldsymbol{\theta}^*)} \right)$$

$$\times \widehat{w}_i(\boldsymbol{\alpha}^*) Y_i(u) \left[ d\Lambda_0^*(u) \exp(\beta_A^* a + \beta_M^* M_i + \boldsymbol{\beta}_C^{*\mathrm{T}} \boldsymbol{C}_i) \right]$$

$$+ \left( 1 - \frac{I(A_i = a^*) \widehat{w}_i(\boldsymbol{\alpha}^*) Y_i(u)}{\pi(u, a^*; \boldsymbol{\alpha}^*)} \right)$$

$$\times \int_m d\Lambda_0^*(u) \exp(\beta_A^* a + \beta_M^* m + \boldsymbol{\beta}_C^{*\mathrm{T}} \boldsymbol{C}_i) \Phi_i(a^*; \boldsymbol{\theta}^*) dm - dQ(t; a, a^*)$$

Based on Equations (3.2.1)–(3.2.4), $\sqrt{n} \left( \widehat{Q}^{TR}\big(t; a, a^*; \widehat{\boldsymbol{\alpha}}, \widehat{\boldsymbol{\theta}}, \widehat{\boldsymbol{\beta}}, \widehat{\Lambda}_0(t; \widehat{\boldsymbol{\beta}})\big) - Q(t; a, a^*) \right)$ can be expressed as follows:

$$\sqrt{n} \left( \widehat{Q}^{TR}\big(t; a, a^*; \widehat{\boldsymbol{\alpha}}, \widehat{\boldsymbol{\theta}}, \widehat{\boldsymbol{\beta}}, \widehat{\Lambda}_0(t; \widehat{\boldsymbol{\beta}})\big) - Q(t; a, a^*) \right) = n^{-1/2} \sum_{i=1}^n \Gamma_i(t, a, a^*) + o_p(1)$$

where

$$\Gamma_i(t, a, a^*) = F_1\big(t; a, a^*; \boldsymbol{\alpha}^*, \boldsymbol{\theta}^*, \boldsymbol{\beta}^*, \Lambda_0^*(u)\big)$$

$$\times E \left[ \frac{\exp(\boldsymbol{\alpha}^{*\mathrm{T}} \boldsymbol{V}_1) \boldsymbol{V}_1 \boldsymbol{V}_1^{\mathrm{T}}}{(1 + \exp(\boldsymbol{\alpha}^{\mathrm{T}} \boldsymbol{V}_1))^2} \right]^{-1} \boldsymbol{V}_{1,i} \left( A_i - \frac{\exp(\boldsymbol{\alpha}^{*\mathrm{T}} \boldsymbol{V}_{1,i})}{1 + \exp(\boldsymbol{\alpha}^{*\mathrm{T}} \boldsymbol{V}_{1,i})} \right)$$

$$+ F_2\big(t; a, a^*; \boldsymbol{\alpha}^*, \boldsymbol{\theta}^*, \boldsymbol{\beta}^*, \Lambda_0^*(u)\big) \mathbb{A}_n^M(\boldsymbol{\theta}^*)^{-1} \mathbb{U}_i^M(\boldsymbol{\theta}^*)$$

$$+ \Psi_1\big(\boldsymbol{\alpha}^*, \boldsymbol{\theta}^*, \boldsymbol{\beta}^*, \Lambda_0^*(t)\big) \mathbb{A}_n^{\widetilde{N}}(\boldsymbol{\beta}^*)^{-1} \mathbb{U}_i^{\widetilde{N}}(\boldsymbol{\beta}^*, t)$$

$$+ \int_0^t \Psi_2(u; \boldsymbol{\alpha}^*, \boldsymbol{\theta}^*, \boldsymbol{\beta}^*) d\Xi_i\big(u; \boldsymbol{\beta}^*, \Lambda_0^*(u)\big) + F_{3,i}\big(t; a, a^*; \boldsymbol{\alpha}^*, \boldsymbol{\theta}^*, \boldsymbol{\beta}^*, \Lambda_0^*(u)\big)$$

$$+ o_p(1)$$

because $\Gamma_i(t, a, a^*), i = 1, \ldots, n$ are independent and identically distributed. We have shown that $\sqrt{n}\left(\hat{Q}^{TR}\left(t; a, a^*; \hat{\boldsymbol{\alpha}}, \hat{\boldsymbol{\theta}}, \hat{\boldsymbol{\beta}}, \hat{\Lambda}_0(t; \hat{\boldsymbol{\beta}})\right) - Q(t; a, a^*)\right)$ is normally asymptotic with mean 0 in $\mathbb{M}_U$ (Appendix Section 3.1). Furthermore, by using the central limit theorem $\sqrt{n}\left(\hat{Q}^{TR}\left(t; a, a^*; \hat{\boldsymbol{\alpha}}, \hat{\boldsymbol{\theta}}, \hat{\boldsymbol{\beta}}, \hat{\Lambda}_0(t; \hat{\boldsymbol{\beta}})\right) - Q(t; a, a^*)\right)$ converge to a normal distribution with mean 0 and variance $E(\Gamma_i(t, a, a^*)^2)$.

## Appendix 4: Procedure for misspecified models

In our first simulation experiment, we had to select misspecified models for exposure, mediator, and outcome variables. We chose a logistic regression model as follows. The first model employed logistic regression:

$$\hat{E}(A|C_1, C_2, M_i^{scaled}) = \text{expit}(\hat{\alpha}_0 + \hat{\alpha}_{C_1}C_1 + \hat{\alpha}_{C_2}C_2 + \hat{\alpha}_M M_i^{scaled}),$$

where $M_i^{scaled} = \frac{M_i^{scaled} - \hat{\mu}_M}{\hat{\sigma}_M}$, with $\hat{\mu}_M, \hat{\sigma}_M$ representing the sample mean and sample standard deviation. Due to model misspecification, the limit values of $\hat{\alpha}_0$ and $\hat{\alpha}_C$ were not equal to the true values; that is $\hat{\alpha}_0 \xrightarrow{p} \alpha_0^* \neq \alpha_{00}$, $\hat{\alpha}_{C_1} \xrightarrow{p} \alpha_{C_1}^* \neq \alpha_{0C_1}$ and $\hat{\alpha}_{C_2} \xrightarrow{p} \alpha_{C_2}^* \neq \alpha_{0C_2}$.

By contrast, the parameters of the $M$ and $\tilde{N}(t)$ models were estimated using linear regression and Cox-type semiparametric regression:

$$\hat{E}(M|A, \boldsymbol{C}) = \theta_0' + \theta_A'A + \theta_{C_1}'C_1 + \theta_{C_2}'C_2$$
$$\hat{E}(\tilde{N}(t)|A, M, \boldsymbol{C}) = \Lambda_0'(t) \exp(\beta_A'A + \beta_M'M + \beta_{C_1}'C_1 + \beta_{C_2}'C_2)$$

where

$$\theta_0' \xrightarrow{p} \theta_{00}$$

$$\theta_A' \xrightarrow{p} \theta_{0A}$$

$$\theta_{C_1}' \xrightarrow{p} \theta_{0C_1}$$

$$\theta_{C_2}' \xrightarrow{p} \theta_{0C_2}$$

$$\Lambda_0'(t) \xrightarrow{p} \Lambda_0(t)$$

$$\beta'_A \xrightarrow{p} \beta_{0A}$$

$$\beta'_M \xrightarrow{p} \beta_{0M}$$

$$\beta'_{C_1} \xrightarrow{p} \beta_{0C_1}$$

$$\beta'_{C_2} \xrightarrow{p} \beta_{0C_2}$$

Next, we introduced measurement errors to the following estimated parameters $\widehat{\boldsymbol{\theta}}, \widehat{\boldsymbol{\beta}}, \widehat{\Lambda}_0(t)$ for iteration $r = 1, \ldots, 1000$, such that

$$\hat{\theta}_0 = \theta'_0$$
$$\hat{\theta}_A = \theta'_A + U_{1,r},$$

$$\hat{\theta}_{C_1} = \theta'_{C_1} + U_{2,r},$$
$$\hat{\theta}_{C_2} = \theta'_{C_2} + U_{3,r},$$
$$\widehat{\Lambda}_0(t) = \Lambda'_0(t),$$
$$\hat{\beta}_A = \beta'_A + U_{4,r},$$
$$\hat{\beta}_M = \beta'_M + U_{5,r},$$
$$\hat{\beta}_{C_1} = \beta'_{C_1},$$
$$\hat{\beta}_{C_2} = \beta'_{C_2},$$

where

$$U_{1,r} \sim \text{Normal}(\mu = -0.2, \sigma = 0.1),$$
$$U_{2,r} \sim \text{Normal}(\mu = 0.8, \sigma = 0.1),$$
$$U_{3,r} \sim \text{Normal}(\mu = 0.8, \sigma = 0.1),$$
$$U_{4,r} \sim \text{Normal}(\mu = -0.05, \sigma = 0.01),$$
$$U_{5,r} \sim \text{Normal}(\mu = 0.05, \sigma = 0.01),$$

As a result, $\widehat{\boldsymbol{\theta}} \xrightarrow{p} \boldsymbol{\theta}^* \neq \boldsymbol{\theta}_0$ and $\widehat{\boldsymbol{\beta}} \xrightarrow{p} \boldsymbol{\beta}^* \neq \boldsymbol{\beta}_0$ are biased.

# Appendix 5: Estimated regression parameters

Model $A$: Logistic regression

$$E(\text{drug}|\text{sex, age, bmi, and cvd\_history}) = \text{expit}(\alpha_0 + \alpha_1 \text{sex} + \alpha_2 \text{age} + \alpha_3 \text{bmi})$$

|  | est. | se(est.) | z value | p-value |
|---|---|---|---|---|
| $\hat{\alpha}_0$ | 3.058 | 0.648 | 4.717 | <0.001 |
| $\hat{\alpha}_1$ | -0.311 | 0.164 | -1.892 | 0.058 |
| $\hat{\alpha}_2$ | -0.032 | 0.006 | -4.944 | <0.001 |
| $\hat{\alpha}_3$ | -0.080 | 0.017 | -4.628 | <0.001 |

Model $M$: Linear regression

$$E(\text{eGFR}|\text{drug, sex, age, bmi, and cvd\_history})$$
$$= \theta_0 + \theta_1 \text{drug} + \theta_2 \text{sex} + \theta_3 \text{age} + \theta_4 \text{bmi} + \theta_5 \text{cvd\_history}$$

|  | est. | se(est.) | t value | p-value |
|---|---|---|---|---|
| $\hat{\theta}_0$ | 205.324 | 10.556 | 19.452 | <0.001 |
| $\hat{\theta}_1$ | -9.275 | 2.956 | -3.138 | 0.002 |
| $\hat{\theta}_2$ | -16.608 | 2.586 | -6.423 | <0.001 |
| $\hat{\theta}_3$ | -1.472 | 0.104 | -14.163 | <0.001 |
| $\hat{\theta}_4$ | -0.323 | 0.258 | -1.254 | 0.210 |
| $\hat{\theta}_5$ | 0.936 | 4.133 | 0.226 | 0.821 |

Recurrent outcome model: proportional mean model

$$E(\tilde{N}(t)|\text{drug, eGFR, sex, age, bmi, and cvd\_history})$$
$$= \Lambda_0(t) \exp(\beta_1 \text{drug} + \beta_2 \text{eGFR} + \beta_3 \text{sex} + \beta_4 \text{age} + \beta_5 \text{bmi} + \beta_6 \text{cvd\_history})$$

|  | est. | exp(est.) | se(est.) | Robust se | z value | p-value |
|---|---|---|---|---|---|---|
| $\hat{\beta}_1$ | -0.501 | 0.606 | 0.229 | 0.262 | -1.915 | 0.056 |
| $\hat{\beta}_2$ | -0.019 | 0.981 | 0.003 | 0.006 | -3.249 | 0.001 |
| $\hat{\beta}_3$ | 0.224 | 1.251 | 0.186 | 0.272 | 0.823 | 0.411 |
| $\hat{\beta}_4$ | 0.001 | 1.001 | 0.008 | 0.015 | 0.088 | 0.930 |
| $\hat{\beta}_5$ | -0.021 | 0.979 | 0.019 | 0.021 | -1.002 | 0.316 |
| $\hat{\beta}_6$ | 0.783 | 2.189 | 0.112 | 0.163 | 4.808 | <0.001 |

$\hat{\Lambda}_0(t)$:

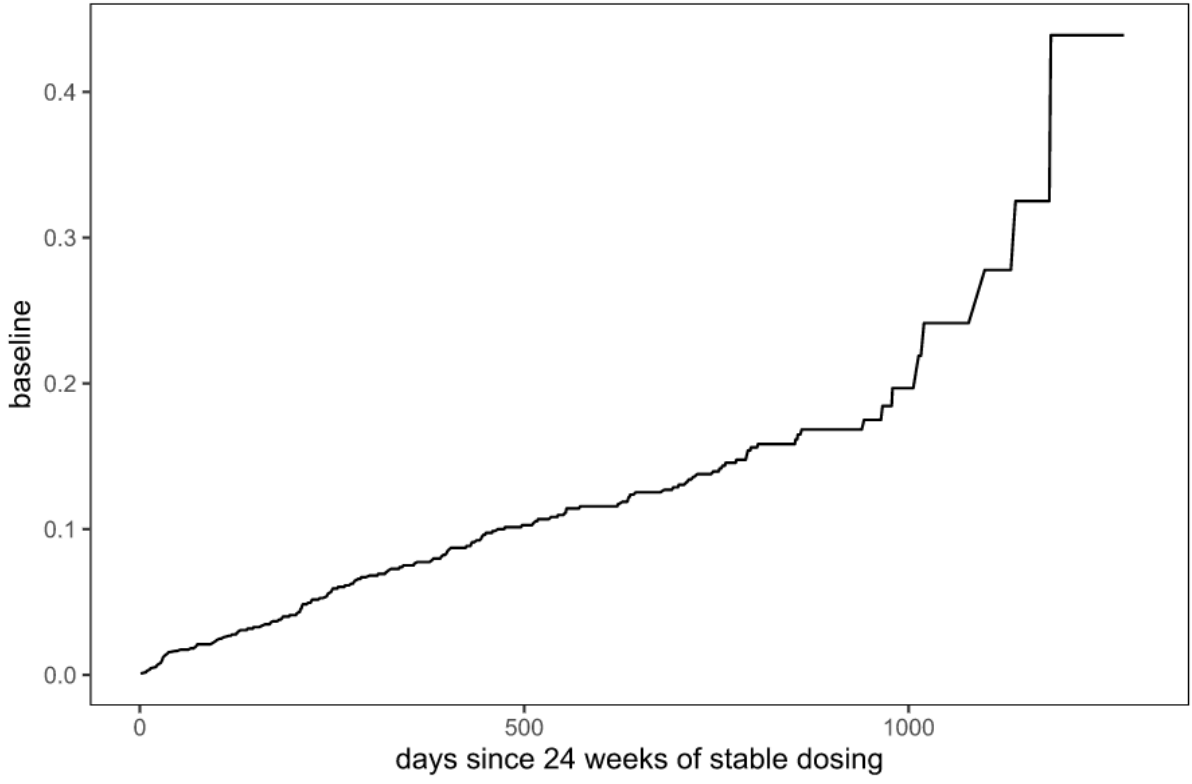